\documentclass[submission, Phys]{SciPost}

\usepackage{placeins}
\usepackage{lipsum}
\usepackage{xcolor}
\usepackage[normalem]{ulem}
\usepackage{comment}
\usepackage{tabularx}
\usepackage{graphicx}
\usepackage{dcolumn}
\usepackage{bm}
\usepackage{subcaption}
\usepackage{float}
\usepackage{booktabs}
\usepackage{dsfont}
\usepackage{enumitem}

\usepackage{bbm}
\usepackage{blindtext}
\usepackage{graphics}
\usepackage{verbatim}   
\usepackage{amsfonts}
\usepackage{amsmath}
\usepackage{amssymb}
\usepackage{adjustbox}
\usepackage{microtype} 
\allowdisplaybreaks 
\usepackage{xspace} 
\usepackage{xparse} 
\usepackage{multirow} 
\usepackage{tabstackengine}
\setstackEOL{\cr}

\usepackage{multirow}
\usepackage{physics2}
\usepackage{hyperref}
\hypersetup{colorlinks=true,breaklinks,linkcolor=blue,urlcolor=blue,citecolor=blue}
\usepackage{orcidlink}

\definecolor{darkgreen}{rgb}{0,0.5,0}
\definecolor{purple}{rgb}{0.6,0,0.5}
\definecolor{orange}{rgb}{1,0.5,0}
\definecolor{darkred}{rgb}{.7,0,0}
\definecolor{darkblue}{rgb}{0,0,.6}
\definecolor{grey}{rgb}{.6,.6,.6}
\definecolor{dimgreen}{rgb}{0.2,0.7,0.2}

\newcommand*{\ndots}{\kern-0.075em.\kern-0.05em.\kern-0.05em.}  
\newcommand*{\nidots}{.\kern-0.05em.\kern-0.05em.} 
\newcommand*{\ncdots}{\kern-0.15em\cdot\kern-0.2em\cdot\kern-0.2em\cdot\kern-0.15em}   

\NewDocumentCommand{\doubleI}{O{}}{\mathbbm{1}_{#1}}
\NewDocumentCommand{\doubleIb}{O{}}{{\overline{\mathbbm{1}}_{#1}}}
\NewDocumentCommand{\doubleIk}{O{}}{\mathbbm{1}^\ks_{\! #1}}
\NewDocumentCommand{\doubleId}{O{}}{\mathbbm{1}^\ds_{\! #1}}
\NewDocumentCommand{\doubleIp}{O{}}{\mathbbm{1}^\ps_{\! #1}}
\NewDocumentCommand{\doubleV}{O{}}{\mathbbm{V}_{\! #1}}
\NewDocumentCommand{\doubleVk}{O{}}{\mathbbm{V}^\ks_{\! #1}}
\NewDocumentCommand{\doubleVd}{O{}}{\mathbbm{V}^\ds_{\! #1}}
\NewDocumentCommand{\doubleVp}{O{}}{\mathbbm{V}^\ps_{\! #1}}
\NewDocumentCommand{\doublev}{o}{{\mathbbm{v}_{#1}}}
\NewDocumentCommand{\doubleVb}{o}{{\overline{\mathbbm{V}}_{\! #1}}}
\NewDocumentCommand{\doubleVt}{o}{{\widetilde{\mathbbm{V}}_{\! #1}}}
\NewDocumentCommand{\doubleVh}{o}{\widehat{{\mathbbm{V}}_{\! #1}}}
\NewDocumentCommand{\doubleW}{o}{\mathbbm{W}_{\! #1}}
\NewDocumentCommand{\doubleWk}{o}{\mathbbm{W}^\ks_{\! #1}}
\NewDocumentCommand{\doubleWd}{o}{\mathbbm{W}^\ds_{\! #1}}
\NewDocumentCommand{\doubleWb}{o}{{\overline{\mathbbm{W}}_{\! #1}}}
\NewDocumentCommand{\doubleWt}{o}{{\widetilde{\mathbbm{V}}_{\! #1}}}
\NewDocumentCommand{\doubleWh}{o}{{\widehat{\mathbbm{V}}_{\! #1}}}

\newcommand{\jvdomit}[1]{}

\newcommand{\TUVienna}{Institute of Solid State Physics, TU Wien, 1040 Vienna, Austria}

\usepackage{physics}
\usepackage{hyperref}
\usepackage{cleveref}
\crefrangeformat{equation}{Eqs.~(#3#1#4) to (#5#2#6)}
\crefformat{equation}{Eq.~(#2#1#3)}
\crefmultiformat{equation}{Eqs.~(#2#1#3)}
{,~(#2#1#3)}{,~(#2#1#3)}{,~(#2#1#3)}
\crefformat{table}{Tab.~#2#1#3}
\crefmultiformat{table}{Tabs.~#2#1#3}
{,~#2#1#3}{,~#2#1#3}{,~#2#1#3}
\crefrangeformat{table}{Tabs.~#3#1#4 to #5#2#6}
\crefformat{figure}{Fig.~#2#1#3}
\crefmultiformat{figure}{Figs.~#2#1#3}
{,~#2#1#3}{,~#2#1#3}{~#2#1#3}
\crefrangeformat{figure}{Figs.~#3#1#4 to #5#2#6}
\crefformat{section}{Sec.~#2#1#3}
\crefmultiformat{section}{Secs.~#2#1#3}
{,~#2#1#3}{,~#2#1#3}{,~#2#1#3}
\crefformat{appendix}{App.~#2#1#3}
\crefmultiformat{appendix}{Apps.~#2#1#3}
{,~#2#1#3}{,~#2#1#3}{,~#2#1#3}
\usepackage{bbold}

\begin{document}
\begin{center}{\Large \textbf{
Stabilizing the parquet problem 
}}\end{center}

\begin{center}
Herbert Eßl\orcidlink{0009-0005-9883-8104}\textsuperscript{1*},
Stefan Rohshap\,\orcidlink{0009-0007-2953-8831}\textsuperscript{1*},
Marcel Gievers\orcidlink{0000-0002-6951-7003}\textsuperscript{1},
Markus Wallerberger\orcidlink{0000-0002-9992-1541}\textsuperscript{1},
Alessandro Toschi\orcidlink{0000-0001-5669-3377}\textsuperscript{1},  
Anna Kauch\,\orcidlink{0000-0002-7669-0090}\textsuperscript{1}
\end{center}

\begin{center}
{\bf 1} \TUVienna
\\
* Authors contributed equally (alphabetical order)
\\
herbert.essl@tuwien.ac.at\\
stefan.rohshap@tuwien.ac.at
\end{center}

\begin{center}
\today
\end{center}

\section*{Abstract}
{\bf
We systematically analyze the stability of the iterative solution of the parquet equations by studying the spectrum of the Jacobian associated with the commonly used damped fixed-point iteration procedure. In this context, we provide an explicit criterion that determines when the physical fixed point of the parquet iteration becomes unstable. Importantly, we demonstrate that misleading convergence issues, observed in parquet calculations at intermediate-to-high interaction values, are not restricted to parameter regions where the two-particle irreducible vertex diverges, but can also arise in absence of vertex divergences. Hence, the misleading convergence issues of parquet-based algorithms are not directly caused by the crossings of two solutions of the (multivalued) Luttinger--Ward functional, that are associated with vertex divergences.  Building on these insights, we apply the controlled stabilization strategy of Phys.\ Rev.\ Lett.\ \textbf{137}, 016502 (2026) to the challenging case of the parquet iteration. This strategy allows the convergence to the physical solution in the instability regimes. We apply this procedure to the zero-point model and the Hubbard model in the atomic limit, where we successfully stabilize the physical solution deep in the non-perturbative regime, even across multiple divergence lines. Further, we compare our stabilization strategy to alternative fixed-point solvers that can be used to solve the parquet problem.
}

\vspace{10pt}
\noindent\rule{\textwidth}{1pt}
\tableofcontents\thispagestyle{fancy}
\noindent\rule{\textwidth}{1pt}
\vspace{10pt}

\section{Introduction}
{Self-consistent methods often suffer from convergence problems, particularly in the  challenging, strong-coupling regime. In a recent paper~\cite{Essl2026}, the authors analyzed the origin of convergence problems in self-consistent schemes and designed a strategy to achieve convergence to the physical solution, based on the analysis of the corresponding Jacobian. Building upon the findings of Ref.~\cite{Essl2026}, 
we here present a practical application of the Jacobian-based strategy to analyze the stability of a fixed-point of the iterative scheme. As example, we choose the self-consistent set of parquet equations~\cite{Dominicis64,Dominicis64-2, Bickers04}, which suffers from misleading convergence problems, but is not directly affected by the multivaluedness of the Luttinger--Ward functional (LWF)~\cite{Kozik2015,Rossi2015,Kim2020}, since the iterative parquet scheme does not rely on an explicit resummation of the skeleton expansion. Thus, the parquet equations represent a different situation from the self-consistent (bold) perturbation expansion, which is directly derived using the skeleton expansion of the LWF~\cite{Kozik2010}.}      

{The parquet formalism~\cite{Dominicis64,Dominicis64-2, Bickers04} is a two-particle self-consistent diagrammatic approach and a method of choice for problems involving a delicate interplay of fluctuations of different electronic degrees of freedom: charge, spin, and orbital, which gives rise to the  rich physics of strongly correlated electron systems}. It allows not only for an almost quantitatively exact description of the correlated electrons in the weak-coupling regime of the Hubbard model \cite{Schafer2021}, but also for diagnostics of fluctuations in terms of physical channels~\cite{Schaefer2021}, as was applied to optical conductivity~\cite{Kauch2020, Pudleiner2019a, Pudleiner2019b} and the self-energy in the intriguing pseudogap regime of the Hubbard model~\cite{Lihm2026}. The numerical advancements in the solution of the self-consistent set of parquet equations~\cite{YangPRE09, Tam2013, Li16,Wentzell2020, victory2019, Eckhardt2020, Krien2020b, Krien2022, Rohshap2025a,Lihm2026}, together with the possibility to use strong-coupling input with \textit{ab initio} obtained parameters, in the framework of the dynamical vertex approximation (D$\Gamma$A)~\cite{Toschi2007}, make the method now applicable for realistic lattice problems beyond weak coupling. However, already in the pioneering applications for lattice systems~\cite{Tam2013}, it was observed that the convergence of the self-consistent scheme is very fragile.

In this work, we systematically investigate the source of instabilities of the iterative parquet scheme by studying the
spectrum of the Jacobian associated with the damped fixed-point iteration.
Interestingly, we discover that the instabilities occur also in parameter regimes that are free from vertex divergences~\cite{Schaefer2013, Gunnarsson2017, Springer2020}. After identifying the exact source of the instability in the Jacobian, we construct a generally applicable stabilization strategy, that is, in its spirit, similar to the heuristic trick of flipping  the sign of a Matsubara frequency applied in Ref.~\cite{Kozik2015}. Here, in a more rigorous and general context, what the proposed scheme essentially does, is flipping the sign of the unstable eigendirections of the Jacobian~\cite{Essl2026}. This scheme leads to a \emph{local} stabilization of the physical fixed point. 
With this approach, we show that it is possible to converge to the physical solution of the parquet equations deep in the non-perturbative region of the Hubbard atom (HA) and the zero-point (ZP) model, given a sufficiently good starting point of the iteration. 

Additionally, we  analyze cases where the physical solution can already be achieved by damping the fixed-point iteration and show (i) how to distinguish them from those cases where elaborate Jacobian-based approaches  are needed, and (ii) how to choose the appropriate damping. 

Further, we investigate alternative fixed-point algorithms with regard to whether the physical fixed point is kept stable at strong coupling. Specifically, we study the Chord method, Anderson acceleration, Broyden's method, and the Jacobian-free Newton--Krylov method, which could provide possible algorithmic alternatives to our proposed stabilization scheme.

Since the instability of the iterative parquet scheme was originally ascribed to vertex divergences, we analyze in detail the relation between the two. We conclude that the divergence of the two-particle irreducible vertex is a sufficient but not a necessary condition for the instability of the  physical fixed point in the damped iteration parquet scheme. In particular, we also present an inverted approach, in which the full vertex, free from divergences, is iterated instead of its potentially divergent (ir)reducible parts. Interestingly, the stability of this approach is almost inverted in comparison to the usual iteration. The number of instabilities is maximal at weak coupling and decreases towards strong coupling; therefore, we dubbed this alternative iteration \emph{strong-coupling iteration}.

This work focuses mainly on the introduction and numerical implementation of the stabilization strategy for the parquet problem, while instabilities in the Jacobian for various self-consistent diagrammatic approaches are investigated in a related work~\cite{Gievers2026Instabilities}.

The paper is organized as follows. We start with the introduction and definitions of the ZP model and the HA (Sec.~\ref{sec:models}) as well as one- and two-particle Green's functions and their relation to the two-particle vertex (Sec.~\ref{sec:def_Gs}). Further, in Sec.~\ref{sec:parquet-equations}, we introduce the parquet equations and the iterative scheme to solve them, including explicit expressions for the Jacobian and analysis of the stability of its eigendirections. In Sec.~\ref{sec:vertex_div}, we give an overview of the studies of vertex divergences and their connection to the branching of the LWF and the misleading convergence. Section~\ref{sec:parquet-sot} is the central part of the manuscript where we present the stabilization strategy: first on a simple example in Sec.~\ref{sec:motivation} and then concretely for the parquet problem in Sec.~\ref{sec:sot}. In the following sections, we apply the stabilization strategy to the ZP model (results in Sec.~\ref{sec:ZP_parquet}) and the HA (Sec.~\ref{sec:HA_parquet}), as well as to the strong-coupling iteration in the ZP model in Sec.~\ref{sec:strong_parquet}. We also investigate alternative routes to reaching the physical solution and compare their convergence speed in Sec.~\ref{sec:method_comparison}. We provide an outlook and describe possible future strategies to overcome the numerical complexity of the stabilization scheme in Sec.~\ref{sec:practical_implementation} and we finalize the discussion with conclusions in Sec.~\ref{sec:conclusion}. 
Further, in \cref{app:defs}, we give the definitions of one- and two-particle quantities, the Bethe--Salpeter equations, the parquet decomposition and the analytical vertices for the ZP model.
In \cref{app:parquet}, we  provide explicit expressions for the iterative maps and the corresponding Jacobian for both the conventional parquet iteration and the strong-coupling iteration. A channel-resolved analysis of the instabilities for the conventional parquet iteration and the strong-coupling iteration is  performed in \cref{app:decoupled_channels} for both the ZP model and the HA. In \cref{app:extended_results_HA}, we  present further numerical details on the calculations for the HA.

\section{Models and formalism}
\label{sec:models_and_formalism}
\subsection{Models}
\label{sec:models}
In this section, we introduce two analytically solvable models that are used to provide a proof of principle for the developed technique, namely the zero-point (ZP) model \cite{Rossi2015} and the Hubbard atom (HA).
However, it is conceptually straightforward to apply our formalism to more general models by simply replacing the Matsubara frequencies with 4-momenta\footnote{Note that also a generalization to multi-orbital models is formally straightforward.}.

\subsubsection{Zero-point model}

Let us start by introducing the zero-point (ZP) model \cite{Rossi2015}, which is an analytically solvable model where the spin of the electrons is the only degree of freedom. The ZP model is often considered in fundamental studies of non-perturbative effects, as it entails the multivalued features of the LWF in a particularly transparent form. 
Physically, the (repulsive) ZP model resembles the coherent potential approximation (CPA) solution of the binary mixture model in infinite dimensions which features a similar metal-insulator transition by increasing interaction at half-filling \cite{Schaefer2016,Essl2026}. Since the ZP model has neither frequency nor momentum dependence, all $N$-particle Green's functions are just complex scalars. The action of the model reads
\begin{align}
\label{eq:S_ZP}
    S_\text{ZP}\left[G_0^{-1}\right] = -\sum_{\sigma} \bar c_{\sigma} G_{0}^{-1}  c_{\sigma} + U \bar c_{\uparrow}c_{\uparrow} \bar c_{\downarrow}c_{\downarrow},
\end{align}
with $G_{0}^{-1}=\delta\mu$ ($\delta\mu\in\mathbb{C}$). 
This model is exactly solvable and we provide all needed quantities for this work in \cref{app:ZP_quantities}.
In the spirit of Ref.~\cite{Kim2020}, where $N$-replica of the ZP model are connected to the HA, we use a complex chemical potential $\delta\mu$, where the imaginary part mimics a Matsubara frequency and a finite real part a chemical potential deviating from particle-hole (ph)-symmetry. For the purpose of this paper, we fix the imaginary part $\mathrm{Im}(\delta\mu)=\pi$ and vary only the real part of $\delta\mu$.

\subsubsection{Hubbard atom}
The Hubbard atom (HA), which represents the strong-coupling limit of the Hubbard model is defined by the following Hamiltonian:
\begin{align} \label{eq:hamiltonian-hubbard-atom}
    \hat{H}_\text{HA} =U (\hat{n}_{\uparrow}-1/2)( \hat{n}_{\downarrow} -1/2)- \delta\mu(\hat{n}_{ \uparrow}+ \hat{n}_{\downarrow}),
\end{align}
where $\hat n_{\sigma} = \hat{c}_{\sigma}^{\dagger} \hat{c}_{\sigma}$, and $\hat{c}_{\sigma}^{(\dagger)}$  is the fermionic annihilation (creation) operator for an electron with spin $\sigma$. The parameter $U$ quantifies the on-site Coulomb repulsion between two electrons, and $\delta\mu$ is the chemical potential deviation from half-filling. The Hubbard atom can be regarded as the limiting case of the Hubbard model in which all hopping amplitudes are set to zero, effectively decoupling the lattice sites. Although this represents a severe simplification, the model captures several essential features of the strong-coupling limit of the full Hubbard model \cite{Thunstroem2018} and is, therefore, frequently used as a benchmark in the development of new numerical tools, such as quantics tensor trains~\cite{Rohshap2025a, Shinaoka2023, Grosso2026}, compact basis representations~\cite{Wallerberger2021}, or the functional renormalization group~\cite{AlEryani2026Functional}. With no inter-site hopping, the only relevant energy scales are $U$, $\delta\mu$ and the temperature $T$. Therefore, we will use the inverse temperature $\beta=1/T=1$ as our energy scale.

\subsection{Definition of one and two-particle Green's functions}\label{sec:def_Gs}
We start by introducing the one-particle Green's function in Matsubara frequency space $G_{\sigma}({\nu})$, which is defined as the Fourier transform of the imaginary time ordered two-point correlation function:
\begin{align}
    G_{\sigma}(\nu) = -\int_0^{\beta}\!\!d\tau\, e^{i \nu \tau} \langle T_{\tau} \hat{c}_{ \sigma}(\tau) \hat{c}_{\sigma}^{\dagger}(0) \rangle \,,
\end{align}
with imaginary time $\tau$ and fermionic Matsubara frequencies $\nu = (2n+1)\pi / \beta, n \in \mathbb{Z}$. 

Further, Fourier transform of the imaginary time-ordered four-point correlator leads to the two-particle Green's function in Matsubara frequencies:
\begin{align}
\label{Eq:2p-G}   G_{\sigma_1\ldots\sigma_4}^{\nu_1\nu_2\nu_3} = \int_0^{\beta}\!\!\!d\tau_1 \int_0^{\beta}\!\!\!d\tau_2\int_0^{\beta}\!\!\!d\tau_3\;e^{i \nu_1 \tau_1+i\nu_2 \tau_2+i \nu_3 \tau_3 
    }  \langle T_{\tau} \hat{c}_{ \sigma_1}(\tau_1) \hat{c}_{\sigma_2}^{\dagger}(\tau_2)\hat{c}_{ \sigma_3}(\tau_3) \hat{c}_{\sigma_4}^{\dagger}(0) \rangle,
\end{align}
with the three fermionic Matsubara frequencies $\nu_1,\nu_2, \nu_3$.

The two-particle Green's function  can be decomposed into so-called \textit{disconnected} and \textit{connected} terms
\begin{align}   G_{\sigma_1\ldots\sigma_4}^{\nu\nu'\omega} = &\;
    G_{\sigma_1}(\nu) G_{\sigma_3}(\nu')\delta_{\omega 0}\,\delta_{\sigma_1\sigma_2} \delta_{\sigma_3\sigma_4}
    - G_{\sigma_1}(\nu) G_{\sigma_2}(\nu+\omega)\delta_{\nu \nu'}\,\delta_{\sigma_1\sigma_4} \delta_{\sigma_2\sigma_3} \nonumber \\
    & -
    G_{\sigma_1}(\nu) G_{\sigma_2}({\nu+\omega}){F}_{\sigma_1\ldots\sigma_4}^{\nu \nu'\omega}G_{\sigma_3}(\nu'+\omega)G_{\sigma_4}(\nu'),
    \label{eq:Fdef1}
\end{align}
where the first two terms are products of one-particle Green's functions, and therefore are disconnected, and the third term is the connected contribution, with  $F$, the full two-particle vertex. Equation~\eqref{eq:Fdef1} defines $F$ and $  G_{\sigma_1\ldots\sigma_4}^{\nu\nu'\omega}$ in the particle-hole (ph) frequency notation, with the following assignment: $\nu=-\nu_1$, $\nu'=-(\nu_1+\nu_2+\nu_3)$, $\omega=\nu_1+\nu_2$, where $\omega$ is a bosonic Matsubara frequency $\omega={2n\pi}/{\beta}$, $n \in \mathbb{Z}$ (for details see App.~\ref{app:defs}).

\subsection{Parquet formalism}\label{sec:parquet-equations}
The parquet formalism establishes a set of exact relations linking different classes of two-particle vertex functions, as well as connecting the self-energy to the full two-particle vertex $F$~\cite{Dominicis64, Dominicis64-2}. A comprehensive introduction to the underlying framework can be found in Refs.~\cite{Bickers04, Rohringer2012, Rohringer18, Held2022, victory2019}.
In the main text, we limit ourselves to exemplary discuss one channel of the parquet formalism, namely the d(ensity) channel which is formulated in the ph-frequency notation and is, in the SU(2) symmetric case, the sum over two spin combinations, i.e., $F_\text{d}=F_{\uparrow\uparrow\uparrow\uparrow}+F_{\uparrow\uparrow\downarrow\downarrow}$. The full vertex  consists of  reducible and irreducible diagrams in a given channel: $F_\mathrm{d} = \Phi_\mathrm{d} + \Gamma_\mathrm{d}$, with $\Phi_\mathrm{d}$, the 2-particle reducible, and $\Gamma_\mathrm{d}$, the 2-particle irreducible vertex in the density channel.
Specifically, the set of coupled equations for the parquet formalism consists of the Bethe--Salpeter equations \eqref{eq:bse-density-main} (BSEs) and parquet decomposition \cref{eq:parquet-density-main}, Schwinger--Dyson equation (SDE) \eqref{eq:sde-main} and the Dyson equation \eqref{eq:dyson-main}:
\begin{align}
    \Phi_\mathrm{d}^{\nu \nu' \omega} &=   \frac{1}{\beta} \sum_{\nu_1} \Gamma_\mathrm{d}^{\nu \nu_1 \omega} G(\nu_1)G(\nu_1+\omega) \left(\Phi_\mathrm{d}^{\nu_1 \nu' \omega}+\Gamma_\mathrm{d}^{\nu_1 \nu' \omega}\right)\quad\text{BSE (d-channel)}, \label{eq:bse-density-main}\\
    \Gamma_\mathrm{d}^{\nu \nu' \omega} &= \Lambda_\mathrm{d}^{\nu \nu' \omega} -\frac{1}{2} \Phi_\mathrm{d}^{\nu (\nu+\omega) (\nu'-\nu)} - \frac{3}{2} \Phi_\mathrm{m}^{\nu (\nu+\omega) (\nu'-\nu)} \nonumber \\ 
    &\quad + \frac{1}{2} \Phi_\mathrm{s}^{\nu \nu' (-\omega -\nu -\nu')} + \frac{3}{2} \Phi_\mathrm{t}^{\nu \nu' (-\omega -\nu -\nu')}\quad\text{parquet decomposition (d-channel)},\label{eq:parquet-density-main}\\
    \Sigma (\nu) &= \frac{U}{\beta}\sum_{\nu'}G(\nu') \nonumber\\
    &\quad- \frac{U}{2\beta^2} \sum_{\nu' \omega} (\Phi_\mathrm{d}^{\nu \nu' \omega}+\Gamma_\mathrm{d}^{\nu \nu' \omega} - \Phi_\mathrm{m}^{\nu \nu' \omega}-\Gamma_\mathrm{m}^{\nu \nu' \omega}) G(\nu') G(\nu'+\omega) G(\nu+\omega)\quad\text{SDE},\label{eq:sde-main}\\
    G(\nu) &= \frac{1}{G^{-1}_{0}(\nu)- \Sigma(\nu)}\quad\text{Dyson equation},\label{eq:dyson-main}
\end{align}
where Eqs.~\eqref{eq:parquet-density-main} and~\eqref{eq:sde-main} also contain vertices reducible ($\Phi$) and irreducible ($\Gamma$) in other channels as well as $\Lambda$, which is the fully irreducible vertex. Further, $G$, $\Sigma$, and $G_0$ are the full Green's function, the self-energy and the non-interacting Green's function. The subscripts $r=\text{d,m,s,t}$ refer to the spin-diagonalized density, magnetic, singlet and triplet channels. A detailed overview of our conventions can be found in Ref.~\cite{Rohshap2025a} and \cref{app:defs,app:parquet}. The parquet formalism itself is exact; in practice, an approximation enters only through the choice of the fully irreducible vertex $\Lambda$. Two common choices are the parquet approximation, where $\Lambda_r$ is replaced by the bare interaction ($\Lambda_\text{d}=U$, $\Lambda_\text{m}=-U$, $\Lambda_\text{s}=2U$, $\Lambda_\text{t}=0$) \cite{Bickers04}, and D$\Gamma$A \cite{Toschi2007}, where $\Lambda$ of the auxiliary impurity model of DMFT is used. In this work, we always use the \emph{exact} $\Lambda$ of the respective model, mimicking the non-perturbative input of D$\Gamma$A. This choice is essential, since such a non-perturbative input is required to investigate the connection between vertex divergences and the instabilities of the parquet iteration.

\subsubsection{Iterative parquet scheme}\label{sec:iterparquet}

Using a fixed input for the fully irreducible vertex $\Lambda$, the non-interacting Green's function $G_0$, the onsite interaction $U$, and initial inputs for $\Phi_r$ and $G$, the set of parquet equations is solved iteratively in the following way: (i) Calculate $\Gamma_r$ with parquet decomposition \cref{eq:parquet-density-main,eq:irreducible_vertices}. (ii) Calculate $\Sigma$ via SDE \eqref{eq:sde-main}. (iii) Calculate $\Phi_r$ via the BSEs \cref{eq:bse-density-main,eq:bse} and $G$ with the Dyson equation \eqref{eq:dyson-main}. (iv) Obtain a self-consistent solution of (i)--(iii). 

\begin{figure}
    \centering
    \includegraphics[scale=0.6]{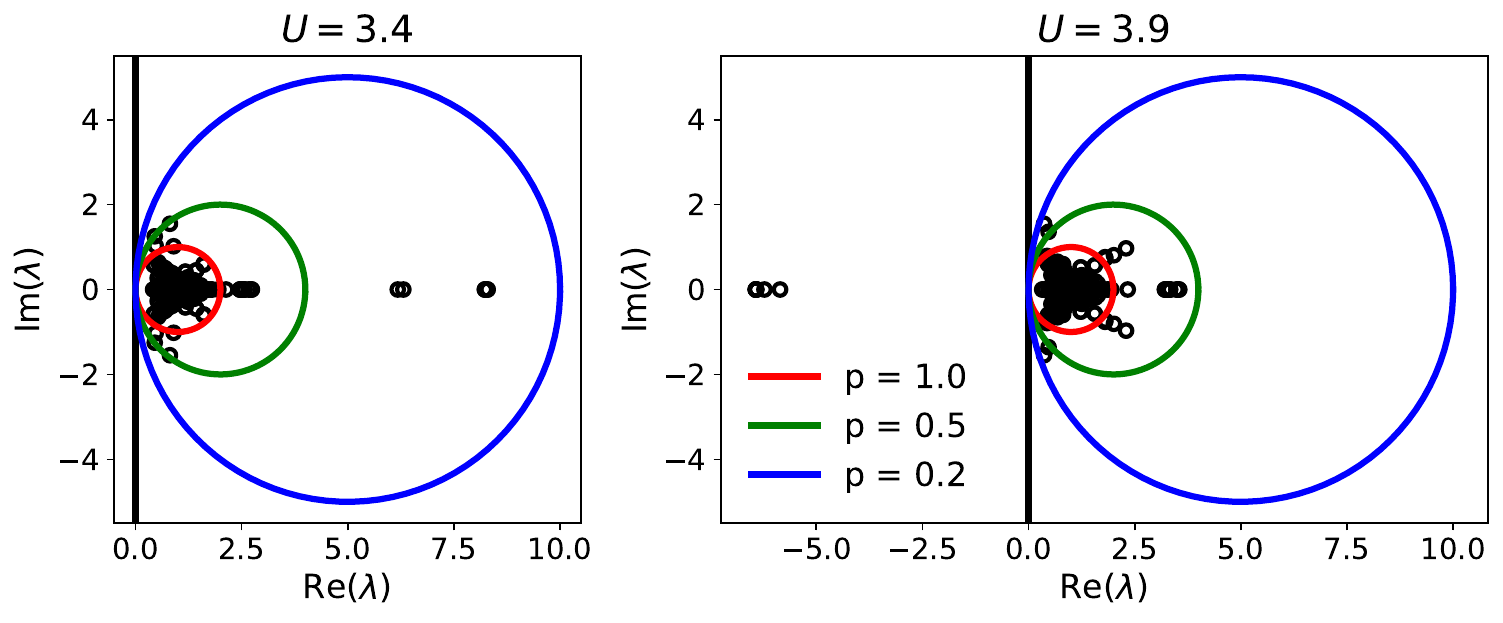}
    \caption{The eigenvalues $\lambda$ (black circles) for the parquet formalism of the HA for two different values of $U$ at ph-symmetry in the complex plane. Colorful circles show the stability regions for given damping parameters $p$. Left: Fixed point can be stabilized with damping, i.e., by decreasing $p$. Right: Fixed point cannot be stabilized with damping since some eigenvalues have a negative real part. The vertical black line marks where $\mathrm{Re}(\lambda)=0$.}
    \label{fig:stab_circle}
\end{figure}

In practice, this self-consistency is often reached by utilizing a damped iterative scheme 
\begin{align}
\label{eq:Psi}
    \Psi^{(n+1)}=p\,\mathcal{F}[\Psi^{(n)}]+(1-p)\Psi^{(n)},
\end{align}
where the iterated abstract state $\Psi=(\Phi_\text{d},\Phi_\text{m},\Phi_\text{s},\Phi_\text{t},G)$ consists of the complex reducible vertices and the one-particle Green's function, and $p\in(0,1]$ is the damping parameter. Further, the iterative map $\mathcal{F}=(f_\text{d},f_\text{m},f_\text{s},f_\text{t},f_\text{G})$ consists of the BSEs~\eqref{eq:bse} and the SDE~\eqref{eq:sde} (see \cref{app:parquet_J}). 
Note that for a formal treatment of the damped iteration, it is important that $\mathcal{F}$ only depends on the iterated quantities $\Phi_\text{d,m,s,t}$ and $G$, and  the fixed inputs $\Lambda_\text{d,m,s,t}$, $U$, and $G_0$.
The iteration of Eq.~\eqref{eq:Psi} leads to the convergence (if convergence is reached at all) to an attractive fixed point $\Psi^*$ of $\mathcal{F}$, i.e., $\mathcal{F}[\Psi^*]=\Psi^*$. As can be already seen on the level of the scalar BSEs~\eqref{eq:ZP_f_d} and \eqref{eq:ZP_f_m} in the ZP model, the BSEs of the density and magnetic channel are quadratic equations in $\Phi_\text{d/m}$ leading to multiple fixed points. Therefore, convergence of the method, \textit{per se}, does not mean that the physical solution is obtained. 
Moreover, in the case of the physical fixed point being repulsive, the solution cannot be obtained by damped iteration. 
However, by investigating the Jacobian of the fixed-point iteration in Eq.~\eqref{eq:Psi}, we can determine whether the physical fixed point is repulsive or attractive. Specifically, the fixed point is repulsive if there exists an eigenvalue of the Jacobian that is greater than $1$ in magnitude.
The corresponding Jacobian reads
\begin{align}
\label{eq:J_general}
    J[\Psi^*]=\fdv{\left(p\mathcal{F}[\Psi]+(1-p)\Psi\right)}{\Psi}\Bigg|_{\Psi^*}=:\mathbb{1}-p\,\Pi[\Psi^*],
\end{align}
where
\begin{align}
\label{eq:dF_dPsi}
    \fdv{\mathcal{F}[\Psi]}{\Psi} = \begin{pmatrix}
        \fdv{f_\text d}{\Phi_\text d}&\fdv{f_\text d}{\Phi_\text m}&\fdv{f_\text d}{\Phi_\text s}&\fdv{f_\text d}{\Phi_\text t}&\fdv{f_\text d}{G}\\
        \fdv{f_\text m}{\Phi_\text d}&\fdv{f_\text m}{\Phi_\text m}&\fdv{f_\text m}{\Phi_\text s}&\fdv{f_\text m}{\Phi_\text t}&\fdv{f_\text m}{G}\\
        \fdv{f_\text s}{\Phi_\text d}&\fdv{f_\text s}{\Phi_\text m}&\fdv{f_\text s}{\Phi_\text s}&\fdv{f_\text s}{\Phi_\text t}&\fdv{f_\text s}{G}\\
        \fdv{f_\text t}{\Phi_\text d}&\fdv{f_\text t}{\Phi_\text m}&\fdv{f_\text t}{\Phi_\text s}&\fdv{f_\text t}{\Phi_\text t}&\fdv{f_\text t}{G}\\
        \fdv{f_\text G}{\Phi_\text d}&\fdv{f_\text G}{\Phi_\text m}&\fdv{f_\text G}{\Phi_\text s}&\fdv{f_\text G}{\Phi_\text t}&\fdv{f_\text G}{G}\\
    \end{pmatrix},
\end{align}
and the matrix $\Pi[\Psi^*]=(\mathbb{1}-\frac{\delta \mathcal{F}}{\delta\Psi}\big|_{\Psi^*})$ is introduced for convenience.
Explicit expressions for the components of the Jacobian and further details can be found in App.~\ref{app:parquet_J}.
Note that, in contrast to continuous-time quantum Monte Carlo methods \cite{Gull2011} or diagrammatic Monte Carlo approaches \cite{Kozik2010,VanHoucke2010}, the form of $\mathcal{F}$ is known analytically for the parquet scheme and therefore the corresponding Jacobian expression can be calculated analytically.
We will denote the eigenvalues of the Jacobian as $1-p\lambda$, where $\lambda$ are the eigenvalues of $\Pi$. 
Evidently, the Jacobian has an eigenvalue with magnitude greater than $1$ if Re$(\lambda)<0$ is fulfilled, independently of the chosen damping $p$. However, if the magnitude of an eigenvalue of  $J$ is greater than $1$ despite all real parts of $\lambda$ being positive, the damping can be increased (smaller $p$) until the fixed point becomes locally stable.
Geometrically, the stability criterion can be represented as a circle in the complex plane of $\lambda$ with radius $1/p$ and center at $1/p$. 
If all eigenvalues $\lambda$ lie within this circle, the fixed point is locally stable. 
In Fig.~\ref{fig:stab_circle}, this is visualized for one stable and one unstable parameter point.

By analyzing the eigenvalues of the Jacobian, the required minimal damping for locally stabilizing the fixed point (if all Re$(\lambda)>0$) can be derived by
\begin{align}
\label{eq:uniform_damping}
    p=\min\left[1,\min_\alpha \left(c\frac{2|\text{Re}(\lambda_\alpha)|}{|\lambda_\alpha|^2}\right)\right],
\end{align}
where $c\in(0,1)$ is an additional factor smaller than $1$ such that the largest eigenvalue is guaranteed to be smaller than $1$ in magnitude. Typically, we find that a smaller $c$ value increases the size of the stability region, i.e., the non-local region for which the fixed point is stable.
With this condition, the damping can be chosen as small as necessary to stabilize the fixed point, while avoiding unnecessary slow convergence due to choosing a too small damping parameter.
From the exact expressions in \cref{app:parquet_J}, it follows that $\Pi=\mathbb{1}$ at $U=0$ (if evaluated at the physical fixed point), which means that at weak coupling the physical fixed point is stable.
However, since $\Pi$ is, in general, a $\kappa$-real matrix (see \cref{app:parquet_J}), i.e., has eigenvalues that are either real or come in complex conjugate pairs, there are three possible ways how $\Pi$ can develop negative eigenvalues: 
\begin{enumerate}[label=(\roman*)]
    \item An eigenvalue of $\Pi$ can feature an odd pole.
    \item A real eigenvalue of $\Pi$ can cross $0$.
    \item The real part of a complex conjugated pair of eigenvalues of $\Pi$ can cross $0$.
\end{enumerate}

\subsection{Connection between misleading convergence, vertex divergences and branching of the LWF}\label{sec:vertex_div}

The term misleading convergence was originally ascribed to the multivaluedness of the LWF that was observed in bold-line (self-consistent) diagrammatic Monte Carlo for the Hubbard model~\cite{Kozik2015}, and also in the two models used in this work, i.e., the ZP model~\cite{Rossi2015,Stan2015,vanHoucke2024} and the HA~\cite{Gunnarsson2017}. The crossing of two self-energies that are constructed with the different branches of the LWF implies a divergence of its second functional derivative with respect to the Green's function and hence the divergence of the two-particle irreducible vertex $\Gamma$~\cite{Gunnarsson2017}. Since after this crossing the perturbative (skeleton) branch of the LWF is no longer the physical one, this also leads to a misleading convergence of the skeleton series~\cite{Kozik2015, Gunnarsson2017, Kim2022,Essl2026,Rossi2015,vanHoucke2024}.

Since the discovery of divergences in the irreducible vertex $\Gamma$~\cite{Schaefer2013}, their formal and physical implications have been intensively discussed: They were shown to signal a failure of self-consistent perturbation theory \cite{Gunnarsson2017,Essl2026} and to constitute an early non-perturbative signature of Mott physics~\cite{Schaefer2016,Pelz2023}. Further, these singularities are connected, via the BSEs, to vanishing eigenvalues of the generalized susceptibilities. Thus, subsequent investigations in the case of the HA~\cite{Thunstroem2018,Essl2024}, the Hubbard model~\cite{Springer2020,Vucicevic2018,Meixner2026,Gunnarsson2016}, and the Anderson impurity model~\cite{Chalupa2021} linked local divergences (depending on specific fermionic Matsubara frequencies) to antisymmetric eigenvectors of the density susceptibility and global divergences to symmetric eigenvectors.
From a physical perspective, vertex divergences with \mbox{(anti-)}symmetric eigenvectors were connected to the suppression (enhancement) of the local density susceptibility~\cite{Reitner2020}. Moreover, the emergence of kinks in the electronic self-energy~\cite{Byczuk2007}, a change in the relaxation mechanism after a quench of the Hubbard interaction~\cite{Eckstein2009}, and the formation of Hubbard subbands in the spectral function~\cite{Schaefer2013}, as well as the formation of local moments~\cite{Chalupa2021, Adler2024, Mazitov2022, Stepanov2022} are connected to vertex divergences.
Further, at strong coupling the vertex divergences are necessary for a diverging charge  susceptibility in the lattice Hubbard model solved by (cluster) dynamical mean-field theory~\cite{Reitner2020,Meixner2026,Meixner2026a}.

In the parquet decomposition, the divergence of the irreducible vertex $\Gamma$ also leads to a divergence of the reducible vertices  $\Phi$ and the fully irreducible vertex $\Lambda$. It is, therefore, natural to assume that this divergence also affects the parquet iteration.
Indeed, it was \emph{empirically} observed that also the parquet equations show misleading convergence at the first vertex divergence \cite{Badr2024,Rohshap2025a}. Recently, it was demonstrated~\cite{Essl2026} that vertex divergences and pseudo-divergences lead to an instability of the iteration for self-consistent perturbation theory. 
In this context, pseudo-divergences represent a formal generalization of the vertex divergences where a complex conjugated pair of eigenvalues of $\chi_0^{-1}\chi$ has zero real part but finite imaginary part \cite{Essl2024,Vucicevic2018,Reitner2024b}.

However, a crucial difference between the parquet equations as presented in this work and self-consistent perturbation theory is the fact that the parquet equations do not represent a fully LWF-derivable theory. The reason is that, although the BSEs~\eqref{eq:bse-density-main} and the parquet decomposition~\eqref{eq:parquet-density-main} can be derived by constructing functional derivatives of the LWF~\cite{Janis1998,Janis1999, Eckhardt2023}, the complementary link between one- and two-particle functions is \emph{not} provided by a self-energy defined through a functional derivative of the LWF but through the exact SDE~\eqref{eq:sde-main}. Using the SDE, which can be rigorously derived from the equation of motion linking the one- and two-particle Green's functions, is numerically significantly better suited since functional integration of the irreducible vertex, that would have to be done otherwise, is generally not possible. A notable exception has been discussed and implemented in Ref.~\cite{Janis2017}, but only for a very special case with a \emph{constant} irreducible vertex. This constitutes a strong approximation and, therefore, to obtain spectral properties of the self-energy, the SDE was used in postprocessing anyway. The relation between the self-energy from the SDE and the one that would come from the derivative of the LWF can be investigated by analyzing the fulfillment of the related  Ward identity. This was done in the parquet approach for the Anderson impurity model~\cite{Chalupa2022} and for the Hubbard model~\cite{Wieser2023}. 
From these studies, one can conclude that the SDE-obtained self-energy numerically differs from the derivative of the LWF. In fact, it seems plausible that they agree only for the exact solution (i.e., in the case where the exact fully irreducible vertex $\Lambda$ is used as an input), as  suggested in Refs.~\cite{Smith1992,Janis2017,Kugler2018}. In practice, to our knowledge, the violation of the related Ward identity is, in general, numerically small for most of hitherto known applications of the parquet equations.
Alternatively, by using a four-particle irreducible functional, it is possible to derive a modified parquet-like  scheme with self-consistently obtained $\Lambda$ (the QUADRILEX method~\cite{Ayral2016}), in which the derivative of the LWF with respect to the one-particle Green's function and the SDE should give the same self-energy.

By investigating the Jacobian of the parquet iteration (\cref{sec:iterparquet,app:parquet_J}), we find that the empirical observation of misleading convergence in the proximity of vertex divergences is not coincidental. In fact, we observe that a vertex divergence leads to an odd pole in the eigenvalues of the Jacobian, which causes unstable eigendirections after the divergence, as described in point (i) at the end of \cref{sec:iterparquet}. Therefore, a vertex divergence implies an unstable physical fixed point. However, the converse is \emph{not} true, i.e., the physical fixed point can also acquire unstable eigendirections that are \emph{not} related to a vertex divergence, as shown in case (ii) and (iii) at the end of \cref{sec:iterparquet}.
In the subsequent \cref{sec:ZP_parquet,sec:HA_parquet}, we will demonstrate this feature on the example of our two toy models.

\section{Stabilization method}\label{sec:parquet-sot}
In this section, we will elaborate on why the iterative parquet scheme fails to converge to the physical solution at strong coupling. Further, we adapt the method that was introduced in Ref.~\cite{Essl2026} to the parquet algorithm to show a possibility on how to stabilize the physical solution. Throughout the paper, we will refer to this scheme as the \emph{stabilization method}.  

\subsection{Minimal examples}\label{sec:motivation}

As the iterative parquet scheme is a high-dimensional fixed-point problem, let us start with a pedagogical motivation. For this reason, we first discuss convergence properties of general fixed-point iterations~\cite{Argyris2017,Strogatz1994}. An equation $x=f(x)$ can be solved via fixed-point iteration,
\begin{equation}
    x^{(n+1)} = f(x^{(n)}),
\end{equation}
for the $(n+1)$-th step of the iteration scheme. Convergence to the fixed point $x^*$ is only possible if a small perturbation $\epsilon>0$ around the fixed point $x^*$ does not push the solution away. Mathematically, we can express this (local) stability of the fixed point through the following condition:
\begin{align}
    \left| f(x^*+\epsilon)-x^* \right| = \left| \epsilon \frac{df}{dx} \big|_{x=x^*} + \mathcal{O}(\epsilon^2) \right| \overset{!}{<} \epsilon.
\end{align}
This condition does not ensure global convergence to the fixed point. If the condition is not satisfied, $x^*$ is locally unstable and the fixed-point iteration cannot converge to $x^*$.

\begin{figure}
    \centering
    \includegraphics[width=0.95\linewidth]{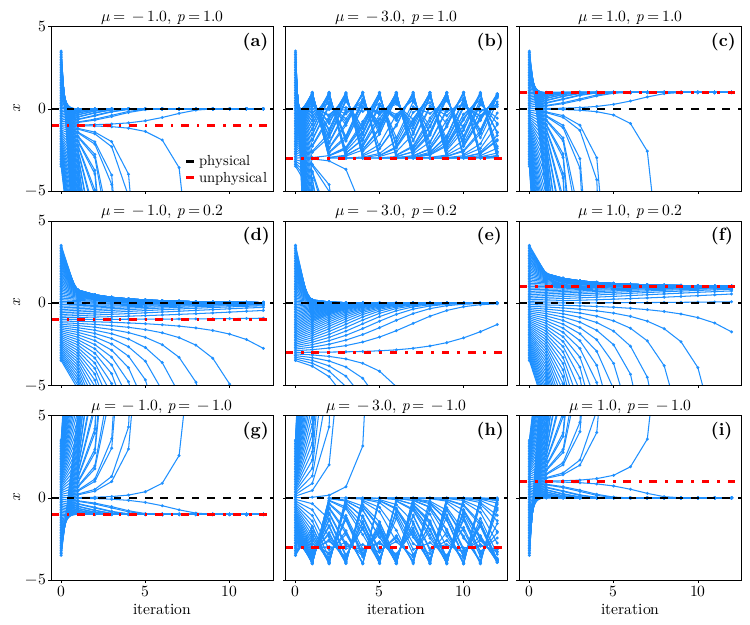}
    \caption{Trajectories of fixed-point iterations with the iterative map $f=x+\mu x - x^2$, Eq.~\eqref{eq:Iterative_map_1}, for different starting points $x_0 \in [-3.49,3.51]$ in steps of $\Delta x_0 = 0.1$. Black (red) dashed lines indicate the (un)physical fixed point $x^*=0 (\mu)$. The first column of panels shows the fixed-point iteration for the equation $\mu=-1$, the middle column for $\mu=-3$, and the third one for $\mu=1$. Panels~(a)--(c) are computed with a damping parameter $p=1.0$, while panels~(d)--(f) with $p=0.2$, which fixes the convergence problems of the case $\mu=-3$. In panel~(i), we apply the stabilization method by switching the sign of the damping parameter ($p=-1.0$) resulting in convergence to the physical fixed point. Switching the sign for the other two cases leads either to a convergence to the unphysical fixed point (g) or to a modification of the oscillation (h).}
    \label{fig:fixed point-iteration}
\end{figure}

Hence, if $\left|\frac{df}{dx}  \right|_{x=x^*}>1$, the correct fixed point is unstable and the fixed-point iteration cannot be converged. Let us now introduce linear mixing with the damping parameter $p \in (0,1]$. Then the fixed-point iteration changes to 
\begin{align} \label{eq:fixed point-with-mixing}
    x^{(n+1)}= g(x^{(n)})=p\,f(x^{(n)})+(1-p)\,x^{(n)}.
\end{align}
In this case, we obtain 
\begin{align} 
    |g(x^*+\epsilon)-x^*| = \left|1+p\left(\frac{df}{dx}  \right|_{x=x^*}-1\right)\bigg| \epsilon =  \begin{cases}
        > \epsilon \ \textrm{ if } \frac{df}{dx}  \big|_{x=x^*}>1 \ \ || \ \ \frac{df}{dx}  \big|_{x=x^*}<\frac{-2+p}{p}  \\
        < \epsilon \ \textrm{ else}.
    \end{cases}    
\end{align}
In total, we distinguish three different cases depending on the value of $\frac{df}{dx}  \big|_{x=x^*}$.
\begin{itemize} 
\item[\textbf{(I)}] If $0 <\big| \frac{df}{dx}  \big|_{x=x^*}<1$, the fixed point is (locally) stable and can be reached with the iterative scheme. 
\item[\textbf{(II)}] In the case of $\frac{df}{dx}  \big|_{x=x^*}<\frac{-2+p}{p}$, the fixed point is unstable. However, by reducing the damping parameter to $p<\frac{2}{\left|\frac{df}{dx}  \right|_{x=x^*}+1}$, the fixed point can be stabilized since $|g(x^*+\epsilon)-x^*| < \epsilon$. 
\item[\textbf{(III)}] If $\frac{df}{dx}  \big|_{x=x^*}>1$, the fixed point is unstable and the fixed point cannot be reached using the damped iterative scheme.
\end{itemize}
\noindent In the following, we showcase this for the simple example of the equation
\begin{equation}
f(x)=x+\mu\, x - x^2,
\label{eq:Iterative_map_1}
\end{equation}
solved with fixed-point iteration, so via 
\begin{equation}
x^{(n+1)} := f(x^{(n)})= x^{(n)} + \mu\, x^{(n)} - (x^{(n)})^2.
\end{equation}
 This system has the two fixed points $x^*=\{0,\mu \}$, where, in analogy to the parquet equations, we define $x^*=0$ as the physical fixed point and $x^*=\mu$ as the unphysical one. For our example, we obtain 
 \begin{equation}
 |f(x^*+\epsilon)-x^*| = \epsilon \left|  \frac{df}{dx}  \right|_{x=x^*}  = |1+\mu - 2x^*| \epsilon.
 \end{equation}
 This lets us distinguish the three cases with respect to $\mu$. \textbf{(I)} In the case of $-2<\mu<0$, the physical fixed point $x^*=0$ is locally stable independent of the damping parameter $p$. \textbf{(II)} If $\mu<-2$, the physical fixed point becomes locally unstable, which can be corrected by reducing $p$. This corresponds to the situation shown in the left panel of \cref{fig:stab_circle}. \textbf{(III)} In the case of $\mu>0$, the physical fixed point becomes unstable, but simultaneously the unphysical one ($x^*=\mu$) becomes locally stable. For the physical fixed point, this corresponds to the situation in the right panel of \cref{fig:stab_circle}. 
This behavior becomes visible in Fig.~\ref{fig:fixed point-iteration}, where we show different trajectories of the fixed-point iteration for different starting points $x_0\in [-3.49,3.51]$, with $\Delta x_0 = 0.1$ being the spacing between different starting points, in the three cases discussed. In panels~(a)--(c), no damping is applied ($p=1.0$) and in panels~(d)--(f) the iteration is damped by $p=0.2$. For the trivial case ($\mu=-1.0$), the physical fixed point is already locally stable for $p=1$ [see panel~(a)]. In panel~(b), we show that for $\mu=-3.0$ and $p=1$ the physical fixed point is locally unstable. This can be corrected by decreasing the damping (see panel~(e) for $p=0.2$). In contrast, for the third case ($\mu=1.0$), the unphysical fixed point is attractive, while the physical one is always repulsive. This cannot be corrected by lowering the damping [see panels~(c) and (f)].

The question remains if there is a way to stabilize the physical fixed point in case \textbf{(III)} [panels~(c) and (f)]. Indeed, this can be achieved by switching the sign of the damping $p$, which is shown in the lower three panels~(g)--(i) of \cref{fig:fixed point-iteration}. 
In panel~(i) of Fig.~\ref{fig:fixed point-iteration}, it becomes visible that switching the sign of $p$ locally stabilizes the previously unstable physical fixed point causing the trajectories to converge towards the physical fixed point. For completeness, we show that switching the sign of $p$ in the other two cases destabilizes the former locally stable physical fixed point [see panel~(g)]. 

We note that in all cases the initial starting point of the iteration may lead to a divergence of the iteration scheme. This is due to the fact that a Jacobian-based analysis only determines whether the fixed point is \emph{locally stable} against small perturbation. 

Let us now discuss the case of a 2-dimensional system of coupled equations. We want to solve the following two coupled equations $f_x=1+bx+\eta y$ and $f_y=1-by+\eta x$ with fixed-point iteration, where $\eta \ll 1$ and $b>1$.  With damping, we can define the fixed-point iteration in the following way:
\begin{align} \label{eq:coupled-fixed point-iteration}
    \Psi^{(n+1)}:= \begin{pmatrix}
        x^{(n+1)} \\
        y^{(n+1)}
    \end{pmatrix} = 
    \begin{pmatrix}
        (1-p)x^{(n)} + p (1+bx^{(n)}+\eta y^{(n)}) \\
        (1-p)y^{(n)} + p (1-by^{(n)}+\eta x^{(n)})
    \end{pmatrix} =: (1-p) \Psi^{(n)} + p\, \mathcal{F}(\Psi^{(n)}), 
\end{align}
where $\mathcal{F}=(f_x,f_y)^T$.
Let us first focus on the case $\eta = 0$, where the equations decouple. In this case, the $x$ direction of the fixed point is unstable and, therefore, switching the sign of $p$ is required, while the $y$ direction around the fixed point is stable for $0<p<\frac{2}{b+1}$. Hence, we need to modify Eq.~\eqref{eq:coupled-fixed point-iteration} as follows:
\begin{align} \label{eq:sot-coupled-fixed point-iteration}
    \Psi^{(n+1)} := \mathcal{G}(\Psi^{(n)}) = (\mathbb{1}-\mathcal{P}) \Psi^{(n)} + \mathcal{P} \mathcal{F}(\Psi^{(n)}), \quad \mathcal{P} = \begin{pmatrix}
        -p &\ 0 \\
        0 &\ p \\
    \end{pmatrix} .
\end{align}
If we now choose $0 <\eta \ll 1$, which weakly couples the two equations, the form of $\mathcal{P}$ in Eq.~\eqref{eq:sot-coupled-fixed point-iteration} is no longer correct. Instead of just calculating  $\frac{df_x}{dx}$, the full Jacobian of the undamped map 
\begin{align}
    J_{p=1} = \frac{\partial \mathcal{F}}{\partial \Psi} = \begin{pmatrix}
        \frac{\partial f_x}{\partial x} &\ \frac{\partial f_x}{\partial y} \\
        \frac{\partial f_y}{\partial x} &\ \frac{\partial f_y}{\partial y}
    \end{pmatrix} =
    \begin{pmatrix}
        b & \phantom{-}\eta \\
        \eta & -b
    \end{pmatrix}
\end{align}
needs to be computed and diagonalized, where the eigenvalues determine the convergence properties. This can be understood in the following way. Let $\boldsymbol{\epsilon}$ be an eigenvector of the Jacobian $J_{p=1}$ associated with the eigenvalue $j$ and $|\boldsymbol{\epsilon}| \ll 1$. Then 
\begin{align}
    \mathcal{F}(\Psi^{*}+ \boldsymbol{\epsilon}) - \Psi^* \approx J_{p=1} \boldsymbol{\epsilon} = j \boldsymbol{\epsilon},
\end{align}
for the fixed point $\Psi^{*}$. 
Hence, if $\exists |j|>1 $, the fixed point is unstable. 

Following the previous discussion, if $j<-1$, the damping parameter $p$ can be reduced, to locally stabilize the direction of the associated eigenvector, while for $j>1$ the sign of the damping parameter $p$ has to be changed. More formally, the matrix $\mathcal{P}$ in Eq.~\eqref{eq:sot-coupled-fixed point-iteration} needs to be adjusted to $\mathcal{P}=   \mathcal{S} \mathcal{D} \mathcal{S}^{-1}$, where $\mathcal{S}$ is the eigenbasis of the Jacobian and $\mathcal{D}$ is a diagonal matrix that has $-p$ as entry if the associated eigenvalue is larger than 1 and $p$ otherwise. This ensures that the sign of the damping gets flipped only in the unstable direction of the associated eigenvalue. 
In the following, we will refer to this procedure as \emph{stabilization method}. 

\begin{figure}
    \centering
    \includegraphics[width=0.95\linewidth]{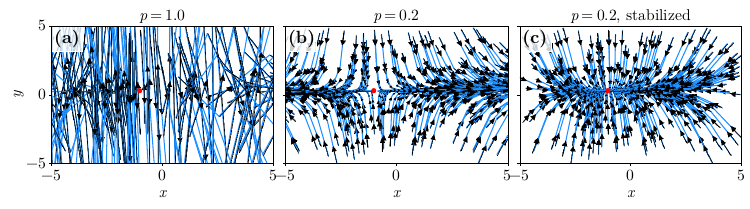}
    \caption{Trajectories of the fixed-point iteration for the map $f_x=1+2x+\eta y$ and $f_y=1-2y+\eta x$ with $\eta =0.1$ for different starting points $(x_0,y_0)$ with $x_0,y_0 \in (-5,5)$ and 40 iterations. In panel~(a), the damping $p=1.0$ leads to unstable directions along both eigenvectors $\mathbf{u_1} = (0.0250,-0.9997)^T, \mathbf{u_2} = (-0.9997,-0.0250)^T$. In panel~(b), the damping $p$ is set to $0.2$ stabilizing  the direction along $\mathbf{u_1}$ (associated with $j_1=-2.0025$). Application of the stabilization method in panel~(c) fixes the stability issues and allows for convergence of the fixed-point iteration to the correct solution $(x^*,y^*)=(-1.03,0.299)$ (red dot).}
    \label{fig:fixed-point-iteration-2d}
\end{figure}

In Fig.~\ref{fig:fixed-point-iteration-2d}, the fixed-point iteration is shown for the coupled equations $f_x=1+bx+\eta y$ and $f_y=1-by+\eta x$ with $b=2,\eta =0.1$. This leads to a Jacobian with the eigenvalues $(j_1,j_2) =(-2.0025,2.0025)$ and the associated eigenvectors $\mathbf{u_1} = (0.0250,-0.9997)^T, \mathbf{u_2} = (-0.9997,-0.0250)^T$. In panel~(a), we take  $p=1$ leading to diverging curves both in $x$ and $y$ directions, since both directions are unstable. The $y$ direction (or more accurately the direction of the eigenvector $\mathbf{u_1}$) becomes stable due to the chosen damping of $p=0.2 < -\frac{2}{j_1-1} \approx \frac{2}{3}$ while the $x$ direction (direction of $\mathbf{u_2}$) remains unstable in panel~(b). In panel~(c), the stabilization method is applied stabilizing the fixed point locally  in the directions of both eigenvectors.

Finally, let us comment on the stability of the fixed point. Since in this specific example only coupled linear equations are considered, local stability of the fixed point simultaneously implies \emph{global} stability, meaning that if the fixed point is indeed locally stable, i.e., $|\mathcal{G}(\Psi^{*}+ \boldsymbol{\epsilon}) - \Psi^*| < |\boldsymbol{\epsilon}|$, then the fixed-point iteration will converge towards the correct fixed point \emph{regardless} of the starting point. In the general case of nonlinear equations (e.g., see the previous example in this section), \emph{local} no longer implies global stability and the starting point matters for convergence.

\subsection{Stabilization method for the parquet iteration} \label{sec:sot}
As indicated by the eigenvalues with negative real parts in the right panel of \cref{fig:stab_circle}, the damped iteration, \cref{eq:Psi}, cannot converge to the physical fixed point. 
To address this issue, Ref.~\cite{Essl2026} introduced a modified iteration scheme capable of stabilizing otherwise unstable fixed points. Importantly, this construction alters only the stability properties of the fixed points, while leaving the fixed points themselves unchanged.
The basic idea of this scheme was already illustrated using the pedagogical example in the previous \cref{sec:motivation}. In this section, we discuss the necessary modifications to extend the method to the case of parquet iterations.

First, the abstract high-dimensional quantities $\Psi$ and $\mathcal{F}$ need to be flattened such that they represent a vector in the same space in which the Jacobian is a matrix (see \cref{app:parquet_J} for details). Second, we replace the scalar damping $p$ by a \emph{stabilization matrix} $\mathcal{P}[\Psi^*]$ which is evaluated at $\Psi^*$
\begin{align}
\label{eq:Psi_tilde}
    \Psi^{(n+1)}=\mathcal{P}[\Psi^*]\cdot\mathcal{F}[\Psi^{(n)}]+(\mathbb{1}-\mathcal{P}[\Psi^*])\cdot\Psi^{(n)},
\end{align}
where $\cdot$ represents a matrix-vector product in the flattened space of the Jacobian.
The stabilization matrix is defined as $\mathcal{P}^{ij}=\sum_{\alpha_1,\alpha_2}\mathcal{S}^{i\alpha_1}\mathcal{D}^{\alpha_1\alpha_2}[\mathcal{S}^{-1}]^{\alpha_2j}$, with $\mathcal{S}$ being the similarity transformation that diagonalizes the Jacobian $J^{ij}=\sum_{\alpha_1,\alpha_2}\mathcal{S}^{i\alpha_1} (1-p\lambda_{\alpha_1} \delta_{\alpha_1\alpha_2})[\mathcal{S}^{-1}]^{\alpha_2j}$ (and therefore also $\Pi$), where the Jacobian is also evaluated at $\Psi^*$. Further, $\mathcal{D}$ is defined as
\begin{align} 
\label{eq:daa}
\mathcal{D}^{\alpha\alpha^\prime} = 
    \begin{cases}
      \phantom{-}p\,\delta_{\alpha\alpha^\prime} & \text{if}\ \text{Re}(\lambda_\alpha)>0 \\
      -p\,\delta_{\alpha\alpha^\prime} & \text{if}\ \text{Re}(\lambda_\alpha)\leq 0
    \end{cases} , 
\end{align}
where the required $p$ can be calculated via \cref{eq:uniform_damping}. Alternatively, $\mathcal{D}$ can be calculated via
\begin{align} 
\label{eq:daa_nonuniform}
\mathcal{D}^{\alpha\alpha^\prime} = 
    \begin{cases}
    \phantom{-}\delta_{\alpha\alpha^\prime}  & \text{if}\ c\frac{2\text{Re}(\lambda_\alpha)}{|\lambda_\alpha|^2}>1 \\
    -\delta_{\alpha\alpha^\prime}  & \text{if}\ c\frac{2\text{Re}(\lambda_\alpha)}{|\lambda_\alpha|^2}<-1 \\
    c\frac{2\text{Re}(\lambda_\alpha)}{|\lambda_\alpha|^2}\delta_{\alpha\alpha^\prime} & \text{else}
    \end{cases},
\end{align}
which corresponds to damping each eigenvector direction differently.
The Jacobian of the stabilized iteration, \cref{eq:Psi_tilde}, reads
$\tilde J=\mathbb{1}-\mathcal{P}\cdot\Pi$ and has the eigenvalues $1-\mathcal{D}^{\alpha\alpha}\lambda_\alpha$. Therefore, all eigenvalues of $\tilde J$ have a magnitude smaller than $1$, if the Jacobian is evaluated at the physical fixed point.

\section{Numerical examples}
We now apply the stabilization method to the parquet iteration of the two exactly solvable models: the ZP model and the HA. Closed expressions for both the corresponding vertices and Jacobian are provided in \cref{app:HA_quantities,app:ZP_quantities,app:parquet_J}. 

\subsection{Parquet iteration for the zero-point model}\label{sec:ZP_parquet}
We apply the stabilization method to the case of the ZP model [\cref{eq:S_ZP}].
To get an initial assessment for the stability of the physical fixed point of the corresponding parquet equations, we calculate the Jacobian \cref{eq:J_general} (for $p=1$) (see \cref{eq:ZP_J_d,eq:ZP_J_m,eq:ZP_J_s,eq:ZP_J_t,eq:ZP_J_G} for the explicit formulas) evaluated at the physical fixed point \cref{eq:G_ZP,eq:Phi_ZP,eq:Lambda_ZP}.

As already illustrated in \cref{sec:motivation}, we can then distinguish between three different cases:
\begin{enumerate}[label=(\Roman*)]
    \item All eigenvalues of $J$ have a magnitude smaller than $1$, which means that the physical fixed point is stable without damping ($p=1$). 
    \item At least one eigenvalue of $J$ has a magnitude greater than $1$, but all eigenvalues have a real part that is smaller than $1$. This means that the physical fixed point can be (locally) stabilized by damping the iterative procedure. 
    \item At least one eigenvalue of $J$ has a real part that is larger than $1$. This means that the physical fixed point is unstable for all damping parameters $p\in(0,1]$, and the fixed point is unstable using a damped iteration scheme. 
    The three scenarios of how the Jacobian can develop such unstable eigendirections, i.e., odd pole, zero crossing or zero crossing with finite imaginary part, are introduced in the end of \cref{sec:iterparquet}.
\end{enumerate}

\begin{figure}[t!]
    \centering
    \includegraphics[scale=0.6]{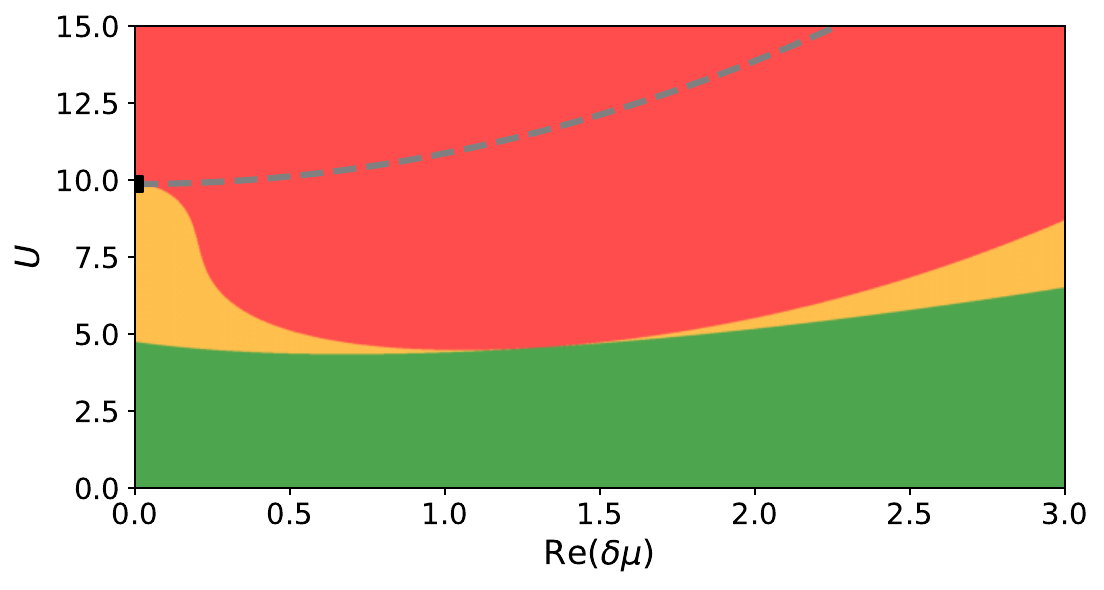}
    \caption{Stability phase space of the parquet formalism with damped iteration, \cref{eq:Psi}, for the ZP model. Color scheme: In the green region, the fixed point is stable without damping ($p=1$); in the orange region, it is unstable for $p=1$ but curable by smaller $p$; in the red region, it is unstable and not curable by a smaller $p$. The vertex divergence, which only appears in the density channel at ($\mathrm{Re}(\delta\mu)=0$, $U=\mathrm{Im}(\delta\mu)^2$), is marked by a black square. The branching of the LWF, i.e., the breakdown of self-consistent perturbation theory, is marked by a gray dashed line.}
    \label{fig:ZP_stability}
\end{figure}

The result of the procedure for different points in the two-dimensional phase space, spanned by $U$ and $\mathrm{Re}(\delta\mu)$, is shown in \cref{fig:ZP_stability}, where the cases (I), (II), and (III) are marked by the colors green, orange, and red. 
We find that at ph-symmetry ($\mathrm{Re}(\delta\mu)=0$), the onset of the unstable (red) region coincides with the occurrence of a vertex divergence (marked by a black square). Out of ph-symmetry, no vertex divergence appears in the exact solution of the ZP model, but we still find an instability. The onset of the instability out of ph-symmetry is caused by an eigenvalue with \emph{zero} real part but \emph{finite} imaginary part and can even appear at lower $U$ values than the one at ph-symmetry.
Further, the branching of the LWF, i.e., the breakdown of self-consistent perturbation theory (gray dashed line), only coincides with the instability if a vertex divergence is present. This again confirms the result of Ref.~\cite{Essl2026}, which shows that the instability of the physical fixed point and breakdown of self-consistent perturbation theory coincide \emph{only} at vertex divergences.
Let us also emphasize that the stability phase space in \cref{fig:ZP_stability} does \emph{not} coincide with the stability phase space that is shown for the ZP model in Ref.~\cite{Essl2026}. This is due to the fact that the stability phase space depends not only on the model that is investigated but also on the self-consistent scheme that is used.

To illustrate the problem of an unstable physical fixed point, we now show a parquet calculation in the ZP model for two representative paths in the phase space, namely $\text{Re}(\delta\mu)=0$ and $\text{Re}(\delta\mu)=1$ for increasing $U$ values.
Specifically, we perform calculations with the conventional damped iteration scheme, \cref{eq:Psi}, and calculations with the stabilized method, \cref{eq:Psi_tilde}, where the damping is determined via \cref{eq:uniform_damping} and the physical solution at the previous $U$ value is chosen as starting point for both schemes. Further, the exact fully irreducible vertices $\Lambda$, \cref{eq:Lambda_ZP}, are used as an input. 

\begin{figure}[t!]
    \centering
    \includegraphics[width=\textwidth]{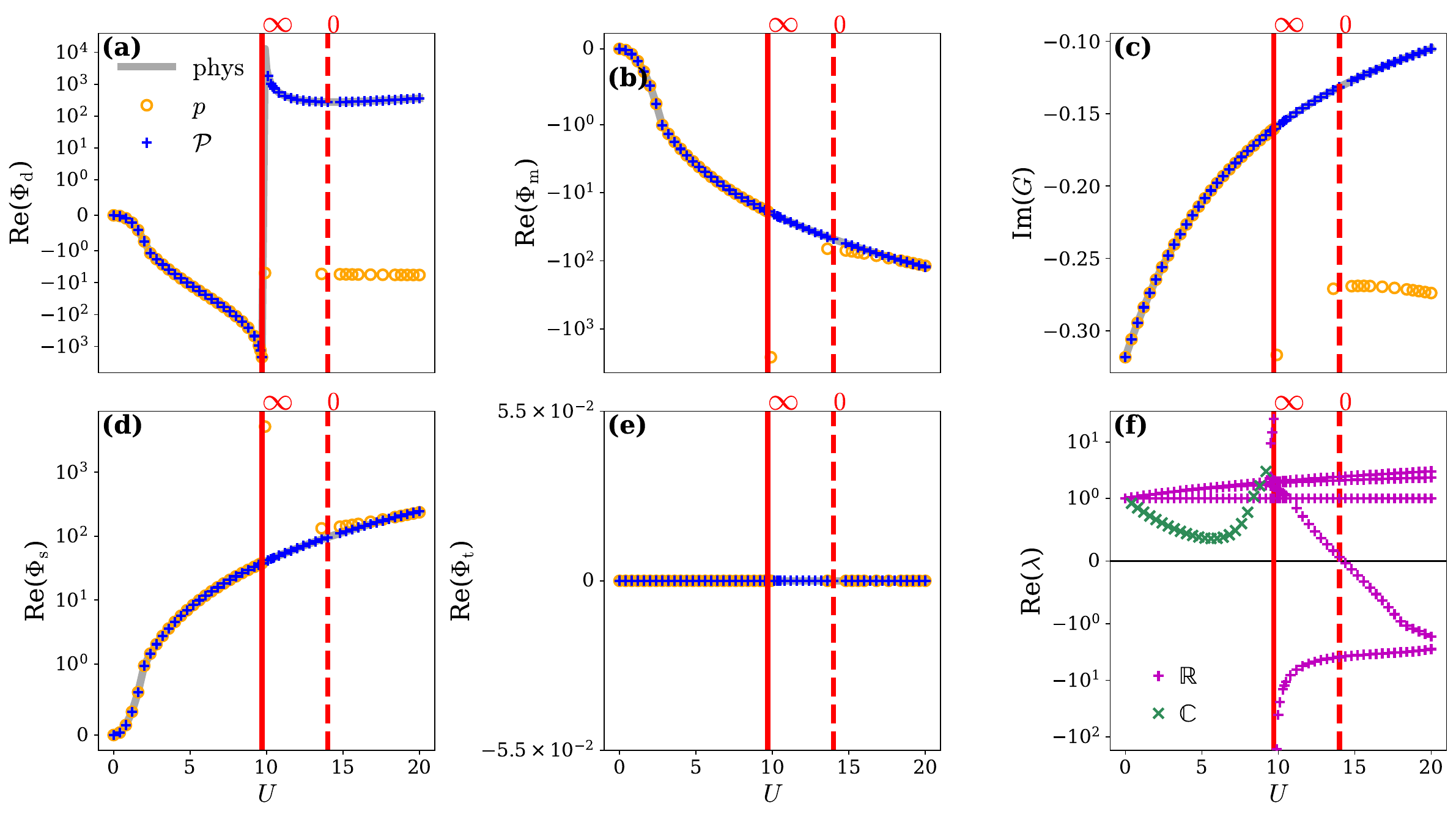}
    \caption{Parquet results for the ZP model with conventional iteration (orange $\circ$ denoted by $p$) and stabilized iteration (blue $+$ denoted by $\mathcal{P}$) for ph-symmetry ($\text{Re}(\delta\mu)=0$), where the red lines mark the instabilities (solid for vertex divergences and dashed for real eigenvalues that go through zero). The physical solution is shown as gray solid line. The starting point for each calculation is the physical fixed point at the previous $U$ value and the damping $p$ is determined by \cref{eq:uniform_damping} with $c=0.1$. In panels~(a)--(e), the channels (d,m,G,s,t) are shown and in panel~(f) the real part of eigenvalues of $\Pi$ are shown, where purely real eigenvalues are marked by magenta $+$ and complex eigenvalues are marked by green $\cross$.}
    \label{fig:ZP_dmu0}
\end{figure}

In \cref{fig:ZP_dmu0}, the result for the path at $\text{Re}(\delta\mu)=0$ is shown. Along this path, two instabilities occur: First an instability that is caused by the vertex divergence [case (i) in \cref{sec:iterparquet}] in the density channel (solid red line) and later an instability that is caused by a real eigenvalue of $\Pi$ that crosses zero [case (ii) in \cref{sec:iterparquet}] marked by a dashed red line. The real eigenvalue that crosses zero is mainly driven by the magnetic channel (see \cref{app:decoupled_channels} for details).
We find that the conventional iteration converges to the physical fixed point until the first instability is reached and after that it shows misleading convergence meaning that the scheme still converges, but to an unphysical fixed point. In contrast to this behavior, the stabilized iteration always converges to the physical fixed point.

\begin{figure}[t!]
    \centering
    \includegraphics[width=\textwidth]{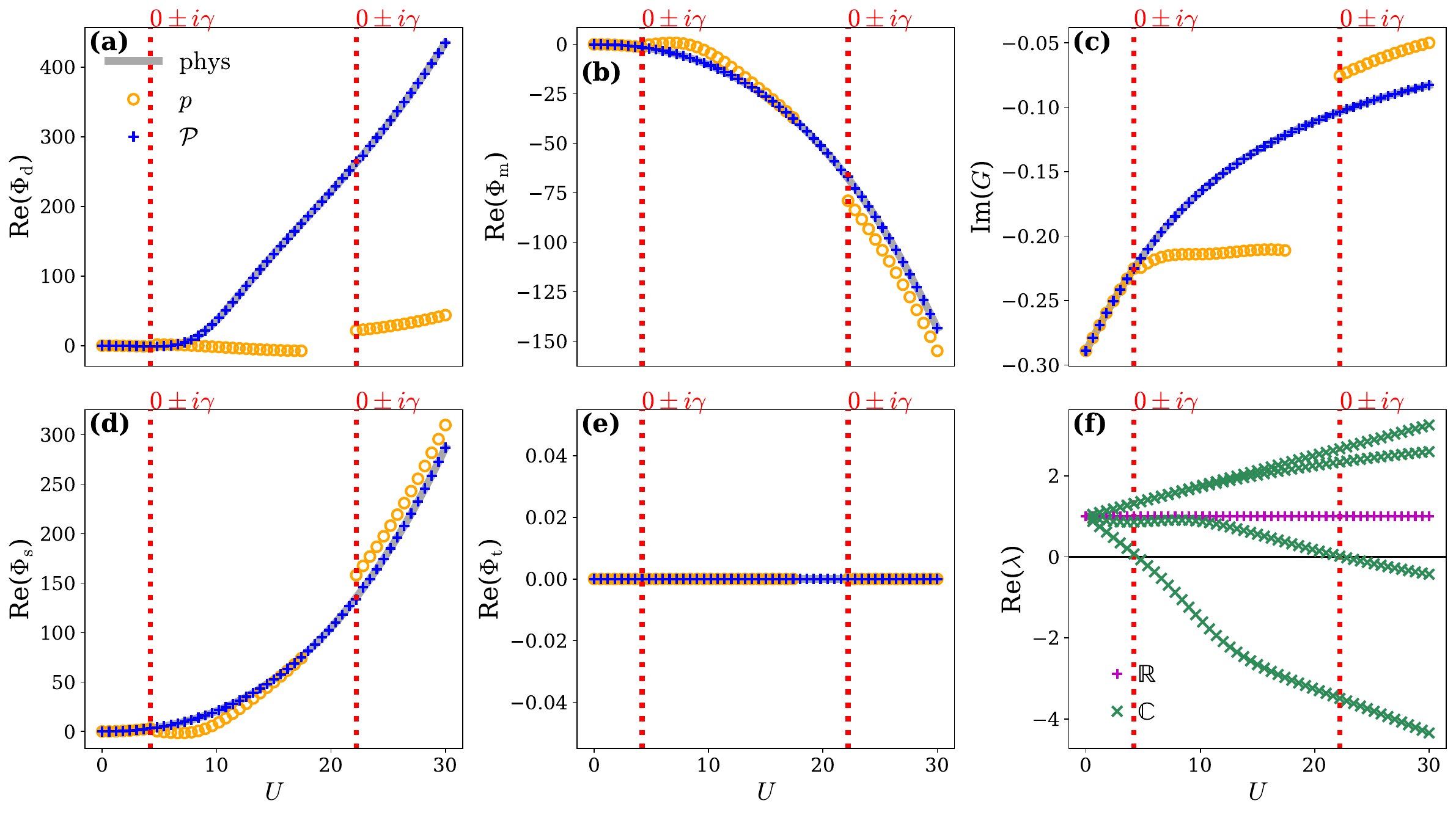}
    \caption{Analogous parquet results for the ZP model as in \cref{fig:ZP_dmu0} but out of ph-symmetry ($\mathrm{Re}(\delta\mu)=1$). The red dotted lines mark the instabilities where the real part of $\lambda$ goes through zero with a finite imaginary part.}
    \label{fig:ZP_dmu1}
\end{figure}

For the path at $\text{Re}(\delta\mu)=1$, the same analysis is shown in \cref{fig:ZP_dmu1}. Two instabilities are found, where the real part of an eigenvalue of $\Pi$ crosses zero, while having a finite imaginary part [case (iii) in \cref{sec:iterparquet}]. As previously, one instability is mainly attributed to the density and the other one to the magnetic channel. Also, in this case, the conventional iteration converges to an unphysical fixed point after the first instability, while the stabilized iteration always converges to the physical fixed point.

\subsection{Parquet iteration for the Hubbard atom}\label{sec:HA_parquet}
Let us now turn to the results for the Hubbard atom [see Hamiltonian in \cref{eq:hamiltonian-hubbard-atom}]. In contrast to the ZP model, the Jacobian (and $\Pi$) becomes frequency dependent and is of size $(4N_f^2N_b+N_f) \times (4N_f^2N_b+N_f)$, with $N_{f(b)}$  being the number of fermionic (bosonic) Matsubara frequencies that are included in the calculation (see App.~\ref{app:parquet_J} for explicit formulas). On a more practical note, we should underline that although the eigenvectors of the Jacobian and $\Pi$ should be the same, they can differ, sometimes even significantly, in numerical calculations. We attribute this mismatch to the fact that the condition number of the eigenbasis can become very large, which might result in problems for the numerical computation of $\mathcal{P}$. For more details on this issue, we refer the reader to App.~\ref{app:condition-number}. 

\begin{figure}
    \centering
    \includegraphics[scale=0.6]{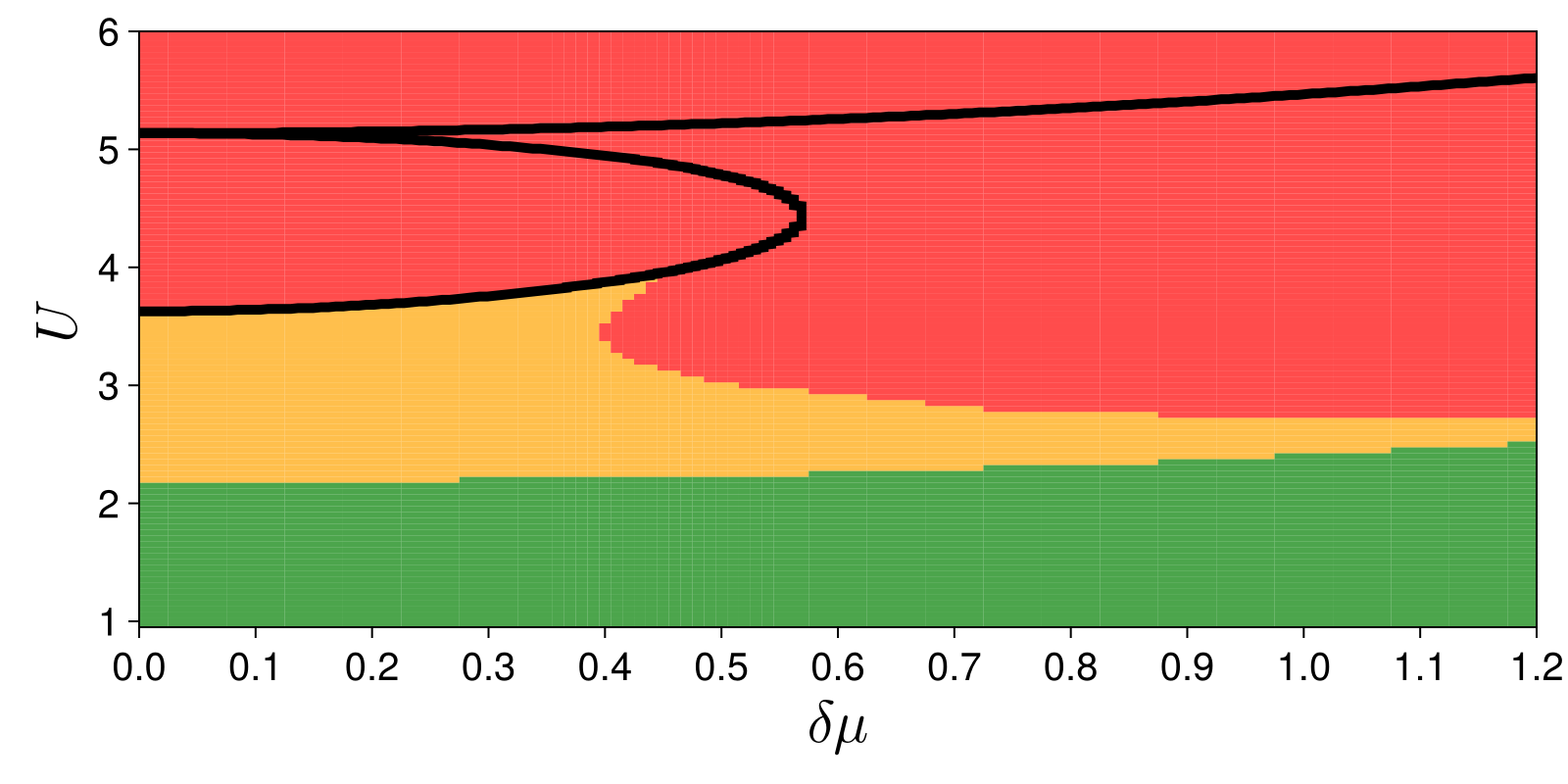}
    \caption{Stability phase space for conventional iteration in the HA. The color code is the same as in \cref{fig:ZP_stability}. All vertex divergence lines that appear in the shown parameter range are marked by black lines. A ``lobe'' structure is built from the first two vertex divergences in the density channel and the third line corresponds to the vertex divergence in the singlet channel.}
    \label{fig:HA_stability}
\end{figure}

Following the discussion of the ZP model, we repeat the stability analysis (\cref{fig:ZP_stability} in the ZP model case) for the HA in \cref{fig:HA_stability}. Even though the ZP model is a very simplified toy model, qualitatively similar stability regions are found in the two models. 
Apart from qualitative agreement of the region where the physical fixed point becomes unstable, close to half-filling, the first instability is caused by a divergence of the irreducible vertex, while far away from half-filling a pair of complex conjugated eigenvalues that has a vanishing real part causes the first instability in both models. In contrast to the ZP model, vertex divergences also occur outside ph-symmetry (black line in \cref{fig:HA_stability}) in the HA case. 
As discussed in \cref{sec:vertex_div}, in Ref.~\cite{Gunnarsson2017}, it was shown that a crossing of two branches of the LWF implies a vertex divergence in the density channel. 
However, it is still unclear where in the phase space this happens once the vertex divergence in the density channel disappears, i.e., for $\delta\mu\gtrsim 0.6$.
The results in \cref{fig:HA_stability} again illustrate that misleading convergence and the branching of the LWF are two \emph{distinct} effects, as shown in Ref.~\cite{Essl2026}.

As discussed before, the onset of the instability region and the vertex divergence coincide close to half-filling. However, the vertex divergence need not be the first instability ($0.4\lesssim \delta\mu\lesssim0.6$) and instabilities can even occur without a vertex divergence in the density channel ($\delta\mu \gtrsim 0.6$).

Another difference is caused by the highly nontrivial frequency structure of the HA resulting in a significantly higher number of unstable eigendirections when increasing $U$ than in the ZP model (see panels~(f) in \cref{fig:HA-Nf-12-Nb-13-U-7,fig:HA-oohf-Nf-12-Nb-13}).

\subsubsection{Half-filling}
\begin{figure}
    \centering
    \includegraphics[width=\linewidth]{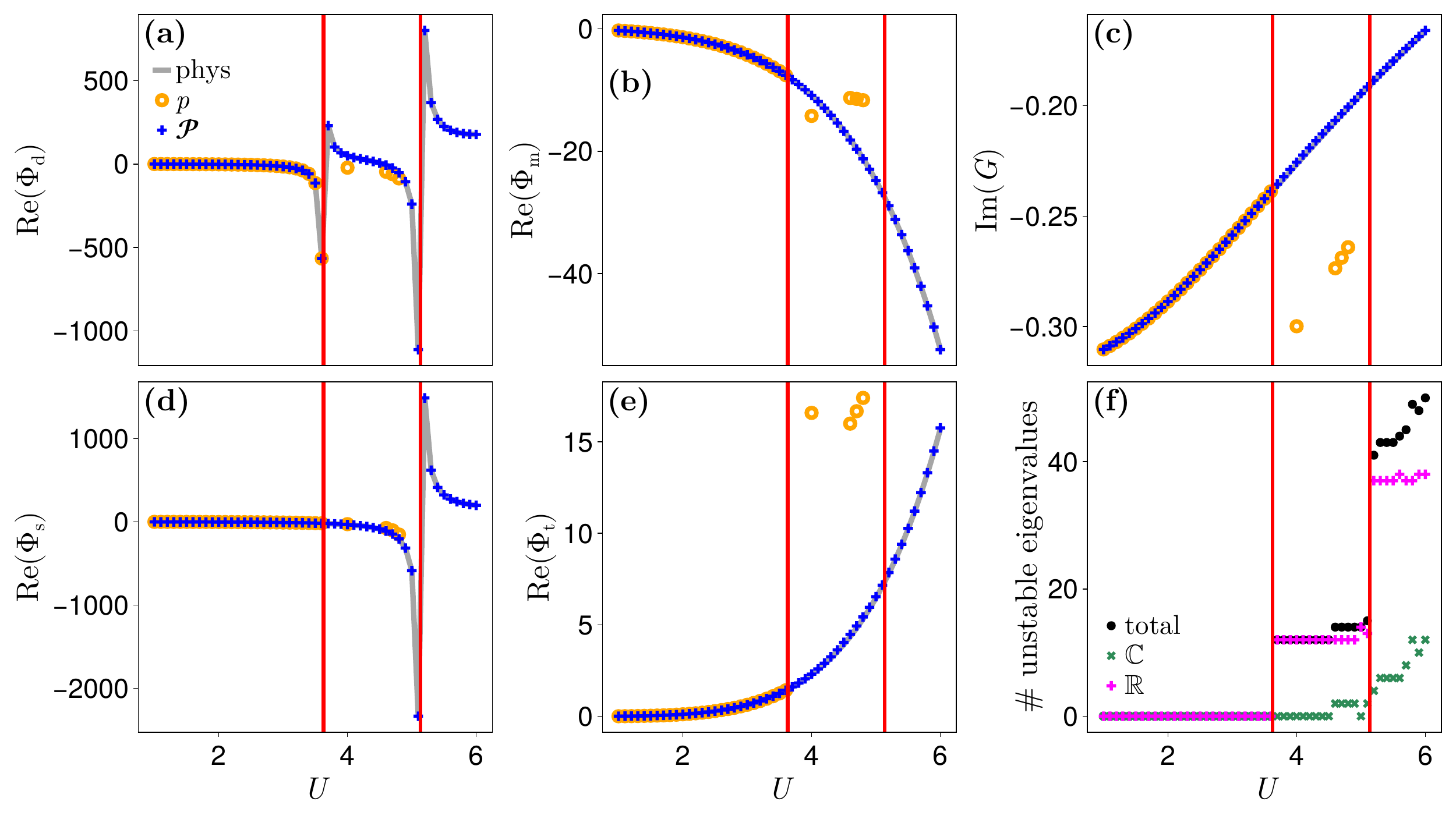}
    \caption{Parquet results for the HA with the conventional iteration (orange $\circ$ denoted by $p$) and the stabilization method (blue $+$ denoted by $\mathcal{P}$) vs.\ exact values of Re[$\Phi_r(\pi/\beta,\pi/\beta,0)$] for the (a) density, (b) magnetic, (d) singlet, and (e) triplet channels, and (c) Im[$G(\pi/\beta)$] for $\delta\mu=0$, $\beta = 1$, $\epsilon = 10^{-6}$, $\Delta U_\mathrm{init}=0.001$, and $N_f=12$, $N_b=13$. The damping for the stabilization method is calculated with \cref{eq:uniform_damping} ($c=0.1$), while a damping $p=0.005$ is used for the conventional iteration. The exact solution is indicated by the gray line. In panel~(f), the number of unstable eigenvalues is shown, where they are also split in number of purely real and complex unstable eigenvalues. Finally, the red solid lines indicate divergences of the irreducible vertex $\Gamma$ in the density channel (first line) and simultaneously in the density and singlet channel (second line).}
    \label{fig:HA-Nf-12-Nb-13-U-7}
\end{figure}

In this subsection, we study the case of half-filling ($\delta\mu = 0$) in greater detail. After initializing the iterative parquet scheme with the exact $\Lambda$, $G_0$, and the iterated properties with the exact ones with a slightly smaller interaction, i.e., $U_\mathrm{init}=U-\Delta U_\mathrm{init}$, the iteration converges to an unphysical solution after the crossing of the first (local, $\omega=0$) divergence of the irreducible vertex in the density channel at $\beta U=2\pi/\sqrt3 $ \cite{Schaefer2013,Thunstroem2018}. We show this behavior in Fig.~\ref{fig:HA-Nf-12-Nb-13-U-7} in the case of $N_f=12$, $N_b=13$, $\Delta U_\mathrm{init}=0.001$, and the maximum normalized error $\epsilon=10^{-6}$, which represents our convergence criterion [\cref{eq:conv_criterion}]. From studying the unstable (negative real part) eigenvalues of $\Pi$ in panel~(f), it can be seen that precisely at the first vertex divergence (first red solid line at $\beta U=2\pi/\sqrt3$), multiple eigenvalues become unstable (pole, type (i) instability in \cref{sec:iterparquet}). 
Specifically, the number of unstable eigenvalues after the first divergence line matches precisely the size of the fermionic frequency box ($N_f=12$). 
Additionally, it can be observed that, at each vertex divergence (red solid lines), a large jump in the number of unstable eigenvalues (pole instabilities) occurs, while the number of unstable eigenvalues can also (smoothly) increase \emph{without} the occurrence of vertex divergences (type (ii) and (iii) instabilities in \cref{sec:iterparquet}). On a more quantitative note, one finds that the number of unstable eigenvalues depends on the size of the fermionic frequency box ($N_f$), as discussed in App.~\ref{app:n-eigenvalues}.

Investigating the iterated properties $G$ and $\Phi_r$ in panels~(a)--(e) shows convergence of the conventional iterative scheme to an unphysical solution starting at the first vertex divergence in the density channel (orange empty circles). After the second divergence simultaneously occurring in the density and singlet irreducible vertices (second red solid line), the scheme stops converging at all. This is likely caused by the unphysical solution being too far away from the initialized iterated properties (exact physical solution at $U_{\mathrm{init}})$ past the second divergence.

On the other hand, the application of our stabilization method (blue pluses) drastically changes this picture.
Here, the parquet equations converge to the correct physical solution (gray solid line) essentially for the whole parameter range since the local stability issues of the physical solution in the case of unstable eigenvalues of $\Pi$ are fixed by the stabilization method. For better visibility of the misleading convergence after the first divergence, we limit our figure to $U=6$ as an upper boundary. In App.~\ref{app:HA-additional-results}, we extend the analysis up to $U=11$. 

\subsubsection{Finite doping}
\begin{figure}
    \centering
    \includegraphics[width=\linewidth]{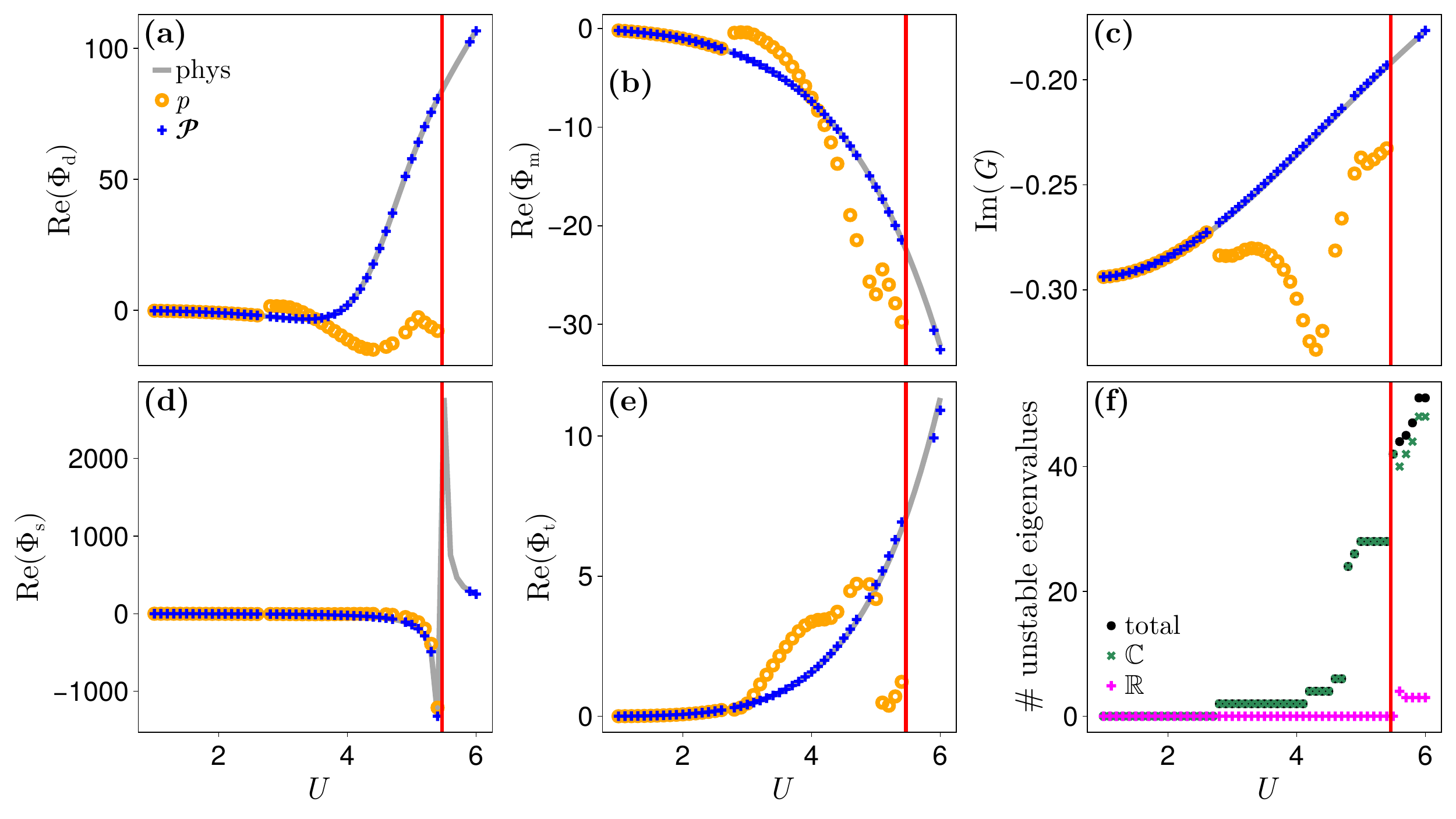}
    \caption{Parquet results for the HA analogous to \cref{fig:HA-Nf-12-Nb-13-U-7} but out of half-filling ($\delta\mu=1$). The red solid line again marks the first vertex divergence, which, for $\delta\mu=1$, is in the singlet channel.}
    \label{fig:HA-oohf-Nf-12-Nb-13}
\end{figure}

We turn now to the case out of half-filling. Keeping all other parameters fixed, we adjust the chemical potential to $\delta \mu =1$. When investigating panel~(f) of Fig.~\ref{fig:HA-oohf-Nf-12-Nb-13}, it can be seen that slightly below $U=3$, a complex conjugated pair of eigenvalues becomes unstable (type (iii) instability), which also matches the start of the red region of the stability phase diagram in Fig.~\ref{fig:HA_stability}. In contrast to the case at half-filling, no divergence of the irreducible vertex occurs at this point and the number of unstable eigenvalues (two) is independent of the size of the fermionic frequency box. 
Due to these complex eigenvalues, the physical fixed point becomes unstable, even in the absence of a vertex divergence, and the parquet equations converge to an unphysical solution for the conventional iteration (see orange empty circles). This misleading convergence continues until the occurrence of the first vertex divergence in $\Gamma_\text{s}$ slightly above $U=5$. Thereafter, convergence of the iterative parquet scheme stops. This is likely due to the same reason as in the half-filling case. The picture changes when the stabilization method is applied (blue pluses). Here, the correct physical fixed point can be found not only past the point, where the first eigenvalues become negative, but also past the first divergence in $\Gamma_\text{s}$.
Moreover, when crossing this vertex divergence, a large jump in the number of negative eigenvalues occurs (type (i) instability), which is again dependent on the size of the fermionic frequency box (see App.~\ref{app:n-eigenvalues}). 
Interestingly, before the vertex divergence around $U=4.8$, another large jump in the number of negative eigenvalues occurs, which depends on the size of the frequency box and seems to be in the vicinity of a pseudo-divergence~\cite{Essl2026} in $\chi_{0,\text{ph}}^{-1}\chi_\text{d}$. We note that the position of the pseudo-divergence and the jump in the number of negative eigenvalues seem to coincide for other values of $\delta \mu$ as well. 

Let us again emphasize that so far the issue of misleading convergence has originally been ascribed to the occurrence of vertex divergences. Here, we have shown that the iterative parquet scheme can start to converge to unphysical solutions at significantly lower interactions than the position of the first divergence of the irreducible vertex. The stabilization method not only provides a clear understanding of the mechanism behind this, but also allows to solve this issue by making the physical fixed point locally stable. 

We conclude this section with noting that convergence depends not only on the damping parameter $p$, but also on the choice of $U_\mathrm{init}$ (see App.~\ref{app:alpha-init}) and the size of the frequency boxes (see App.~\ref{app:box-size}).

\subsection{Strong-coupling iteration}\label{sec:strong_parquet}
Even though the divergences of the irreducible vertex have physical implications (see \cref{sec:vertex_div}), they are directly visible only through a diagrammatic decomposition, i.e., the quantities that diverge are the two-particle reducible vertex $\Phi_r$, the two-particle irreducible vertex $\Gamma_{r}$, and the fully irreducible vertex $\Lambda_r$. The quantities that are directly related to the physical response of the system, i.e., the generalized susceptibility $\chi$ or the full vertex $F$, display a regular behavior.

Inspired by this consideration, we will now proceed to construct a parquet iteration that iterates the full vertices $F$ instead of the reducible vertices $\Phi$, as is usually done.
The main difference is that the BSEs need to be inverted to get the irreducible vertex $\Gamma_r$ as a functional of $F$ and $G$:
\begin{align}
\label{eq:Gamma_F}
    \Gamma_r[F,G] = \frac{F_r}{1\mp \chi_{0,r}F_r},
\end{align}
where minus is used for $r=\text{d,m,t}$ and plus for $r=\text{s}$.
By performing this inversion, we assumed that the geometric series, which is summed up iteratively in usual parquet equations [see \cref{eq:it_maps}], converges and can be manipulated to obtain $\Gamma[F,G]$. Finally, we then use the parquet decomposition, \cref{eq:parquet}, to relate $\Gamma[F,G]$ to a new $F$ (see \cref{app:parquet_J_F} for details).
Now, in order to investigate the stability of the physical fixed point for this iteration, we perform the same analysis as in \cref{fig:ZP_stability} but for the Jacobian $J^F$ of the ZP model where $F$ instead of $\Phi$ is iterated. 
\begin{figure}[!t]
    \centering
    \includegraphics[scale=0.6]{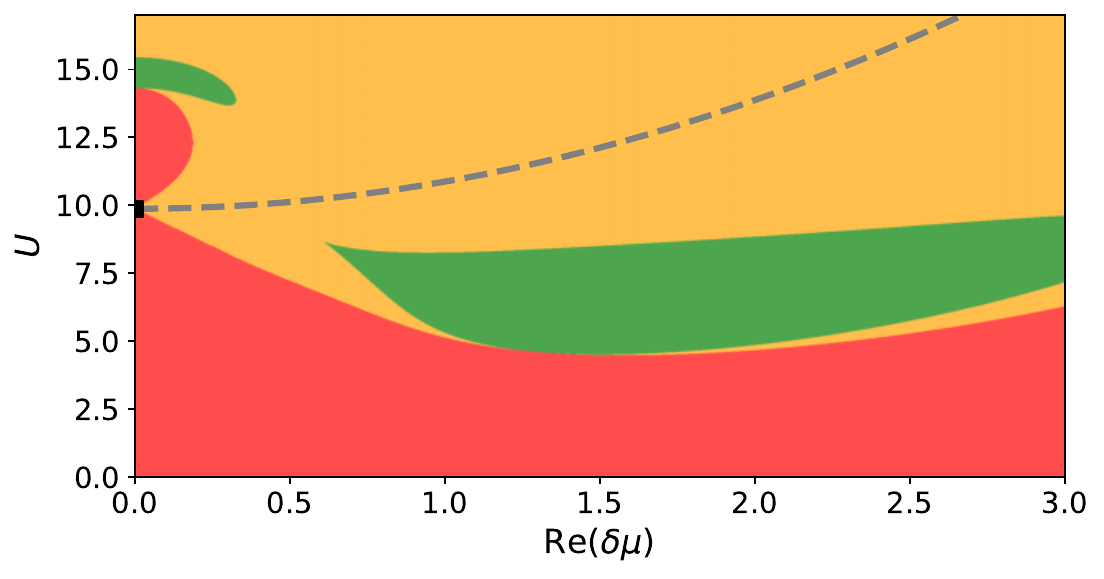}
    \caption{Stability phase space for the strong-coupling iteration in the ZP model. The presentation is analogous to \cref{fig:ZP_stability}.}
    \label{fig:ZP_F_stability}
\end{figure}
We find that, in a fundamentally different fashion w.r.t.\ the usual parquet iteration, the physical fixed point is unstable for weak coupling and becomes stable with increasing interaction. Because of this and the fact that we use the already resumed geometric series for the BSEs, we coin this iteration scheme \emph{strong-coupling iteration}.

To confirm these stability properties, we now conduct the strong-coupling iteration, once with damped iteration and once with the stabilized iteration.

\begin{figure}
    \centering
    \includegraphics[width=\textwidth]{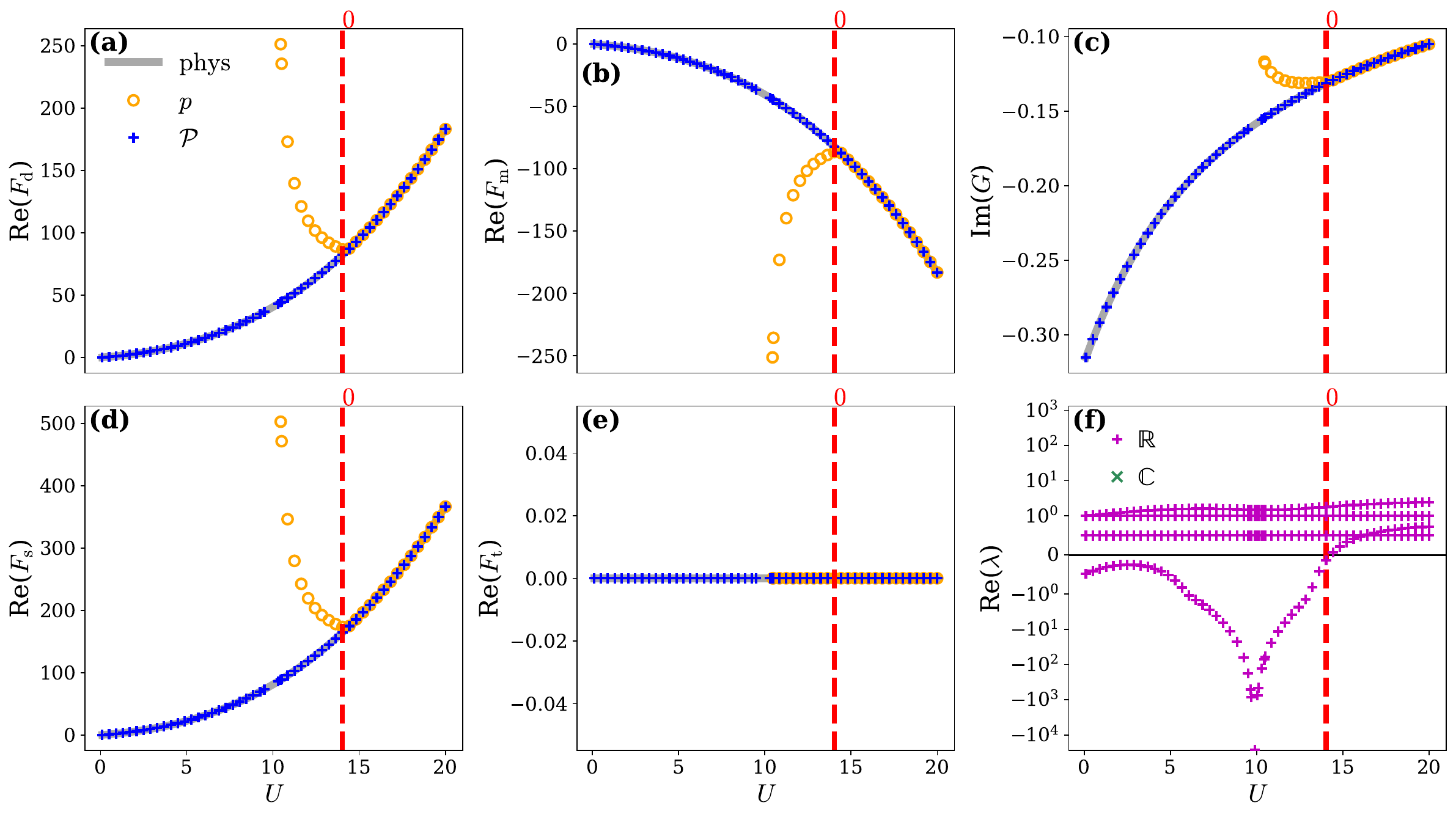}
    \caption{Parquet results from the strong-coupling iteration for the ZP model with damped iteration (orange $\circ$ denoted by $p$) and stabilized iteration (blue $+$ denoted by $\mathcal{P}$) for ph-symmetry ($\text{Re}(\delta\mu)=0$). The red lines mark the instabilities. Note that only an instability that is related to a real eigenvalue that crosses zero is present (red dashed line). The starting point for each calculation is the physical fixed point at the previous $U$ value and the damping $p$ is determined by \cref{eq:uniform_damping} with $c=0.1$. In panels~(a)--(e), the channels (d,m,G,s,t) are shown and in panel~(f) the real part of eigenvalues of $\Pi^F$ are shown, where purely real eigenvalues are marked by magenta $+$ and complex eigenvalues are marked by green $\cross$. Note that for ph-symmetry only real eigenvalues appear in $\Pi^F$, which is in contrast to the $\Phi$ iteration in \cref{fig:ZP_dmu0}.}
    \label{fig:ZP_F_dmu0}
\end{figure}

With the damped iteration scheme for $\delta\mu=0$ and increasing $U$, we find, as expected, that the physical fixed point is unstable for weak coupling and becomes stable for strong coupling, while one always converges to the physical solution when using the stabilization method. 
By investigating the eigenvalues $\lambda^F$ of $\Pi^F$ (which is defined via the Jacobian $J^F=\mathbb{1}-p\,\Pi^F$), we find that there is only one instability, corresponding to a real eigenvalue that crosses zero.

Interestingly, we find that most of the instabilities are driven by the coupling between the channels, i.e., the channel off-diagonal terms in $\Pi^F$, while only a small instability close to the vertex divergence is associated with the density channel alone (see \cref{app:decoupled_channels}). 
As seen in the panel~(f) of \cref{fig:ZP_F_dmu0}, the vertex divergence also leads to a divergence of one eigenvalue of $\Pi^F$, but in contrast to the iteration in $\Phi$, this pole is even and, therefore, does not change the stability of the physical fixed point. This behavior can be understood when looking at, e.g., \cref{eq:J_F_dd}, where the denominator is squared such that it becomes an \emph{even}, instead of an odd pole.
The analogous results out of ph-symmetry are shown in \cref{fig:ZP_F_dmu1} in \cref{app:parquet_J_F}.

With its almost inverted stability properties, i.e., a maximal number of unstable eigendirections at weak coupling and a decrease of this number towards strong coupling, this alternative scheme reveals, in a very transparent fashion, that the stability properties of an iterative scheme can highly depend on how the equations are iterated.

\begin{figure}[!t]
    \centering
    \includegraphics[width=0.8\textwidth]{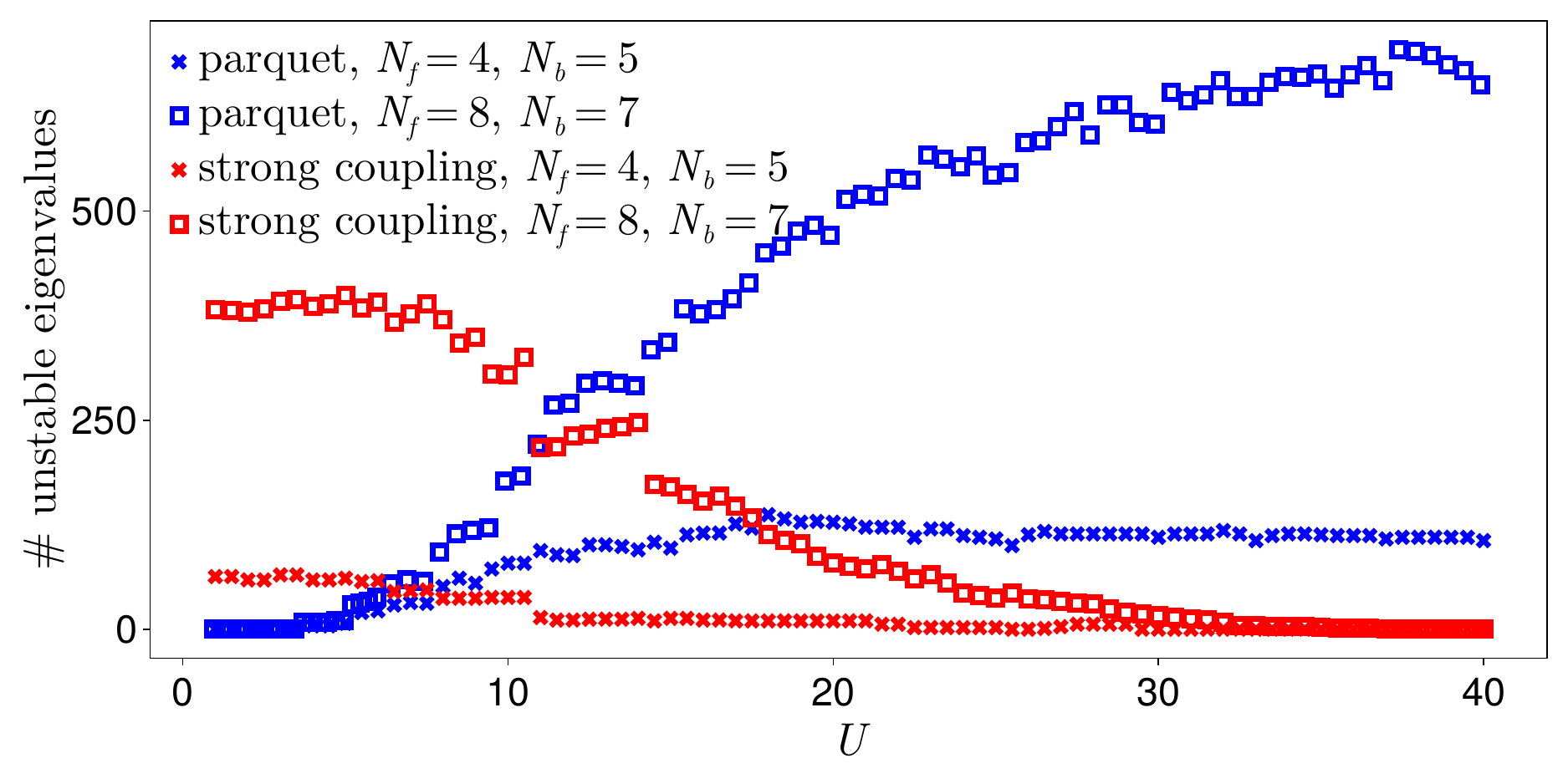}
    \caption{Number of unstable eigenvalues, i.e., $\#\mathrm{Re}(\lambda^{(F)})<0$, at half-filling in the HA for increasing interaction $U$, for the conventional parquet iteration (blue) and the strong-coupling iteration (red) for two different frequency boxes (different markers).}
    \label{fig:strong_it_HA}
\end{figure}

However, the maximal number of unstable eigendirections in the HA is formally infinite due to the infinite number of Matsubara frequencies. Therefore, the physical fixed point of the strong-coupling iteration is never stable with respect to the damped iteration; if it is stable at strong coupling, this is an effect of the box truncation. 
To visualize this behavior, \cref{fig:strong_it_HA} presents the number of unstable eigenvalues, i.e., $\#\mathrm{Re}(\lambda^{(F)})<0$, for the usual parquet iteration and the strong-coupling iteration at half-filling with increasing $U$.
As expected, the number of unstable eigenvalues increases with $U$ for the conventional parquet iteration, while it decreases for the strong-coupling iteration. Further, one sees the described effect of the finite frequency box for the strong-coupling iteration, i.e., that, for an infinitely large box, one would always obtain unstable eigenvalues.

\subsection{Comparison of different fixed-point iteration schemes}\label{sec:method_comparison}
As discussed in the previous sections, the  problem of misleading convergence in the parquet equations stems from the fact that the physical fixed point is locally unstable  with regard to the conventional damped iteration when entering the strong-coupling regime. 
One possibility to stabilize this physical fixed point is the introduced stabilization method. However, there also exist other iterative methods in which the physical fixed point is kept stable for high interaction values.

One possibility is the \emph{Newton--Raphson method}~\cite{Suli2003}, which can be used to find the root of $\mathcal{H}(\Psi)$
\begin{align}
\label{eq:root_eq}
    \mathcal{H}(\Psi) = \Psi - \mathcal{F}(\Psi).
\end{align}
Note that the Jacobian of $\mathcal{H}$ is simply $\Pi$, which is defined in Eq.~\eqref{eq:J_general}.

Using the (damped) Newton--Raphson method, $\Psi$ is updated by
\begin{align}
\label{eq:Newton}
    \Psi^{(n+1)} = \Psi^{(n)} - p\left[\Pi\left(\Psi^{(n)}\right)\right]^{-1}\cdot\mathcal{H}(\Psi^{(n)}).
\end{align}
Since for $p=1$, the Jacobian of Eq.~\eqref{eq:Newton} is zero at all fixed points of $\mathcal{F}$, all fixed points of $\mathcal{F}$ are locally stable with respect to the Newton--Raphson method. To which fixed point one converges is then only determined by the starting point of the iteration.

The Newton--Raphson method converges faster than damped iteration and therefore also than the stabilization method (quadratic vs.\ linear order of convergence), but it requires the calculation of $\Pi$ in each iteration while for the stabilization method the calculation of $\Pi$ is necessary only once per parameter point. 

To overcome this drawback, instead of the Newton--Raphson method, we resort, for our calculations, to a quasi-Newton method, namely the \emph{Chord method}~\cite{Kelly1995}. For the Chord method, $\Pi\left(\Psi^{(n)}\right)$ in Eq.~\eqref{eq:Newton} is replaced by $\Pi\left(\overline{\Psi}\right)$, where $\overline{\Psi}$ is chosen such that $\Pi\left(\overline{\Psi}\right)$ approximates $\Pi\left(\Psi^{(n)}\right)$ well enough. To compare to our proof-of-principle calculation, we use $\overline{\Psi}=\Psi_\text{phys}$. 

\begin{figure}[!t]
    \centering
    \includegraphics[scale=0.5]{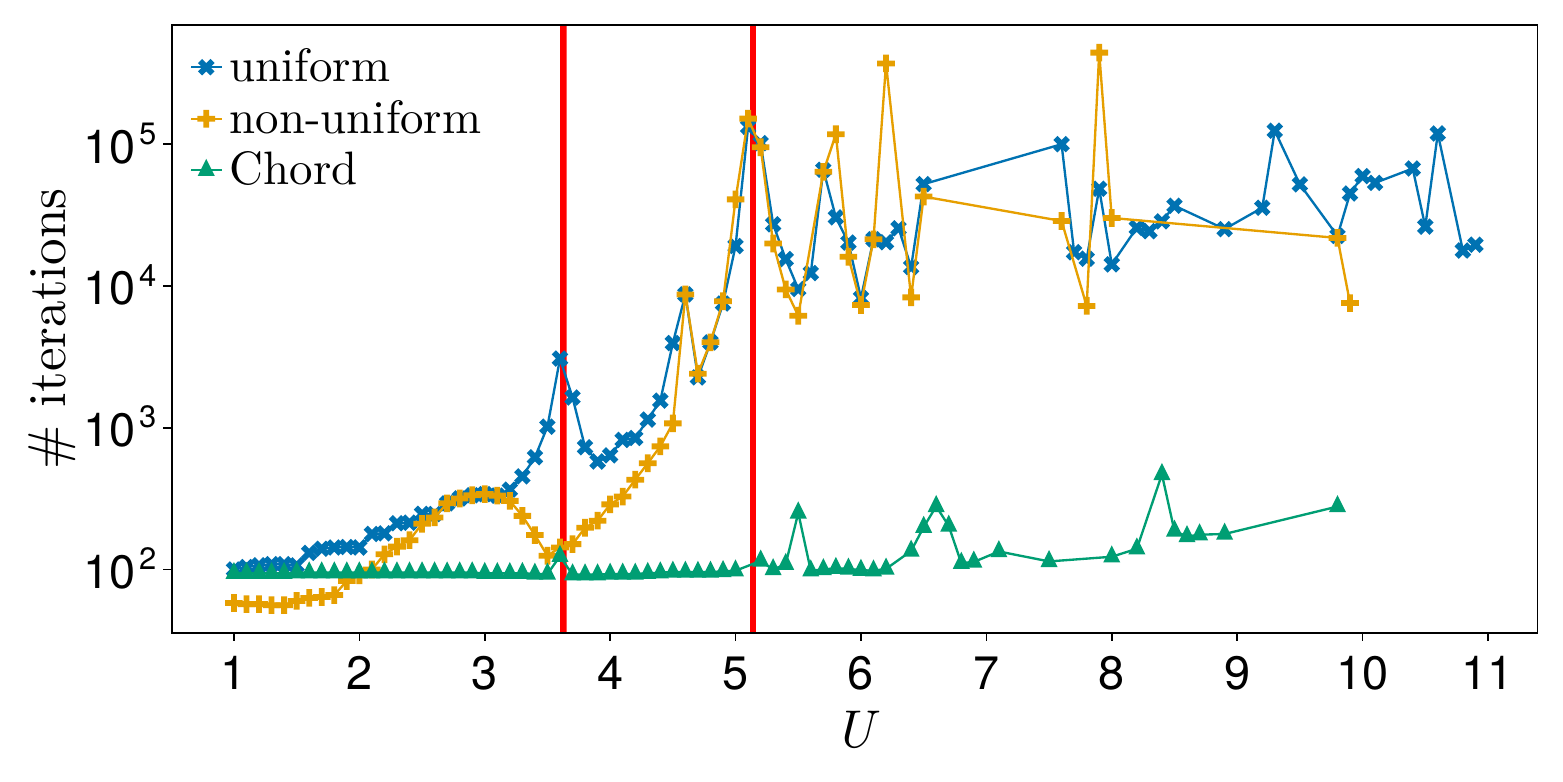}
    \caption{Number of iterations for different iteration schemes over interaction $U$ for ph-symmetry in the HA. To take into account problems with misleading convergence, only calculations where the methods converged to the physical solution are shown. The first two vertex divergences are marked with solid red lines. The (non-)uniform damping data is calculated with the stabilized iteration where the damping is calculated by \cref{eq:uniform_damping} [\cref{eq:daa_nonuniform}] with $c=0.1$. For the Chord method, a uniform damping $p=0.1$ is used. Convergence is defined via \cref{eq:conv_criterion} with $\epsilon=10^{-6}$, identically for all methods.}
    \label{fig:method_comp_it}
\end{figure}

Since the computational cost per iteration is approximately the same for the stabilized iteration and the Chord method, we compare the performances of the considered methods by inspecting the number of iterations the algorithm needs. To explicitly take the problem of misleading convergence into account, only runs where the method converged to the physical fixed point are shown in Fig.~\ref{fig:method_comp_it}. For all methods, convergence is declared once the maximum relative change between consecutive iterations of the iterated quantities $\Psi=(\Phi_\text{d},\Phi_\text{m},\Phi_\text{s},\Phi_\text{t},G)$ drops below a threshold $\epsilon$, i.e.,
\begin{align}
\label{eq:conv_criterion}
    \max_{X\in\{\Phi_\text{d},\Phi_\text{m},\Phi_\text{s},\Phi_\text{t},G\}}\frac{\max\left|X^{(n)}-X^{(n-1)}\right|}{\max\left|X^{(n-1)}\right|}<\epsilon,
\end{align}
where the maxima in the numerator and denominator are taken over the elements inside the frequency box, and we use $\epsilon=10^{-6}$ identically for all methods.

The results are the following: We find that the Chord algorithm either converges to the physical fixed point or not at all, i.e., we do not find a misleading convergence for the parameters we used. However, this is most likely a consequence of our initial values, which are close to the physical fixed point; how this transfers to more realistic situations remains to be investigated. Further, we find that the Chord method converges much faster than the stabilized iteration scheme, whereas the corresponding interaction interval of convergence shrinks a bit w.r.t. the stabilization method. One reason for the faster convergence is that a much lower damping is needed for the Chord algorithm in comparison to the stabilized iteration. 
By comparison with identical damping for the Chord and the stabilization method, we find that the advantage in convergence speed is only minor and that the convergence of the Chord algorithm is still restricted to slightly lower values of $U$ than for the stabilized iteration.

In spite of its fast convergence, we should stress that, in general, one potential drawback of the Chord method is that, as for the Newton--Raphson method, all fixed points are stable. 
Therefore, the convergence to the physical or an unphysical fixed point is only controlled by the starting point of the algorithm. 
This is especially problematic in the cases where a physical fixed point crosses an unphysical one, which plausibly happens when a real eigenvalue of $\Pi$ crosses zero \cite{Essl2026}. For the stabilized iteration such cases are less problematic since (at least locally) only the physical fixed point is stable \cite{Essl2026}. 
However, it could happen that these problems can be mitigated by using a better starting point for the iteration as outlined in \cref{sec:practical_implementation}.

\begin{figure}[!t]
    \centering
    \includegraphics[scale=0.5]{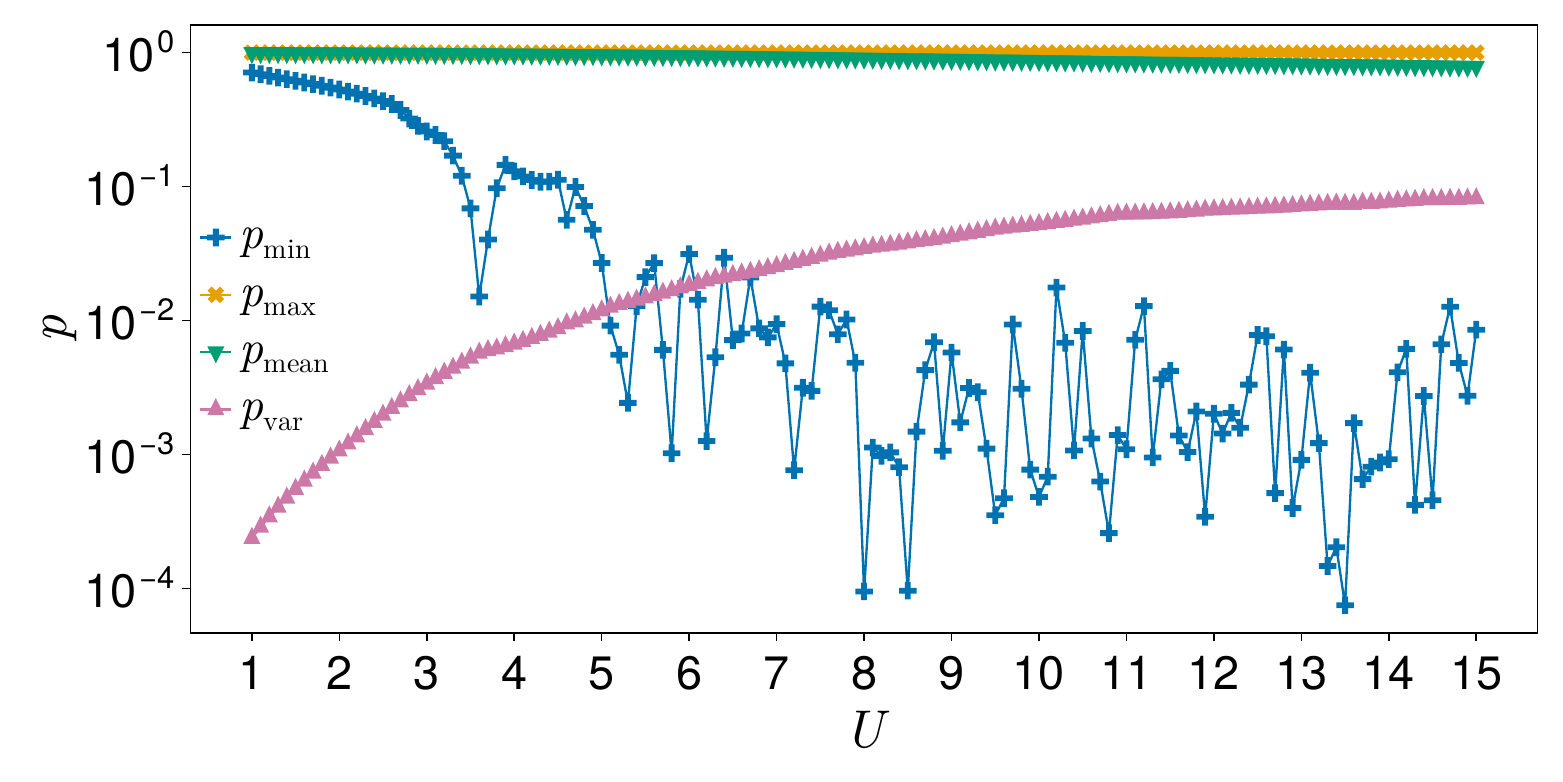}
    \caption{Minimum, maximum, mean and variance for the absolute value of the damping values for different eigenvector directions which are calculated with \cref{eq:daa_nonuniform} ($c=1$) are shown against $U$ at ph-symmetry for the HA.}
    \label{fig:p_values}
\end{figure}

By comparing the uniform damping, \cref{eq:daa}, in which the same damping parameter is applied to all eigendirections according to \cref{eq:uniform_damping}, with the non-uniform damping scheme, where each eigendirection is assigned an individual damping value determined by \cref{eq:daa_nonuniform}, we observe two main features.
First, the non-uniform scheme does not provide a significant speed-up in convergence. Second, it appears to be less stable than the uniform damping scheme. This suggests that the eigenvector directions which have to be damped the most constitute the bottleneck for the convergence speed. 
To investigate this further, we calculate the damping for each eigenvalue direction of the physical solution, \cref{eq:daa_nonuniform}, and plot the minimum, maximum, mean and variance of the absolute values of these damping values for the path at half-filling in \cref{fig:p_values}.
We find that the $p$ value for the eigendirection that requires the strongest damping decreases with increasing $U$, but seems to saturate around $10^{-2}{-}10^{-4}$ for large $U$ values. Further, since $p_\text{max}=1$ for all $U$ values, there are always eigendirections that do not require damping. Moreover, the mean value has no strong dependence on $U$ and is close to $1$, which suggests that most of the directions do not require much damping. However, since the variance increases, there are more and more directions that need damping for larger $U$ values.

Further, we present a selection of alternative algorithms that can be used to obtain a self-consistent solution of the parquet problem but do not have the same cost per iteration as the stabilized iteration and the Chord method.
Specifically, we choose to investigate three different algorithms:
\begin{itemize}
    \item \emph{Anderson acceleration} \cite{anderson1965} which is a quasi-Newton (multisecant) method \cite{fang2009} where the information of the $m$ previous iterations is used to approximate the Jacobian at the current iteration. It is reported that Anderson acceleration can converge to fixed points that are unstable with regards to damped iteration \cite{pollock2021}. However, when and how exactly Anderson acceleration is able to converge to these fixed points represents still a subject of active research \cite{fang2009,walker2011,pollock2021,yang2022}. 
    \item \emph{Broyden's method} \cite{Broyden1965} is a quasi-Newton method where the exact Jacobian is not calculated at every iteration, but is instead determined iteratively by a finite-difference approximation. As for the Newton--Raphson method, every root of the problem is a \emph{locally} stable fixed point for Broyden's method. Therefore, only the initial value of the iteration determines the solution to which one converges.
    \item \emph{Jacobian-free Newton--Krylov method} (JFNK method) \cite{Knoll2004}, which is a generalized minimal residual method (GMRES) \cite{Saad1986} that approximates the solution of the root equation by a vector in a Krylov subspace. This Krylov subspace is built by the action of the Jacobian matrix on a vector; therefore, the Jacobian matrix never needs to be stored. Again every root of the problem is a \emph{locally} stable fixed point for the JFNK method.
\end{itemize}

To conduct a preliminary study of these methods on the parquet problem, we calculate a self-consistent solution of the HA at half-filling, where the Anderson acceleration is implemented using the \texttt{NLsolve.jl} package~\cite{Mogensen2020} and the \texttt{NonlinearSolve.jl} package~\cite{Pal2024} is used for the implementation of Broyden's method and the JFNK method. For all methods, the same convergence criterion, \cref{eq:conv_criterion}, and initial values are used as for the previous calculations.

\begin{figure}[!t]
    \centering
    \includegraphics[width=\textwidth]{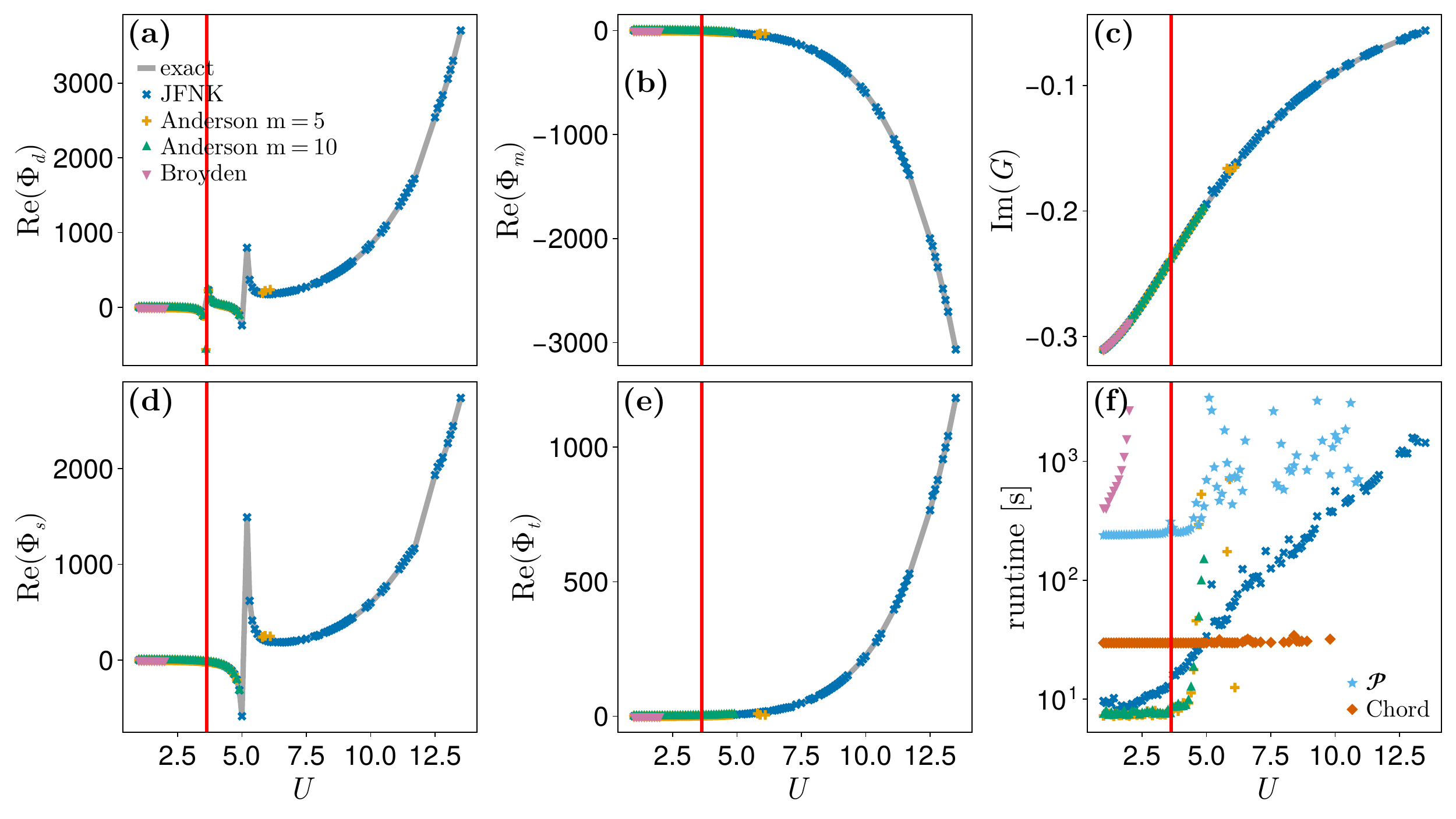}
    \caption{Converged parquet results for the HA for the Jacobian-free Newton--Krylov (JFNK) method, Anderson acceleration (with different memory depths $m$) and Broyden's method vs.\ exact values of Re[$\Phi_r(\pi/\beta,\pi/\beta,0)$] for the (a) density, (b) magnetic, (d) singlet, and (e) triplet channels, and (c) Im[$G(\pi/\beta)$] for $\delta\mu=0$, $\beta = 1$ and $N_f=12$, $N_b=13$. For all calculations, the convergence criterion \cref{eq:conv_criterion} with $\epsilon = 10^{-6}$, an initial value with $\Delta U_\mathrm{init}=0.001$, and a uniform damping $p=0.1$ are used. In panel~(f), the runtime per $U$ value in seconds is shown (all runs are performed on four CPU cores on an \href{https://www.amd.com/en/products/processors/server/epyc/4th-generation-9004-and-8004-series/amd-epyc-9654.html}{AMD EPYC\texttrademark\, 9654 96-Core Processor}); for comparison the runtimes of the stabilized iteration [\cref{fig:HA-Nf-12-Nb-13}] and the Chord method [\cref{fig:method_comp_it}] are included as well. The first vertex divergence, which is also the first instability for the damped iteration, is marked by a red line.}
    \label{fig:solver_comp}
\end{figure}

In \cref{fig:solver_comp}, we find that both Broyden's method and the JFNK method either converge to the physical fixed point or not at all. As for the Chord method, this is most likely a consequence of the initial values used. How this transfers to more realistic situations represents an interesting path for future studies. 
Broyden's method converges only for very small $U$ values and ceases to converge even before the first vertex divergence, which is also the first instability of the damped iteration (marked by a red line in \cref{fig:solver_comp}).
Interestingly, the JFNK method converges to the physical fixed point up to very large $U$ values. This is especially intriguing since JFNK is less memory intensive than Broyden's method, because the Jacobian matrix is never built explicitly.
We find that the number of Krylov vectors that are calculated each iteration increases with increasing $U$. Specifically, for $U=1,\;5$ and $10$ the average number of Krylov vectors per iteration is $11,\;210$ and $1757$ which is still small in comparison to the full space  $4N_f^2N_b+N_f=7500$.

Our computations using Anderson acceleration even converge to the physical fixed point after the first vertex divergence; however, they cease to converge to the physical fixed point after the second vertex divergence.
Further, we find for $m=5/10$ four/one data point(s) that exhibit(s) misleading convergence. This demonstrates again that it is not clear \textit{a priori} whether Anderson acceleration exhibits misleading convergence or converges to the physical fixed point.

To compare the performance of the different solvers, panel~(f) of
\cref{fig:solver_comp} shows the runtime until convergence at each $U$ value. We use the runtime
rather than the number of iterations, since a single iteration does not have
the same computational complexity for the different algorithms and the
iteration count is therefore not a meaningful measure of comparison here. 
The absolute values of the runtime are evidently dependent on the machine. However, their relative magnitude still provides a useful
indication of the respective performances. For completeness, panel~(f) also shows the runtimes of the calculations with the stabilized iteration, whose results are presented in \cref{fig:HA-Nf-12-Nb-13}, and the Chord method [see \cref{fig:method_comp_it}].
Specifically, we find that the ``setup'' time, i.e., the calculation of the Jacobian and its diagonalization/inversion, is substantial for the stabilized iteration and the Chord method in comparison to the time needed for the iteration. This can be seen by the large offset at small $U$ values. Especially, the runtime of the Chord method, which has a smaller number of iterations than the stabilized iteration [see \cref{fig:method_comp_it}], is almost flat over $U$. This makes the algorithm very fast for large $U$ values. Further, we want to emphasize that the comparison of the runtime of the stabilized iteration with highly optimized code of the other algorithms might not be fair at this stage.
For Broyden's method, we find that it is the slowest among these algorithms in the
(small) interaction range, where it converges at all, whereas the Anderson
acceleration is the fastest at low $U$ values, while the JFNK method is the most stable for large $U$ values.

We want to conclude with emphasizing that the parquet iteration requires some form of stabilization when performing calculations at strong coupling. Whether the introduced stabilization method or alternative schemes such as the Chord method, Anderson acceleration with large depth $m$, or the JFNK method provide the most stable and efficient solution might be problem specific and has to be systematically investigated in the future.

\section{Route to a realistic implementation}\label{sec:practical_implementation}
The structure of stability/instability regions for the HA and the much simpler ZP model is qualitatively very similar (see \cref{fig:ZP_stability,fig:HA_stability}). This lets us expect that these features may also be qualitatively similar in more complicated situations. One situation we envisage is parquet D$\Gamma$A \cite{Toschi2007} for the Hubbard model on a 2D/3D lattice. Here, the fully irreducible vertex $\Lambda$ of the model, i.e., the input to parquet equations, is approximated by the fully irreducible vertex of the auxiliary impurity model of the dynamical mean-field theory (DMFT) \cite{georges1996}. Therefore, we believe that our investigation might provide useful general insights on the instability problem well beyond the model specifics of the HA and the ZP model.

A very important remark for the future applications of our approach is, however, due here:
For the proof-of-principle calculations in Secs.~\ref{sec:ZP_parquet} and \ref{sec:HA_parquet}, the evaluation of the Jacobian is performed at the physical fixed point that we want to calculate. Evidently, this is not possible for more advanced calculations (e.g., parquet D$\Gamma$A for the Hubbard model), since their fixed point is not known. However, we can still apply the stabilization method by starting at a low enough $U$ value such that all eigendirections are stable and thereafter use the Jacobian evaluated at the converged result of the previous $U$ value for the subsequent, slightly larger $U$ value. By doing so, at a certain point we need to change the signs of eigenvalues that we suspect will become unstable at the next $U$ value, i.e., the ones that either have very small (but positive) real part or approach a pole, as it was also discussed in Ref.~\cite{Essl2026}.

One can improve our basic scheme by not simply exploiting the converged solution at the previous $U$ value but by using predictors \cite{allgower2003}. Predictors are used in numerical continuation methods and use the previous solution (or a longer history) to predict a better starting point for the current $U$ value. Then not only the predictor can be used as a starting point but also the Jacobian for the current $U$ calculation can be evaluated at this prediction.
For the ZP model, we tested secant predictors and arclength predictors \cite{allgower2003} and found that secant predictors perform better.

A further, crucial problem for practical calculations are the vertex divergences: They lead to an odd pole in the reducible vertices $\Phi$ which, therefore, no longer continuously evolve with increasing $U$. As a result, the computed $\Phi$ slightly before the divergence is not continuously connected to the physical $\Phi$ slightly after the divergence. Therefore, the previously computed $\Phi$ might be too far from the physical fixed point to use it as a starting point for the next $U$ value.

While solving this problem would go beyond the scope of the present work, we outline here four different future directions that can be followed to possibly overcome this problem:

First, since vertex divergences usually vanish for large enough doping and/or magnetic field \cite{Essl2024,Vucicevic2018,Reitner2024}, one could try to bypass the vertex divergence by performing calculations on a path in parameter space in the vicinity of the singularity without crossing it and then use the converged solution on this close path as a starting point for the parameter point beyond the vertex divergence. This was successfully tested for the ZP model.

Second, as the full vertex $F$ evolves (excluding phase transitions) continuously in $U$, one may try to use the introduced \emph{strong-coupling iteration} in combination with stabilized iteration (\cref{sec:strong_parquet}). 

Third, in order to have vertex divergences in the parquet calculation a non-perturbative input for the fully irreducible vertices $\Lambda$ is needed, e.g., the $\Lambda$ of DMFT in the case of D$\Gamma$A. The pole structure of the $\Phi$ calculated by parquet is then essentially given by the local $\Phi$ of DMFT. One could therefore hope that it is sufficient to start from the local $\Phi$ of DMFT in order to capture the correct behavior near the vertex divergences. 

Finally, another promising direction could be the finite-difference parquet method that was introduced in Ref.~\cite{Lihm2026}. This method uses the difference between the parquet calculation and a reference system in order to eliminate the vertex divergences. Combination of the finite-difference parquet method with our stabilization scheme may thus lead to a practical and efficient solution of this problem on the whole phase space of the model. This also includes investigating to which extent the convergence of the finite-difference parquet method is due to the fact that Ref.~\cite{Lihm2026} uses Anderson acceleration together with a preconditioned fixed-point iteration to solve the self-consistent equations.

Another issue that arises in realistic calculations is the sheer size of the reducible vertices if also momenta and/or multiple orbitals need to be considered. In fact, the storage needed to compute the Jacobian can become quickly too large in realistic lattice calculations. One possible way to mitigate this problem is considering the Jacobian only on a subgrid of the full parquet calculation. This should be sufficient since, empirically, the unstable eigenvalues are connected to low Matsubara frequencies. However, as discussed in detail in \cref{app:n-eigenvalues}, this procedure can only be carried out for channels that do \emph{not} exhibit vertex divergences (or pseudo-divergences in $\chi_0^{-1}\chi$) since the divergences lead to a grid-dependent number of unstable eigenvalues.
Therefore, a more promising route is to exploit the quantics tensor trains (QTT)~\cite{Oseledets2009,Shinaoka2023,Ritter2024, Rohshap2025a, Rohshap2025b, Rohshap2025c, Grosso2026} to compress the vertices and perform the parquet calculation in this compressed space, as tested in Ref.~\cite{Rohshap2025a}. 
Formulating the stabilization method, or a less memory-intensive method such as the JFNK method, in this compressed space could pave the way for realistic calculations.

At the same time, the single-boson exchange (SBE) formalism \cite{Krien2020b,Gievers2022Multiloop} can be used to reduce the size of the quantities used in parquet calculations. This is due to the fact that the 4-point functions can be calculated on a smaller grid compared to conventional parquet calculations. Further, Ref.~\cite{Gievers2026Instabilities} suggests that instabilities occur even in the \emph{embedded multi-boson exchange} \cite{Kiese2024b} approach, in which the divergences of the irreducible vertex do not appear.

\section{Conclusion}\label{sec:conclusion}
By investigating the local stability of the physical fixed point of the parquet iteration, we were able to clarify the origin of the misleading convergence in the parquet iteration. 
In contrast to self-consistent perturbation theory where the misleading convergence is a result of \emph{both} the multivaluedness of the LWF and the unstable fixed point \cite{Kozik2015,Rossi2015,vanHoucke2024,Essl2026}, the parquet iteration is only affected by the instability of the physical fixed point. 
This is due to the fact that, differently from the self-consistent perturbation theory, the parquet formalism does not rely on the validity of the skeleton expansion. With our investigation, we showed that the divergence of the irreducible vertex $\Gamma$ is a sufficient but not a necessary condition for the instability of the physical fixed point with regard to the damped iteration. 
We also show that, by introducing a strong-coupling iteration scheme for solving the parquet equations, where the iterated object is not $\Phi$ but the full vertex $F$, the stability properties of the iteration, solved by damped iteration, change. The physical fixed point becomes unstable at weak coupling, but the number of unstable eigenvalues decreases with increasing $U$.

Our investigation of the instabilities of the parquet scheme can be directly used in practical computations to calculate the minimal damping value that is necessary to converge to the physical fixed point. Furthermore, building on the findings of Ref.~\cite{Essl2026}, we introduce the \emph{stabilized iteration} for the parquet equations by using their Jacobian. This approach \emph{locally} stabilizes the physical fixed point and allows, therefore, convergence to the physical fixed point in regimes where this can no longer be achieved by increasing the damping of the iteration. 

We provide a proof of principle for all these algorithmic features by means of the Hubbard atom and the zero-point model. In a related work \cite{Gievers2026Instabilities}, the same models have been used to study the stabilization of various self-consistent  diagrammatic approaches based on the single-boson exchange formalism.

Further, we also compare the \emph{stabilized iteration} to a selection of alternative algorithms that can be used to solve the parquet problem. Specifically, we investigate Anderson acceleration, the Chord method, Broyden's method, and the Jacobian-free Newton--Krylov method. In our preliminary comparison, the Chord method is the most efficient at strong coupling, and the Jacobian-free Newton--Krylov method converges up to the largest $U$. For both, however, every fixed point is locally stable, so that the selected solution is determined solely by the initial condition. The stabilized iteration, by contrast, stabilizes the physical fixed point specifically. Which algorithm to use in realistic calculations is a subject for future investigations.

Although our methodology is general and applicable to realistic calculations, such as D$\Gamma$A, the numerical cost of computing the Jacobian for momentum-dependent vertices is currently not manageable. Vertex compression techniques~\cite{Rohshap2025a, Wallerberger2021, Badr2024}  might mitigate this cost. 
However, future work is needed on how to obtain a sufficiently good starting point for the iteration.

\section*{Acknowledgments}
The authors thank Samuel Badr, Jan von Delft, Fabian Kugler, Jae-Mo Lihm, and Nepomuk Ritz for insightful discussions. Further, we thank Hugo Strand for suggesting, in his referee report, the Jacobian-free Newton--Krylov method, which turned out to be a competitive fixed-point iteration for the parquet problem. This work was funded in part  by the Austrian Science Fund (FWF) projects through Grant DOI 10.55776/P36332,  10.55776/V1018, 10.55776/PIN4372024 and 10.55776/I5487. For open access purposes, the authors have applied a CC BY public copyright license to any author-accepted manuscript version arising from this submission. Calculations have been partly performed using Austrian Scientific Computing (ASC). We acknowledge the use of large language models (LLMs) for assistance in the preparation of this manuscript, particularly for text editing and language refinement. The authors take full responsibility for the content.

\section*{Data and code availability}
A dataset containing all numerical data and scripts for calculating and plotting these data is publicly available in the TU Wien Research Data repository \cite{data}.

\appendix

\section{Definitions}
\label{app:defs}

\subsection{One-particle quantities}
We start by introducing the one-particle Green's function in Matsubara frequency space $G_{\sigma}({\nu})$, which is defined as the Fourier transform of the imaginary-time-ordered two-point correlation function:
\begin{align}
    G_{\sigma}(\nu) = -\int_0^{\beta}\!\!d\tau\, e^{i \nu \tau} \langle T_{\tau} \hat{c}_{ \sigma}(\tau) \hat{c}_{\sigma}^{\dagger}(0) \rangle \,,
\end{align}
with imaginary time $\tau$ and fermionic Matsubara frequencies $\nu = (2n+1)\pi / \beta, n \in \mathbb{Z}$. The Dyson equation,
\begin{align}
   G_{\sigma}(\nu) = \frac{1}{G^{-1}_{0,\sigma}(\nu)- \Sigma_{\sigma}(\nu)},
   \label{eq:dyson}
\end{align}
connects the one-particle Green's function with the non-interacting Green's function of the HA,
\begin{align}
    G_{0,\sigma}(\nu) = \frac{1}{i\nu + U/2 +\delta\mu},
\end{align}
and the self-energy $\Sigma_{\sigma}(\nu)$.

\subsection{Two-particle quantities}
Fourier transformation of the imaginary-time-ordered four-point correlator leads to the two-particle Green's function in Matsubara frequencies:
\begin{align}
    G_{\sigma_1\ldots\sigma_4}^{\nu_1\nu_2\nu_3} = \int_0^{\beta}\!\!\!d\tau_1 \int_0^{\beta}\!\!\!d\tau_2\int_0^{\beta}\!\!\!d\tau_3\;e^{i \nu_1 \tau_1+i\nu_2 \tau_2+i \nu_3 \tau_3 
    }  \langle T_{\tau} \hat{c}_{ \sigma_1}(\tau_1) \hat{c}_{\sigma_2}^{\dagger}(\tau_2)\hat{c}_{ \sigma_3}(\tau_3) \hat{c}_{\sigma_4}^{\dagger}(0) \rangle,
    \label{eq:G2}
\end{align}
with the three fermionic Matsubara frequencies $\nu_1,\nu_2, \nu_3$. In general, it is useful to move to the Bethe--Salpeter picture (so-called \emph{channel native} picture), where the two-particle quantities are expressed in terms of two fermionic frequencies $\nu, \nu'$ and one bosonic Matsubara frequency $\omega$. This allows for a native parametrization of the Bethe--Salpeter equations (BSEs) (see Sec.~\ref{sec:bse_eq}) in the particle-hole (ph)  ($\omega=\nu_1+\nu_2$), the particle-particle (pp)  ($\omega=\nu_1+\nu_3$), and the transversal particle-hole ($\overline{\mathrm{ph}}$) channels ($\omega=\nu_2+\nu_3$). Following Refs.~\cite{Thunstroem2018, Rohshap2025a}, we use the following frequency convention in the ph\ and pp\ channels
\begin{align}
    \label{eq:ph_pp_notation}
    \textbf{ph: } &\nu_1 = -\nu  \qquad \qquad  &\textbf{pp: } \nu_1 &= -\nu \\
    &\nu_2 = \nu+\omega   &\nu_2 &= -(\nu'+\omega) \nonumber \\
    &\nu_3 = -(\nu'+\omega)  &\nu_3 &= \nu + \omega \nonumber 
\end{align}
where, in the following, we will note the pp\ notation explicitly by a subscript, while the quantities are in ph\ notation if we do not write a subscript.
Although this channel-native description simplifies the BSEs, the parquet equations mix vertex functions of different channel parametrizations. This makes frequency transformations between different channels necessary (see Apps.~A and C in Ref.~\cite{Rohshap2025a} for more details).

The two-particle Green's function, parametrized in the ph\ frequency convention has the following form
\begin{align}    G_{\sigma_1\ldots\sigma_4}^{\nu\nu'\omega} = &\;
    G_{\sigma_1}(\nu) G_{\sigma_3}(\nu')\delta_{\omega 0}\,\delta_{\sigma_1\sigma_2} \delta_{\sigma_3\sigma_4}
    - G_{\sigma_1}(\nu) G_{\sigma_2}(\nu+\omega)\delta_{\nu \nu'}\,\delta_{\sigma_1\sigma_4} \delta_{\sigma_2\sigma_3} \nonumber \\
    & -
    G_{\sigma_1}(\nu) G_{\sigma_2}({\nu+\omega}){F}_{\sigma_1\ldots\sigma_4}^{\nu \nu'\omega}G_{\sigma_3}(\nu'+\omega)G_{\sigma_4}(\nu').
    \label{eq:Fdef2}
\end{align}
It consists of two disconnected terms and a connected contribution, where the full two-particle vertex $F$ parametrizes two-particle scatterings in the system. In the pp\ frequency parametrization, we can write the two-particle Green's function as follows:
\begin{align}
    G_{\sigma_1\ldots\sigma_4, \mathrm{pp}}^{\nu\nu'\omega} = &\;
    G_{\sigma_1}(\nu) G_{\sigma_3}(\nu')\delta_{\omega+\nu+\nu', 0}\,\delta_{\sigma_1\sigma_2} \delta_{\sigma_3\sigma_4}
    - G_{\sigma_1}(\nu) G_{\sigma_2}(-\nu-\omega)\delta_{\nu \nu'}\,\delta_{\sigma_1\sigma_4} \delta_{\sigma_2\sigma_3} \nonumber \\
    & -
    G_{\sigma_1}(\nu) G_{\sigma_2}({-\nu-\omega}){F}_{\sigma_1\ldots\sigma_4,\mathrm{pp}}^{\nu \nu'\omega}G_{\sigma_3}(-\nu'-\omega)G_{\sigma_4}(\nu').
    \label{eq:Fdef3}
\end{align}
Using $\mathrm{SU}(2)$ symmetry of the discussed models, we can limit ourselves to the computation of two-particle 
vertices for selected spin combinations only, namely ${\uparrow\uparrow\uparrow\uparrow}$ and ${\uparrow\uparrow\downarrow\downarrow}$ (the other combinations are either zero due to spin conservation or can be obtained  by using  $\mathrm{SU}(2)$ symmetry, e.g., $G_{\uparrow\downarrow\downarrow\uparrow} =G_{\uparrow\uparrow\uparrow\uparrow} - G_{\uparrow\uparrow\downarrow\downarrow}$).
Furthermore, since $G:=G_\uparrow=G_\downarrow$, we can neglect spin indices from the one-particle objects $G_0$, $G$, and $\Sigma$.

Introducing the following spin-diagonalized conventions leads to a decoupling of the BSEs in the spin variable:
\begin{subequations}
\begin{align}
    F_\mathrm{d} &= F_{\uparrow \uparrow\uparrow\uparrow} +F_{\uparrow \uparrow\downarrow\downarrow},  \\
    F_\mathrm{m} &= F_{\uparrow \uparrow\uparrow\uparrow} -F_{\uparrow\uparrow \downarrow\downarrow},  \\
    F_\mathrm{s} &= F_{\uparrow\uparrow\uparrow \uparrow,\mathrm{pp}} -F_{\uparrow \uparrow\downarrow\downarrow,\mathrm{pp}},  \\
    F_\mathrm{t} &= F_{\uparrow \uparrow\uparrow\uparrow,\mathrm{pp}} \label{eq:spins},
\end{align}
\end{subequations}
with the d(ensity) and m(agnetic) channels in ph-notation and s(inglet) and t(riplet) channels in pp-notation.

\subsubsection{Bethe--Salpeter equations}
\label{sec:bse_eq}

The BSEs are the two-particle analogue of the one-particle Dyson equation \eqref{eq:dyson}, where the one-particle Green's function is related to its one-particle irreducible part --- the self-energy. However, the definition of two-particle reducibility is more complex than in the one-particle case since it can be defined with respect to the different channels resulting in multiple BSEs. 
Here, the two-particle reducible vertices $\Phi_r$ in channel $r$ and the corresponding two-particle irreducible vertices $\Gamma_r$ are related in the following way, where the chosen parametrizations lead to decoupled equations in bosonic frequency $\omega$ and spin components d\text{(ensity)}, m\text{(agnetic)}, s\text{(inglet)}, t\text{(riplet)}:
\begin{subequations}
\begin{align}
    &\Phi_\mathrm{d}^{\nu \nu' \omega} =  - \frac{1}{\beta^2} \sum_{\nu_1 \nu_2} \Gamma_\mathrm{d}^{\nu \nu_1 \omega} \chi_{0,\mathrm{ph}}^{\nu_1 \nu_2 \omega} F_\mathrm{d}^{\nu_2 \nu' \omega}, \label{eq:bse-density} \\
    & \Phi_\mathrm{m}^{\nu \nu' \omega} =  - \frac{1}{\beta^2} \sum_{\nu_1 \nu_2} \Gamma_\mathrm{m}^{\nu \nu_1 \omega} \chi_{0,\mathrm{ph}}^{\nu_1 \nu_2 \omega} F_\mathrm{m}^{\nu_2 \nu' \omega}; \\
    & \Phi_\mathrm{s}^{\nu \nu' \omega} =  + \frac{1}{\beta^2} \sum_{\nu_1 \nu_2} F_\mathrm{s}^{\nu \nu_1 \omega} \chi_{0,\mathrm{pp}}^{\nu_1 \nu_2 \omega} \Gamma_\mathrm{s}^{\nu_2 \nu' \omega},\\
    & \Phi_\mathrm{t}^{\nu \nu' \omega} =  - \frac{1}{\beta^2} \sum_{\nu_1 \nu_2} F_\mathrm{t}^{\nu \nu_1 \omega} \chi_{0,\mathrm{pp}}^{\nu_1 \nu_2 \omega} \Gamma_\mathrm{t}^{\nu_2 \nu' \omega}, 
\end{align}
\label{eq:bse}
\end{subequations}
with the bare susceptibilities
\begin{subequations}
\label{eq:bubble_HA}
\begin{align}
    &\chi_{0,\mathrm{ph}}^{\nu \nu' \omega}=-\beta G(\nu)G(\nu+\omega) \delta_{\nu \nu'},\\
    &\chi_{0,\mathrm{pp}}^{\nu \nu' \omega}=-\frac{\beta}{2}G(\nu)G(-\nu-\omega) \delta_{\nu \nu'}.
\end{align}
\end{subequations}
The full vertex $F$ contains all two-particle reducible and irreducible contributions and, thus, 
\begin{align}
\label{eq:F_gamma_phi}
    F^{\nu \nu' \omega}_r=\Gamma^{\nu \nu' \omega}_r+\Phi^{\nu \nu' \omega}_r.
\end{align}

\subsubsection{\label{sec:parquet}Parquet decomposition}

The contributions from the different BSEs are mixed by the parquet equation (see Ref.~\cite{Rohshap2025a} for details): 
\begin{align}
F = \Lambda + \Phi_{\mathrm{ph}} + \Phi_{\overline{\mathrm{ph}}} + \Phi_{\mathrm{pp}},
\label{eq:parquet_ph}
\end{align}
where $\Phi_{\mathrm{ph} / \overline{\mathrm{ph}}}$ denote contributions coming from $\Phi_\mathrm{d}$ or $\Phi_\mathrm{m}$ in the respective {ph} and {$\overline{\mathrm{ph}}$} parametrizations and $\Phi_{\mathrm{pp}}$ contributions from $\Phi_\mathrm{s}$ or $\Phi_\mathrm{t}$. $\Lambda$ represents the fully two-particle irreducible vertex containing all diagrams that are not two-particle reducible in any of the channels, and can therefore not be generated by the BSEs. Since we chose the channel-native parametrization, the parquet equations have to be transformed into d,m,s,t-channels. This leads to the following set of equations: 
\begin{subequations}
\begin{align}
    F_\mathrm{d}^{\nu \nu' \omega} \!&=\! \Lambda_\mathrm{d}^{\nu \nu' \omega} \!+\!  \Phi_\mathrm{d}^{\nu \nu' \omega} \!-\!\tfrac{1}{2} \Phi_\mathrm{d}^{\nu (\nu+\omega) (\nu'-\nu)}\!-\! \tfrac{3}{2} \Phi_\mathrm{m}^{\nu (\nu+\omega) (\nu'-\nu)} \!+\! \tfrac{1}{2} \Phi_\mathrm{s}^{\nu \nu' (-\omega -\nu -\nu')} \!+\! \tfrac{3}{2} \Phi_\mathrm{t}^{\nu \nu' (-\omega -\nu -\nu')} \,, \\
    F_\mathrm{m}^{\nu \nu' \omega} \!&=\! \Lambda_\mathrm{m}^{\nu \nu' \omega} \!+\!     \Phi_\mathrm{m}^{\nu \nu' \omega} \!-\!\tfrac{1}{2} \Phi_\mathrm{d}^{\nu (\nu+\omega) (\nu'-\nu)} \!+\! \tfrac{1}{2} \Phi_\mathrm{m}^{\nu (\nu+\omega) (\nu'-\nu)}
    \!-\! \tfrac{1}{2} \Phi_\mathrm{s}^{\nu \nu' (-\omega -\nu -\nu')} \!+\! \tfrac{1}{2} \Phi_\mathrm{t}^{\nu \nu' (-\omega -\nu -\nu')} \,,\\
    F_\mathrm{s}^{\nu \nu' \omega} \!&=\! \Lambda_\mathrm{s}^{\nu \nu' \omega} \!+\! \Phi_\mathrm{s}^{\nu \nu' \omega}\!+\!\tfrac{1}{2} \Phi_\mathrm{d}^{\nu \nu' (-\omega-\nu-\nu')} \!-\! \tfrac{3}{2} \Phi_\mathrm{m}^{\nu \nu' (-\omega-\nu-\nu')}
    \!+\! \tfrac{1}{2} \Phi_\mathrm{d}^{\nu (-\nu'-\omega) (\nu' -\nu)} \!-\! \tfrac{3}{2} \Phi_\mathrm{m}^{\nu (-\nu'-\omega) (\nu' -\nu)} \,,\\
    F_\mathrm{t}^{\nu \nu' \omega} \!&=\! \Lambda_\mathrm{t}^{\nu \nu' \omega} \!+\! \Phi_\mathrm{t}^{\nu \nu' \omega} \!+\! \tfrac{1}{2} \Phi_\mathrm{d}^{\nu \nu' (-\omega-\nu-\nu')} \!+\! \tfrac{1}{2} \Phi_\mathrm{m}^{\nu \nu' (-\omega-\nu-\nu')}
    \!-\! \tfrac{1}{2} \Phi_\mathrm{d}^{\nu (-\nu'-\omega) (\nu' -\nu)} \!-\! \tfrac{1}{2} \Phi_\mathrm{m}^{\nu (-\nu'-\omega) (\nu' -\nu)} \,,
\end{align}
\label{eq:parquet}
\end{subequations}
with the $r$-channel component $\Lambda_r$ of the fully irreducible vertex. 

\subsubsection{Schwinger--Dyson equation}

Full self-consistency among one- and two-particle properties in the iterative parquet scheme is achieved by including the Schwinger--Dyson equation (SDE) relating the two-particle vertex $F$ to the one-particle self-energy
\begin{align}
    \Sigma (\nu) = \frac{U}{\beta}\sum_{\nu'}G(\nu' ) - \frac{U}{2\beta^2} \sum_{\nu' \omega} (F_\mathrm{d}^{\nu \nu' \omega} - F_\mathrm{m}^{\nu \nu' \omega}) G(\nu') G(\nu'+\omega) G(\nu+\omega)
    \,,\label{eq:sde}
\end{align}
where $\frac{1}{\beta} \sum_{\nu}G(\nu)e^{i0^-} =n/2$ is the average particle density per spin.

\subsubsection{Exact expressions for the two-particle quantities in HA}\label{app:HA_quantities}

While two-particle Green's functions for the HA can be calculated analytically \cite{Pairault1998,Essl2024}, the vertex $\Gamma$ (and by extension $\Phi$ and $\Lambda$) requires the inversion of an infinite-dimensional matrix [\cref{eq:bse}]. Consequently, analytical expressions for $\Gamma$ are only available in the presence of ph-symmetry \cite{Thunstroem2018}.

Since in the ZP model all quantities are complex scalars, we can also calculate the vertices $\Gamma$, $\Phi$ and $\Lambda$ analytically which is done in \cref{app:ZP_quantities}.

\subsection{Calculation of two-particle quantities for the ZP model}\label{app:ZP_quantities}
Due to the fact that the ph and pp channel are related via the frequency shift $\chi_\text{ph}^{\nu\nu^\prime\omega}=\chi_\text{pp}^{\nu\nu^\prime(\nu+\nu^\prime+\omega)}$ \cite{rohringer2013}, there is no difference between the ph and pp channel for the generalized susceptibilities in the ZP model. Therefore, we drop this label in the following.
The one-particle Green's function can be explicitly calculated by evaluating the path integral in \cref{eq:S_ZP} (see Ref.~\cite{Rossi2015}):
\begin{align}
\label{eq:G_ZP}
    G_\sigma[G_0]=\frac{G_{0,-\sigma}^{-1}}{(G_{0,\sigma}^{-1}G_{0,-\sigma}^{-1}-U)}\quad\text{with}\quad G_{0,\sigma}^{-1}=\delta\mu+\sigma h,
\end{align}
\begin{subequations}
while the generalized susceptibilities can be calculated via the functional derivative:
\begin{align}
    \chi_{\sigma\sigma^\prime} &= \fdv{G_\sigma}{G_{0,\sigma^\prime}^{-1}}\\
    \Rightarrow \chi_{\uparrow\uparrow} &= -\frac{G_{0,\downarrow}^{-2}}{(G_{0,\uparrow}^{-1}G_{0,\downarrow}^{-1}-U)^2},\\
    \chi_{\uparrow\downarrow} &= -\frac{U}{(G_{0,\uparrow}^{-1}G_{0,\downarrow}^{-1}-U)^2}.
\end{align}
\end{subequations}
For $h=0$, we consider again four BSE channels for these generalized susceptibility (conventions from Ref.~\cite{Thunstroem2018}):
\begin{subequations}
\begin{align}
    \chi_\text{d} &= \chi_{\uparrow\uparrow} +\chi_{\uparrow\downarrow},\\
    \chi_\text{m} &= \chi_{\uparrow\uparrow} -\chi_{\uparrow\downarrow},\\
    \chi_\text{s} &= (-\chi_{\uparrow\uparrow} +2\chi_{\uparrow\downarrow} -2\chi_{0,\text{pp}})/4,\\
    \chi_\text{t} &= (\chi_{\uparrow\uparrow} +2\chi_{0,\text{pp}})/4,
\end{align}
\end{subequations}
where the bubble terms are
\begin{subequations}
\begin{align}
    \chi_{0,\text{ph}}&=-G^2=\chi_{0,\text{d}}=\chi_{0,\text{m}},\\
    \chi_{0,\text{pp}}&=-G^2/2=\chi_{0,\text{s}}=\chi_{0,\text{t}}.
\end{align}
\end{subequations}
The two-particle irreducible vertex $\Gamma$ is then calculated by inverting the BSEs \eqref{eq:bse}: $\Gamma_r=\chi_r^{-1}\mp \chi_{0,r}^{-1}$ where minus is taken for $r=\mathrm{d},\mathrm{m},\mathrm{t}$ and plus for $r=\mathrm{s}$:
\begin{subequations}
    \label{eq:Gamma_ZP}
    \begin{align}
    \label{eq:Gamma_d}
    \Gamma_\text{d} &=\frac{U(U-\delta\mu^2)^2}{\delta\mu^2(U+\delta\mu^2)},\\
    \label{eq:Gamma_m}
    \Gamma_\text{m} &=U\left(\frac{U}{\delta\mu^2}-1\right),\\
    \label{eq:Gamma_s}
    \Gamma_\text{s} &=2U\left(1-\frac{U}{\delta\mu^2}\right),\\
    \label{eq:Gamma_t}
    \Gamma_\text{t} &=0.
\end{align}
\end{subequations}
From this, the full vertex $F$ is calculated by $F_r=-\frac{\chi_r\mp \chi_{0,r}}{\chi_{0,r}^2}$:
\begin{subequations}
    \label{eq:F_ZP}
    \begin{align}
    \label{eq:F_d}
    F_\text{d} &=\frac{U(U-\delta\mu^2)^2}{\delta\mu^4},\\
    \label{eq:F_m}
    F_\text{m} &=-\frac{U(U-\delta\mu^2)^2}{\delta\mu^4},\\
    \label{eq:F_s}
    F_\text{s} &=\frac{2U(U-\delta\mu^2)^2}{\delta\mu^4},\\
    \label{eq:F_t}
    F_\text{t} &=0.
\end{align}
\end{subequations}
With this, the fully irreducible vertices $\Lambda$ can be calculated by:
\begin{subequations}
    \label{eq:Lambda_ZP}
    \begin{align}
    \label{eq:lambda_d}
    \Lambda_\text{d} &= \Gamma_\text{d}/2-3\Gamma_\text{m}/2+\Gamma_\text{s}/2+3\Gamma_\text{t}/2-2F_\text{d}=\frac{U(-2U^3+U\delta\mu^4+\delta\mu^6)}{\delta\mu^4(U+\delta\mu^2)},\\
    \label{eq:lambda_m}
    \Lambda_\text{m} &= 3\Gamma_\text{m}/2-\Gamma_\text{d}/2-\Gamma_\text{s}/2+\Gamma_\text{t}/2-2F_\text{m}=\frac{-U(-2U^3+U\delta\mu^4+\delta\mu^6)}{\delta\mu^4(U+\delta\mu^2)}=-\Lambda_\mathrm{d},\\
    \label{eq:lambda_s}
    \Lambda_\text{s} &= \Gamma_\text{s}+\Gamma_\text{d}-3\Gamma_\text{m}-2F_\text{s}=\frac{2U(-2U^3+U\delta\mu^4+\delta\mu^6)}{\delta\mu^4(U+\delta\mu^2)}=2\Lambda_\mathrm{d},\\
    \label{eq:lambda_t}
    \Lambda_\text{t} &= \Gamma_\text{t}-2F_\text{t}=0.
\end{align}
\end{subequations}
Finally, we can also calculate the reducible vertices $\Phi_r=F_r-\Gamma_r$:
\begin{subequations}
    \label{eq:Phi_ZP}
    \begin{align}
    \label{eq:Phi_d}
    \Phi_\text{d} &=\frac{U^2(U-\delta\mu^2)^2}{\delta\mu^4(U+\delta\mu^2)},\\
    \label{eq:Phi_m}
    \Phi_\text{m} &=-\frac{U^2(U-\delta\mu^2)}{\delta\mu^4},\\
    \label{eq:Phi_s}
    \Phi_\text{s} &=\frac{2U^2(U-\delta\mu^2)}{\delta\mu^4},\\
    \label{eq:Phi_t}
    \Phi_\text{t} &=0.
\end{align}
\end{subequations}

\section{Parquet iteration}\label{app:parquet}
\subsection{Iterative maps and Jacobian}
\label{app:parquet_J}
In order to construct the Jacobian \cref{eq:J_general}, we need to clarify the functional dependence of $\mathcal{F}=(f_\text{d},f_\text{m},f_\text{s},f_\text{t},f_\text{G})$ in \cref{eq:Psi} in such a way that it only depends on $\Lambda_\text{d,m,s,t}$, $G_0$, and $U$ as fixed inputs, and on $\Phi_\text{d,m,s,t}$ and $G$, which are iterated.
The equations that are used to construct the iterative maps are the BSEs \eqref{eq:bse} and the Dyson equation \eqref{eq:dyson}, in which the parquet equation \eqref{eq:parquet} and the SDE \eqref{eq:sde} are inserted.
The iterative maps read:
\begin{subequations}
\label{eq:it_maps}
\begin{align}
    f_\text{d}^{\nu \nu' \omega}[\Phi_\text{d,m,s,t},G;\Lambda_\text{d}]&=\frac{1}{\beta}\sum_{\nu_1} \Gamma_\text{d}^{\nu \nu_1 \omega}[\Phi_\text{d,m,s,t};\Lambda_\text{d}] G_{\nu_1}G_{\nu_1+\omega} \left(\Phi_\text{d}^{\nu_1 \nu' \omega}+\Gamma_\text{d}^{\nu_1 \nu' \omega}[\Phi_\text{d,m,s,t};\Lambda_\text{d}]\right),\\
    f_\text{m}^{\nu \nu' \omega}[\Phi_\text{d,m,s,t},G;\Lambda_\text{m}]&=\frac{1}{\beta}\sum_{\nu_1} \Gamma_\text{m}^{\nu \nu_1 \omega}[\Phi_\text{d,m,s,t};\Lambda_\text{m}] G_{\nu_1}G_{\nu_1+\omega} \left(\Phi_\text{m}^{\nu_1 \nu' \omega}+\Gamma_\text{m}^{\nu_1 \nu' \omega}[\Phi_\text{d,m,s,t};\Lambda_\text{m}]\right),\\
    f_\text{s}^{\nu \nu' \omega}[\Phi_\text{d,m,s},G;\Lambda_\text{s}]&=-\frac{1}{2\beta}\sum_{\nu_1}\bigg(\Phi_\text{s}^{\nu \nu_1 \omega}+\Gamma_\text{s}^{\nu \nu_1 \omega}[\Phi_\text{d,m};\Lambda_\text{s}]\bigg) G_{\nu_1}G_{-\nu_1-\omega} \Gamma_\text{s}^{\nu_1 \nu \omega}[\Phi_\text{d,m};\Lambda_\text{s}],\\
    f_\text{t}^{\nu \nu' \omega}[\Phi_\text{d,m,t},G;\Lambda_\text{t}]&=\frac{1}{2\beta}\sum_{\nu_1}\bigg(\Phi_\text{t}^{\nu \nu_1 \omega}+\Gamma_\text{t}^{\nu \nu_1 \omega}[\Phi_\text{d,m};\Lambda_\text{t}]\bigg) G_{\nu_1}G_{-\nu_1-\omega} \Gamma_\text{t}^{\nu_1 \nu \omega}[\Phi_\text{d,m};\Lambda_\text{t}],\\
    f_\text{G}^{\nu}[\Phi_\text{d,m,s,t},G;\Lambda_\text{d,m},G_0]&=\left(G_{0,\nu}^{-1}-\Sigma_\nu[\Phi_\text{d,m,s,t},G;\Lambda_\text{d,m}]\right)^{-1},
\end{align}
\end{subequations}
where we indicate the dependence on iterated and fixed quantities by a semicolon.
The one- and two-particle irreducible vertices via the parquet equation \eqref{eq:parquet} and the SDE \eqref{eq:sde} read:
\begin{subequations}
\label{eq:irreducible_vertices}
\begin{align}
    \Gamma_\mathrm{d}^{\nu \nu' \omega}[\Phi_\text{d,m,s,t};\Lambda_\mathrm{d}] &= \Lambda_\mathrm{d}^{\nu \nu' \omega} -\frac{1}{2} \Phi_\mathrm{d}^{\nu (\nu+\omega) (\nu'-\nu)} - \frac{3}{2} \Phi_\mathrm{m}^{\nu (\nu+\omega) (\nu'-\nu)} \\ \nonumber
    &\phantom{=} + \frac{1}{2} \Phi_\mathrm{s}^{\nu \nu' (-\omega -\nu -\nu')} + \frac{3}{2} \Phi_\mathrm{t}^{\nu \nu' (-\omega -\nu -\nu')},  \\ 
    \Gamma_\mathrm{m}^{\nu \nu' \omega}[\Phi_\text{d,m,s,t};\Lambda_\mathrm{m}] &= \Lambda_\mathrm{m}^{\nu \nu' \omega} -\frac{1}{2} \Phi_\mathrm{d}^{\nu (\nu+\omega) (\nu'-\nu)} + \frac{1}{2} \Phi_\mathrm{m}^{\nu (\nu+\omega) (\nu'-\nu)} \\ \nonumber
    &\phantom{=} - \frac{1}{2} \Phi_\mathrm{s}^{\nu \nu' (-\omega -\nu -\nu')} + \frac{1}{2} \Phi_\mathrm{t}^{\nu \nu' (-\omega -\nu -\nu')}, \\
    \Gamma_\mathrm{s}^{\nu \nu' \omega}[\Phi_\text{d,m};\Lambda_\mathrm{s}] &= \Lambda_\mathrm{s}^{\nu \nu' \omega} +\frac{1}{2} \Phi_\mathrm{d}^{\nu \nu' (-\omega-\nu-\nu')} - \frac{3}{2} \Phi_\mathrm{m}^{\nu \nu' (-\omega-\nu-\nu')} \\ \nonumber
    &\phantom{=} + \frac{1}{2} \Phi_\mathrm{d}^{\nu (-\nu'-\omega) (\nu' -\nu)} - \frac{3}{2} \Phi_\mathrm{m}^{\nu (-\nu'-\omega) (\nu' -\nu)}, \\
    \Gamma_\mathrm{t}^{\nu \nu' \omega}[\Phi_\text{d,m};\Lambda_\mathrm{t}] &= \Lambda_\mathrm{t}^{\nu \nu' \omega} + \frac{1}{2} \Phi_\mathrm{d}^{\nu \nu' (-\omega-\nu-\nu')} + \frac{1}{2} \Phi_\mathrm{m}^{\nu \nu' (-\omega-\nu-\nu')} \\ \nonumber
    &\phantom{=} - \frac{1}{2} \Phi_\mathrm{d}^{\nu (-\nu'-\omega) (\nu' -\nu)} - \frac{1}{2} \Phi_\mathrm{m}^{\nu (-\nu'-\omega) (\nu' -\nu)},
\end{align}   
\end{subequations}
\begin{align}
\label{eq:SDE}
\begin{split}
    \Sigma_\nu[\Phi_\text{d,m,s,t},G;\Lambda_\text{d,m},G_0]&=\frac{U}{\beta}\sum_{\nu_1}G_{\nu_1} \\
    -\frac{U}{2\beta^2}\sum_{\nu_1,\omega}&\left(\Phi_\mathrm{d}^{\nu\nu_1\omega}+\Gamma_\mathrm{d}^{\nu\nu_1\omega}[\Phi_\text{d,m,s,t};\Lambda_\mathrm{d}]+\Phi_\mathrm{m}^{\nu\nu_1\omega}+\Gamma_\mathrm{m}^{\nu\nu_1\omega}[\Phi_\text{d,m,s,t};\Lambda_\mathrm{m}]\right)\\
    &\hspace{5mm} G_{\nu_1}G_{\nu_1+\omega}G_{\nu+\omega}.
\end{split}
\end{align}
Given the iterative maps \cref{eq:it_maps}, we can calculate the coefficients of the Jacobian analytically. For completeness, we provide these analytical equations in the following, where we sum over repeated fermionic indices $\nu_i$.
Starting with the map for the density channel, we obtain:
\begin{subequations}
\label{eq:J_d}
\begin{align}
    \frac{\delta f_\mathrm{d}^{\nu \nu' \omega}}{\delta \Phi_\mathrm{d}^{\tilde \nu \tilde \nu' \tilde \omega}} &= -\frac{1}{\beta^2} \Big[ -\frac{1}{2} \delta_{\nu , \tilde \nu} \delta_{\nu + \omega, \tilde \nu'} \chi_{0,\text{ph}}^{(\nu + \tilde \omega) \nu_2 \omega} \left( \Gamma_\mathrm{d}^{\nu_2 \nu' \omega} + \Phi_\mathrm{d}^{\nu_2 \nu' \omega} \right)  \label{eq:J_dd}\\
    &\phantom{=}+ \Gamma_\mathrm{d}^{\nu \nu_1 \omega} \chi_{0,\text{ph}}^{\nu_1 \tilde \nu \omega} \left( \delta_{\nu', \tilde \nu'} \delta_{\omega, \tilde \omega} - \frac{1}{2} \delta_{\tilde \nu + \omega, \tilde \nu'} \delta_{\nu' - \tilde \nu, \tilde \omega} \right) \Big]  \nonumber, \\
    \frac{\delta f_\mathrm{d}^{\nu \nu' \omega}}{\delta \Phi_\mathrm{m}^{\tilde \nu \tilde \nu' \tilde \omega}} &= \frac{3}{2\beta^2} \Big[ \delta_{\nu , \tilde \nu} \delta_{\nu + \omega, \tilde \nu'} \chi_{0,\text{ph}}^{(\nu + \tilde \omega) \nu_2 \omega} \left( \Gamma_\mathrm{d}^{\nu_2 \nu' \omega} + \Phi_\mathrm{d}^{\nu_2 \nu' \omega} \right)  \\
    &\phantom{=}+ \Gamma_\mathrm{d}^{\nu \nu_1 \omega} \chi_{0,\text{ph}}^{\nu_1 \tilde \nu \omega}  \delta_{\tilde \nu + \omega, \tilde \nu'} \delta_{\nu' - \tilde \nu, \tilde \omega}  \Big]  \nonumber, \\
    \frac{\delta f_\mathrm{d}^{\nu \nu' \omega}}{\delta \Phi_\mathrm{s}^{\tilde \nu \tilde \nu' \tilde \omega}} &= -\frac{1}{2\beta^2} \Big[ \delta_{\nu, \tilde \nu} \delta_{-\omega - \nu -\tilde \nu', \tilde \omega} \chi_{0,\text{ph}}^{\tilde \nu' \nu_2 \omega} \left( \Gamma_\mathrm{d}^{\nu_2 \nu' \omega} + \Phi_\mathrm{d}^{\nu_2 \nu' \omega} \right) \\
    &\phantom{=}+ \delta_{\nu', \tilde \nu'} \delta_{-\omega - \tilde \nu - \nu', \tilde \omega} \Gamma_\mathrm{d}^{\nu \nu_1 \omega} \chi_{0,\text{ph}}^{\nu_1 \tilde \nu \omega} \Big] \nonumber, \\
    \frac{\delta f_\mathrm{d}^{\nu \nu' \omega}}{\delta \Phi_\mathrm{t}^{\tilde \nu \tilde \nu' \tilde \omega}} &= -\frac{3}{2\beta^2} \Big[ \delta_{\nu, \tilde \nu} \delta_{-\omega - \nu -\tilde \nu', \tilde \omega} \chi_{0,\text{ph}}^{\tilde \nu' \nu_2 \omega} \left( \Gamma_\mathrm{d}^{\nu_2 \nu' \omega} + \Phi_\mathrm{d}^{\nu_2 \nu' \omega} \right) \\
    &\phantom{=}+ \delta_{\nu', \tilde \nu'} \delta_{-\omega - \tilde \nu - \nu', \tilde \omega} \Gamma_\mathrm{d}^{\nu \nu_1 \omega} \chi_{0,\text{ph}}^{\nu_1 \tilde \nu \omega} \Big] \nonumber, \\
    \fdv{f_\mathrm{d}^{\nu \nu' \omega}}{G_{\tilde\nu}} &= \frac{1}{\beta}\Big(\Gamma_\mathrm{d}^{\nu\tilde\nu\omega}G(\tilde\nu+\omega)F_\mathrm{d}^{\tilde\nu\nu^\prime\omega} +\Gamma_\mathrm{d}^{\nu,\tilde\nu-\omega,\omega}G(\tilde\nu-\omega)F_\mathrm{d}^{\tilde\nu-\omega,\nu^\prime,\omega}\Big).
\end{align}
\end{subequations}
For the magnetic channel, we obtain:
\begin{subequations}
\label{eq:J_m}
\begin{align}
    \frac{\delta f_\mathrm{m}^{\nu \nu' \omega}}{\delta \Phi_\mathrm{d}^{\tilde \nu \tilde \nu' \tilde \omega}} &= \frac{1}{2\beta^2} \Big[ \delta_{\nu , \tilde \nu} \delta_{\nu + \omega, \tilde \nu'} \chi_{0,\text{ph}}^{(\nu + \tilde \omega) \nu_2 \omega} \left( \Gamma_\mathrm{m}^{\nu_2 \nu' \omega} + \Phi_\mathrm{m}^{\nu_2 \nu' \omega} \right)  \\
    &\phantom{=}+ \Gamma_\mathrm{m}^{\nu \nu_1 \omega} \chi_{0,\text{ph}}^{\nu_1 \tilde \nu \omega}  \delta_{\tilde \nu + \omega, \tilde \nu'} \delta_{\nu' - \tilde \nu, \tilde \omega}  \Big]  \nonumber ,\\
    \frac{\delta f_\mathrm{m}^{\nu \nu' \omega}}{\delta \Phi_\mathrm{m}^{\tilde \nu \tilde \nu' \tilde \omega}} &= -\frac{1}{\beta^2} \Big[ \frac{1}{2} \delta_{\nu , \tilde \nu} \delta_{\nu + \omega, \tilde \nu'} \chi_{0,\text{ph}}^{(\nu + \tilde \omega) \nu_2 \omega} \left( \Gamma_\mathrm{m}^{\nu_2 \nu' \omega} + \Phi_\mathrm{m}^{\nu_2 \nu' \omega} \right)  \\
    &\phantom{=}+ \Gamma_\mathrm{m}^{\nu \nu_1 \omega} \chi_{0,\text{ph}}^{\nu_1 \tilde \nu \omega} \left( \delta_{\nu', \tilde \nu'} \delta_{\omega, \tilde \omega} + \frac{1}{2} \delta_{\tilde \nu + \omega, \tilde \nu'} \delta_{\nu' - \tilde \nu, \tilde \omega} \right) \Big] \nonumber ,\\
    \frac{\delta f_\mathrm{m}^{\nu \nu' \omega}}{\delta \Phi_\mathrm{s}^{\tilde \nu \tilde \nu' \tilde \omega}} &= \frac{1}{2\beta^2} \Big[ \delta_{\nu, \tilde \nu} \delta_{-\omega - \nu -\tilde \nu', \tilde \omega} \chi_{0,\text{ph}}^{\tilde \nu' \nu_2 \omega} \left( \Gamma_\mathrm{m}^{\nu_2 \nu' \omega} + \Phi_\mathrm{m}^{\nu_2 \nu' \omega} \right) \\
    &\phantom{=}+ \delta_{\nu', \tilde \nu'} \delta_{-\omega - \tilde \nu - \nu', \tilde \omega} \Gamma_\mathrm{m}^{\nu \nu_1 \omega} \chi_{0,\text{ph}}^{\nu_1 \tilde \nu \omega} \Big] \nonumber ,\\
    \frac{\delta f_\mathrm{m}^{\nu \nu' \omega}}{\delta \Phi_\mathrm{t}^{\tilde \nu \tilde \nu' \tilde \omega}} &= -\frac{1}{2\beta^2} \Big[ \delta_{\nu, \tilde \nu} \delta_{-\omega - \nu -\tilde \nu', \tilde \omega} \chi_{0,\text{ph}}^{\tilde \nu' \nu_2 \omega} \left( \Gamma_\mathrm{m}^{\nu_2 \nu' \omega} + \Phi_\mathrm{m}^{\nu_2 \nu' \omega} \right) \\
    &\phantom{=}+ \delta_{\nu', \tilde \nu'} \delta_{-\omega - \tilde \nu - \nu', \tilde \omega} \Gamma_\mathrm{m}^{\nu \nu_1 \omega} \chi_{0,\text{ph}}^{\nu_1 \tilde \nu \omega} \Big] \nonumber ,\\
    \fdv{f_\mathrm{m}^{\nu \nu' \omega}}{G_{\tilde\nu}} &= \frac{1}{\beta}\Big(\Gamma_\mathrm{m}^{\nu\tilde\nu\omega}G(\tilde\nu+\omega)F_\mathrm{m}^{\tilde\nu\nu^\prime\omega} +\Gamma_\mathrm{m}^{\nu,\tilde\nu-\omega,\omega}G(\tilde\nu-\omega)F_\mathrm{m}^{\tilde\nu-\omega,\nu^\prime,\omega}\Big).
\end{align}
\end{subequations}
For the singlet channel, we obtain:
\begin{subequations}
\label{eq:J_s}
\begin{align} 
    \frac{\delta f_\mathrm{s}^{\nu \nu' \omega}}{\delta \Phi_\mathrm{d}^{\tilde \nu \tilde \nu' \tilde \omega}} &= \frac{1}{2\beta^2} \Big[ \delta_{\nu, \tilde \nu} \delta_{-\omega - \nu - \tilde \nu', \tilde \omega} \left( \chi_{0,\text{pp}}^{\tilde \nu' \nu_2 \omega} + \chi_{0,\text{pp}}^{(-\tilde \nu'-\omega) \nu_2 \omega} \right) \Gamma_\mathrm{s}^{\nu_2 \nu' \omega} \\
    &\phantom{=}+ \left( \delta_{\nu', \tilde \nu'} \delta_{-\omega -\tilde \nu - \nu', \tilde \omega} + \delta_{-\nu' -\omega, \tilde \nu'} \delta_{\nu' - \tilde \nu, \tilde \omega} \right) \left( \Gamma_\mathrm{s}^{\nu \nu_1 \omega} + \Phi_\mathrm{s}^{\nu \nu_1 \omega} \right) \chi_{0,\text{pp}}^{\nu_1 \tilde \nu \omega}  \Big], \nonumber \\
    \frac{\delta f_\mathrm{s}^{\nu \nu' \omega}}{\delta \Phi_\mathrm{m}^{\tilde \nu \tilde \nu' \tilde \omega}}&= -\frac{3}{2\beta^2} \Big[ \delta_{\nu, \tilde \nu} \delta_{-\omega - \nu - \tilde \nu', \tilde \omega} \left( \chi_{0,\text{pp}}^{\tilde \nu' \nu_2 \omega} + \chi_{0,\text{pp}}^{(-\tilde \nu'-\omega) \nu_2 \omega} \right) \Gamma_\mathrm{s}^{\nu_2 \nu' \omega} \\
    &\phantom{=}+ \left( \delta_{\nu', \tilde \nu'} \delta_{-\omega -\tilde \nu - \nu', \tilde \omega} + \delta_{-\nu' -\omega, \tilde \nu'} \delta_{\nu' - \tilde \nu, \tilde \omega} \right) \left( \Gamma_\mathrm{s}^{\nu \nu_1 \omega} + \Phi_\mathrm{s}^{\nu \nu_1 \omega} \right) \chi_{0,\text{pp}}^{\nu_1 \tilde \nu \omega}  \Big], \nonumber \\
    \frac{\delta f_\mathrm{s}^{\nu \nu' \omega}}{\delta \Phi_\mathrm{s}^{\tilde \nu \tilde \nu' \tilde \omega}} &= \frac{1}{\beta^2} \delta_{\nu, \tilde \nu} \delta_{\omega, \tilde \omega} \chi_{0,\text{pp}}^{\tilde \nu' \nu_2 \omega} \Gamma_\mathrm{s}^{\nu_2 \nu' \omega} ,\label{eq:J_ss} \\
    \frac{\delta f_\mathrm{s}^{\nu \nu' \omega}}{\delta \Phi_\mathrm{t}^{\tilde \nu \tilde \nu' \tilde \omega}} &= 0,\\
    \fdv{f_\mathrm{s}^{\nu \nu' \omega}}{G_{\tilde\nu}} &= -\frac{1}{2\beta}\Big(F_\mathrm{s}^{\nu\tilde\nu\omega}G(-\tilde\nu-\omega)\Gamma_\mathrm{s}^{\tilde\nu\nu^\prime\omega} +F_\mathrm{s}^{\nu,-\tilde\nu-\omega,\omega}G(-\tilde\nu-\omega)\Gamma_\mathrm{s}^{-\tilde\nu-\omega,\nu^\prime,\omega}\Big).
\end{align}
\end{subequations}
For the triplet channel, we obtain:
\begin{subequations}
\label{eq:J_t}
\begin{align}
    \frac{\delta f_\mathrm{t}^{\nu \nu' \omega}}{\delta \Phi_\mathrm{d}^{\tilde \nu \tilde \nu' \tilde \omega}} &= -\frac{1}{2\beta^2} \Big[ \delta_{\nu, \tilde \nu} \delta_{-\omega - \nu - \tilde \nu', \tilde \omega} \left( \chi_{0,\text{pp}}^{\tilde \nu' \nu_2 \omega} - \chi_{0,\text{pp}}^{(-\tilde \nu'-\omega) \nu_2 \omega} \right) \Gamma_\mathrm{t}^{\nu_2 \nu' \omega} \\
    &\phantom{=}+ \left( \delta_{\nu', \tilde \nu'} \delta_{-\omega -\tilde \nu - \nu', \tilde \omega} - \delta_{-\nu' -\omega, \tilde \nu'} \delta_{\nu' - \tilde \nu, \tilde \omega} \right) \left( \Gamma_\mathrm{t}^{\nu \nu_1 \omega} + \Phi_\mathrm{t}^{\nu \nu_1 \omega} \right) \chi_{0,\text{pp}}^{\nu_1 \tilde \nu \omega}  \Big] \nonumber, \\
    \frac{\delta f_\mathrm{t}^{\nu \nu' \omega}}{\delta \Phi_\mathrm{m}^{\tilde \nu \tilde \nu' \tilde \omega}} &= -\frac{1}{2\beta^2} \Big[ \delta_{\nu, \tilde \nu} \delta_{-\omega - \nu - \tilde \nu', \tilde \omega} \left( \chi_{0,\text{pp}}^{\tilde \nu' \nu_2 \omega} - \chi_{0,\text{pp}}^{(-\tilde \nu'-\omega) \nu_2 \omega} \right) \Gamma_\mathrm{t}^{\nu_2 \nu' \omega} \\
    &\phantom{=}+ \left( \delta_{\nu', \tilde \nu'} \delta_{-\omega -\tilde \nu - \nu', \tilde \omega} - \delta_{-\nu' -\omega, \tilde \nu'} \delta_{\nu' - \tilde \nu, \tilde \omega} \right) \left( \Gamma_\mathrm{t}^{\nu \nu_1 \omega} + \Phi_\mathrm{t}^{\nu \nu_1 \omega} \right) \chi_{0,\text{pp}}^{\nu_1 \tilde \nu \omega}  \Big] \nonumber, \\
    \frac{\delta f_\mathrm{t}^{\nu \nu' \omega}}{\delta \Phi_\mathrm{s}^{\tilde \nu \tilde \nu' \tilde \omega}} &= 0, \\
    \frac{\delta f_\mathrm{t}^{\nu \nu' \omega}}{\delta \Phi_\mathrm{t}^{\tilde \nu \tilde \nu' \tilde \omega}} &= -\frac{1}{\beta^2} \delta_{\nu, \tilde \nu} \delta_{\omega, \tilde \omega} \chi_{0,\text{pp}}^{\tilde \nu' \nu_2 \omega} \Gamma_\mathrm{t}^{\nu_2 \nu' \omega}, \\
    \fdv{f_\mathrm{t}^{\nu \nu' \omega}}{G_{\tilde\nu}} &= \frac{1}{2\beta}\Big(F_\mathrm{t}^{\nu\tilde\nu\omega}G(-\tilde\nu-\omega)\Gamma_\mathrm{t}^{\tilde\nu\nu^\prime\omega} +F_\mathrm{t}^{\nu,-\tilde\nu-\omega,\omega}G(-\tilde\nu-\omega)\Gamma_\mathrm{t}^{-\tilde\nu-\omega,\nu^\prime,\omega}\Big).
\end{align}
\end{subequations}
For the Green's function, we obtain:
\begin{subequations}
\label{eq:J_G}
\begin{align}
    \fdv{f_\text{G}^\nu}{\Phi_\mathrm{d}^{\tilde \nu \tilde \nu' \tilde \omega}}&=-\frac{U}{2\beta^2}G^2(\nu) G(\tilde\nu^\prime)G(\tilde\nu'+\tilde\omega) G(\tilde\nu+\tilde\omega)\delta_{\nu\tilde\nu},\\
    \fdv{f_\text{G}^\nu}{\Phi_\mathrm{m}^{\tilde \nu \tilde \nu' \tilde \omega}}&=-\frac{U}{2\beta^2}G^2(\nu)\Big( -G(\tilde\nu^\prime)G(\tilde\nu'+\tilde\omega) G(\tilde\nu+\tilde\omega)\delta_{\nu\tilde\nu} - 2G(\nu+\tilde\omega) G(\tilde\nu'+\tilde\omega) G(\tilde\nu^\prime) \delta_{\nu\tilde\nu}\Big),\\
    \fdv{f_\text{G}^\nu}{\Phi_\mathrm{s}^{\tilde \nu \tilde \nu' \tilde \omega}}&=-\frac{U}{2\beta^2}G^2(\nu)G(\tilde\nu') G(-\nu-\tilde\omega) G(-\tilde\nu^\prime-\tilde\omega) \delta_{\nu\tilde\nu},\\
    \fdv{f_\text{G}^\nu}{\Phi_\mathrm{t}^{\tilde \nu \tilde \nu' \tilde \omega}}&=-\frac{U}{2\beta^2}G^2(\nu)G(\tilde\nu') G(-\nu-\tilde\omega) G(-\tilde\nu^\prime-\tilde\omega) \delta_{\nu\tilde\nu}=\fdv{f_\text{G}^\nu}{\Phi_\mathrm{s}^{\tilde \nu \tilde \nu' \tilde \omega}},\\
    \fdv{f_\text{G}^\nu}{G_{\tilde\nu}} &=  - \frac{G_{\nu}^2U}{2\beta^2} \bigg(\sum_{\omega} (F_\mathrm{d}^{\nu \tilde\nu \omega}-F_\mathrm{m}^{\nu \tilde\nu \omega}) G(\tilde\nu+\omega) G(\nu+\omega)\\ \nonumber
    &\phantom{=  - \frac{G_{\nu}^2U}{2\beta^2}}\quad +\sum_{\omega} (F_\mathrm{d}^{\nu, \tilde\nu-\omega ,\omega}-F_\mathrm{m}^{\nu, \tilde\nu-\omega, \omega}) G(\tilde\nu-\omega) G(\nu+\omega)\\ \nonumber
    &\phantom{=  - \frac{G_{\nu}^2U}{2\beta^2}}\quad +\sum_{\nu_1} (F_\mathrm{d}^{\nu, \nu_1 ,\tilde\nu-\nu}-F_\mathrm{m}^{\nu, \nu_1 ,\tilde\nu-\nu}) G(\nu_1) G(\nu_1+\tilde\nu-\nu)\bigg)\\ \nonumber
    &\phantom{=}+G_{\nu}^2U/\beta.
\end{align}
\end{subequations}
where \cref{eq:bubble_HA,eq:F_gamma_phi,eq:irreducible_vertices} can be used to restore the functional dependence of the Jacobian. Note that $\fdv{\Sigma_\nu}{G_{\nu^\prime}}\neq\Gamma$ since the parquet equations are not fully LWF derivable. Mathematically, $\fdv{\Sigma_\nu}{G_{\nu^\prime}}\neq\Gamma$ does not even hold for the exact solution since we treat $\Phi$ and $G$ as independent variables in this algorithm. 
To obtain matrices, the frequencies of the iterative maps $f$ and $\Phi_\text{d,m,s,t}/G$ are considered as one super-index each and then the blocks are used to build the Jacobian as shown in \cref{eq:J_general}.
Since for $U=0$ all vertices $\Phi$, $\Gamma$, and $F$ vanish, the $\Pi$ matrix is equal to the identity. Therefore, all channels are decoupled at $U=0$ and the physical fixed point is stable and only becomes unstable when increasing the interaction.

Further, we show that the Jacobian is a $\kappa$-real matrix \cite{Hill1992}. To show this property of the Jacobian, we recall that, under complex conjugation, it holds that $\left(X^{\nu\nu^\prime\omega}_{\sigma\sigma\sigma^\prime\sigma^\prime,\text{ph/pp}}\right)^*=X^{-\nu^\prime,-\nu,-\omega}_{\sigma^\prime\sigma^\prime\sigma\sigma,\text{ph/pp}}$ where $X=\Gamma,\chi,\Phi$. Further, it holds that $\left[G(\nu)\right]^*=G(-\nu)$ \cite{rohringer2013,Essl2024}.

With these two relations, it is clear that, under the index-permutation $\nu,\nu^\prime,\omega\rightarrow-\nu^\prime,-\nu,-\omega$, each block in \cref{eq:J_general}, i.e., \cref{eq:J_d,eq:J_m,eq:J_s,eq:J_t,eq:J_G}, is complex conjugated which is the definition of a $\kappa$-real matrix. As a consequence, eigenvalues of the Jacobian are either real or come in complex conjugated pairs \cite{Hill1992}.
Note that this also holds for the Hubbard model since complex conjugation leads to the relations $\left(X^{\mathbf{k},\mathbf{k^\prime},\mathbf{q}}_{\sigma\sigma\sigma^\prime\sigma^\prime,\text{ph/pp}}(\nu,\nu^\prime,\omega)\right)^*=X^{\mathbf{k^\prime},\mathbf{k},\mathbf{q}}_{\sigma^\prime\sigma^\prime\sigma\sigma,\text{ph/pp}}(-\nu^\prime,-\nu,-\omega)$ and $\left[G(\mathbf{k},\nu)\right]^*=G(\mathbf{k},-\nu)$, which can again be viewed as an index permutation.

\subsection{Jacobian for the ZP model}\label{app:J_ZP}
For the ZP model, we can use \cref{eq:J_d,eq:J_m,eq:J_s,eq:J_t,eq:J_G} but we can just neglect the frequency dependence. For completeness, we also give all the equations in the following.

The iterative map for the density channel reads
\begin{align}
\label{eq:ZP_f_d}
\begin{split}
    f_\text{d} &= -\Gamma_\text{d}\chi_{0,\text{ph}}(\Gamma_\text{d}+\Phi_\text{d})\\
    \text{with } \Gamma_\text{d}&=\Lambda_\text{d}-\Phi_\text{d}/2-3\Phi_\text{m}/2+\Phi_\text{s}/2+3\Phi_\text{t}/2,
\end{split}
\end{align}
while the Jacobian coefficients for the density channel read
\begin{subequations}
\label{eq:ZP_J_d}
\begin{align}
    \fdv{f_\text{d}}{\Phi_\text{d}} &= \chi_{0,\text{ph}}\Phi_\text{d}/2,\\
    \fdv{f_\text{d}}{\Phi_\text{m}} &= 3\chi_{0,\text{ph}}(\Gamma_\text{d}+\Phi_\text{d}/2),\\
    \fdv{f_\text{d}}{\Phi_\text{s}} &= -\chi_{0,\text{ph}}(\Gamma_\text{d}+\Phi_\text{d}/2),\\
    \fdv{f_\text{d}}{\Phi_\text{t}} &= -3\chi_{0,\text{ph}}(\Gamma_\text{d}+\Phi_\text{d}/2),\\
    \fdv{f_\text{d}}{G} &= 2\Gamma_\text{d}G(\Gamma_\text{d}+\Phi_\text{d}).
\end{align}
\end{subequations}
The iterative map for the magnetic channel reads
\begin{align}
\begin{split}
\label{eq:ZP_f_m}
    f_\text{m} &= -\Gamma_\text{m}\chi_{0,\text{ph}}(\Gamma_\text{m}+\Phi_\text{m})\\
    \text{with } \Gamma_\text{m}&=\Lambda_\text{m}-\Phi_\text{d}/2+\Phi_\text{m}/2-\Phi_\text{s}/2+\Phi_\text{t}/2,
\end{split}
\end{align} 
while the Jacobian coefficients for the magnetic channel read
\begin{subequations}
\label{eq:ZP_J_m}
\begin{align}
    \fdv{f_\text{m}}{\Phi_\text{d}} &= \chi_{0,\text{ph}}(\Gamma_\text{m}+\Phi_\text{m}/2),\\
    \fdv{f_\text{m}}{\Phi_\text{m}} &= -\chi_{0,\text{ph}}(2\Gamma_\text{m}+\Phi_\text{m}/2),\\
    \fdv{f_\text{m}}{\Phi_\text{s}} &= \chi_{0,\text{ph}}(\Gamma_\text{m}+\Phi_\text{m}/2),\\
    \fdv{f_\text{m}}{\Phi_\text{t}} &= -\chi_{0,\text{ph}}(\Gamma_\text{m}+\Phi_\text{m}/2),\\
    \fdv{f_\text{m}}{G} &= 2\Gamma_\text{m}G(\Gamma_\text{m}+\Phi_\text{m}).
\end{align}
\end{subequations}
The iterative map for the singlet channel reads
\begin{align}
\label{eq:ZP_f_s}
    f_\text{s} = \Gamma_\text{s}\chi_{0,\text{pp}}(\Gamma_\text{s}+\Phi_\text{s})\text{ with } \Gamma_\text{s}=\Lambda_\text{s}+\Phi_\text{d}-3\Phi_\text{m},
\end{align}
while the Jacobian coefficients for the singlet channel read
\begin{subequations}
\label{eq:ZP_J_s}
\begin{align}
    \fdv{f_\text{s}}{\Phi_\text{d}} &= \chi_{0,\text{pp}}(2\Gamma_\text{s}+\Phi_\text{s}),\\
    \fdv{f_\text{s}}{\Phi_\text{m}} &= -3\chi_{0,\text{pp}}(2\Gamma_\text{s}+\Phi_\text{s}),\\
    \fdv{f_\text{s}}{\Phi_\text{s}} &= \chi_{0,\text{pp}}\Gamma_\text{s},\\
    \fdv{f_\text{s}}{\Phi_\text{t}} &= 0,\\
    \fdv{f_\text{s}}{G} &= -\Gamma_\text{s}G(\Gamma_\text{s}+\Phi_\text{s}).
\end{align}
\end{subequations}
The iterative map for the triplet channel reads
\begin{align}
\label{eq:ZP_f_t}
    f_\text{t} = -\Gamma_\text{t}\chi_{0,\text{pp}}(\Gamma_\text{t}+\Phi_\text{t})\text{ with } \Gamma_\text{t}=\Lambda_\text{t}=0,
\end{align}
while the Jacobian coefficients for the triplet channel read
\begin{subequations}
\label{eq:ZP_J_t}
\begin{align}
    \fdv{f_\text{t}}{\Phi_\text{d}} &= 0,\\
    \fdv{f_\text{t}}{\Phi_\text{m}} &= 0,\\
    \fdv{f_\text{t}}{\Phi_\text{s}} &= 0,\\
    \fdv{f_\text{t}}{\Phi_\text{t}} &= -\chi_{0,\text{pp}}\Gamma_\text{t},\\
    \fdv{f_\text{t}}{G} &= \Gamma_\text{t}G(\Gamma_\text{t}+\Phi_\text{t})=0.
\end{align}
\end{subequations}
The iterative map for the Green's function reads
\begin{align}
\label{eq:ZP_f_G}
    f_\text{G} = 1/(G_0^{-1}-\Sigma)\quad\text{with}\quad \Sigma=UG-\frac{U}{2}G^3(F_\text{d}-F_\text{m}),
\end{align}
while the Jacobian coefficients for the Green's function read
\begin{subequations}
\label{eq:ZP_J_G}
\begin{align}
    \fdv{f_\text{G}}{\Phi_\text{d}} &= -UG^5/2,\\
    \fdv{f_\text{G}}{\Phi_\text{m}} &= 3UG^5/2,\\
    \fdv{f_\text{G}}{\Phi_\text{s}} &= -UG^5/2,\\
    \fdv{f_\text{G}}{\Phi_\text{t}} &= -UG^5/2,\\
    \fdv{f_\text{G}}{G} &= G^2(U-3UG^2(F_\text{d}-F_\text{m})/2).
\end{align}
\end{subequations}
Note that, strictly speaking, the Jacobian of the ZP model is not $\kappa$-real since there is no frequency dependence. However, if one considers two decoupled ZP models with complex-conjugated chemical potentials that mimic two Matsubara frequencies $\nu$ and $-\nu$, one again ends up with a $\kappa$-real matrix.

\subsection{Strong-coupling iteration}\label{app:parquet_J_F}
As outlined in \cref{sec:strong_parquet}, we adapt the parquet formalism in such a way that we can iterate $F$ instead of $\Phi$. 
To achieve that, we use the BSEs \eqref{eq:bse} to obtain the irreducible vertex $\Gamma$ as a functional of $F$ and $G$, i.e., 
$\Gamma_r[F,G]=\frac{F_r}{1\mp\chi_{0,r}F_r}$, which gives for the derivatives $\fdv{\Gamma_r}{F_{r^\prime}}=\frac{\delta_{rr^\prime}}{(1\mp\chi_{0,r}F_r)^2}$, $\fdv{\Gamma_{d/m}}{G}=-\frac{2F_{d/m}^2G}{(1-\chi_{0,d/m}F_{d/m})^2}$ and $\fdv{\Gamma_{s/t}}{G}=\mp\frac{F_{s/t}^2G}{(1\mp\chi_{0,s/t}F_{s/t})^2}$.
For the iterative maps, we use the parquet decomposition, \cref{eq:parquet,eq:Lambda_ZP}.
In the following, we will give the explicit equations for the ZP model. 
The iterative map for the density channel reads
\begin{align}
    f_\text{d}^F &= \frac{1}{2} (-\Lambda_\text{d} + \Gamma_\text{d}/2-3\Gamma_\text{m}/2+\Gamma_\text{s}/2+3\Gamma_\text{t}/2),
\end{align}
while the Jacobian coefficients for the density channel read
\begin{subequations}
\label{eq:J_F_d}
\begin{align}
\label{eq:J_F_dd}
    \fdv{f^F_\text{d}}{F_\text{d}} &= \frac{1}{4}\frac{1}{(1-\chi_{0,\text{ph}}F_\text{d})^2},\\
    \fdv{f^F_\text{d}}{F_\text{m}} &= -\frac{3}{4}\frac{1}{(1-\chi_{0,\text{ph}}F_\text{m})^2},\\
    \fdv{f^F_\text{d}}{F_\text{s}} &= \frac{1}{4}\frac{1}{(1+\chi_{0,\text{pp}}F_\text{s})^2},\\
    \fdv{f^F_\text{d}}{F_\text{t}} &= \frac{3}{4}\frac{1}{(1-\chi_{0,\text{pp}}F_\text{t})^2},\\
    \fdv{f^F_\text{d}}{G} &= \frac{1}{2}\left(-\frac{F_\text{d}^2G}{(1-\chi_{0,\text{ph}}F_\text{d})^2}+3\frac{F_\text{m}^2G}{(1-\chi_{0,\text{ph}}F_\text{m})^2}+\frac{1}{2}\frac{F_\text{s}^2G}{(1+\chi_{0,\text{pp}}F_\text{s})^2}-\frac{3}{2}\frac{F_\text{t}^2G}{(1-\chi_{0,\text{pp}}F_\text{t})^2}\right).
\end{align}
\end{subequations}
The iterative map for the magnetic channel reads
\begin{align}
    f_\text{m}^F &= \frac{1}{2} (-\Lambda_\text{m} - \Gamma_\text{d}/2+3\Gamma_\text{m}/2-\Gamma_\text{s}/2+\Gamma_\text{t}/2),
\end{align}
\begin{subequations}
\label{eq:J_F_m}
\begin{align}
    \fdv{f^F_\text{m}}{F_\text{d}} &= -\frac{1}{4}\frac{1}{(1-\chi_{0,\text{ph}}F_\text{d})^2},\\
    \fdv{f^F_\text{m}}{F_\text{m}} &= \frac{3}{4}\frac{1}{(1-\chi_{0,\text{ph}}F_\text{m})^2},\\
    \fdv{f^F_\text{m}}{F_\text{s}} &= -\frac{1}{4}\frac{1}{(1+\chi_{0,\text{pp}}F_\text{s})^2},\\
    \fdv{f^F_\text{m}}{F_\text{t}} &= \frac{1}{4}\frac{1}{(1-\chi_{0,\text{pp}}F_\text{t})^2},\\
    \fdv{f^F_\text{m}}{G} &= \frac{1}{2}\left(\frac{F_\text{d}^2G}{(1-\chi_{0,\text{ph}}F_\text{d})^2}-3\frac{F_\text{m}^2G}{(1-\chi_{0,\text{ph}}F_\text{m})^2}-\frac{1}{2}\frac{F_\text{s}^2G}{(1+\chi_{0,\text{pp}}F_\text{s})^2}-\frac{1}{2}\frac{F_\text{t}^2G}{(1-\chi_{0,\text{pp}}F_\text{t})^2}\right).
\end{align}
\end{subequations}
The iterative map for the singlet channel reads
\begin{align}
    f_\text{s}^F &= \frac{1}{2} (-\Lambda_\text{s} + \Gamma_\text{d}-3\Gamma_\text{m}+\Gamma_\text{s}),
\end{align}
while the Jacobian coefficients for the singlet channel read
\begin{subequations}
\label{eq:J_F_s}
\begin{align}
    \fdv{f^F_\text{s}}{F_\text{d}} &= \frac{1}{2}\frac{1}{(1-\chi_{0,\text{ph}}F_\text{d})^2},\\
    \fdv{f^F_\text{s}}{F_\text{m}} &= -\frac{3}{2}\frac{1}{(1-\chi_{0,\text{ph}}F_\text{m})^2},\\
    \fdv{f^F_\text{s}}{F_\text{s}} &= \frac{1}{2}\frac{1}{(1+\chi_{0,\text{pp}}F_\text{s})^2},\\
    \fdv{f^F_\text{s}}{F_\text{t}} &= 0\\
    \fdv{f^F_\text{s}}{G} &= \left(-\frac{F_\text{d}^2G}{(1-\chi_{0,\text{ph}}F_\text{d})^2}+3\frac{F_\text{m}^2G}{(1-\chi_{0,\text{ph}}F_\text{m})^2}+\frac{1}{2}\frac{F_\text{s}^2G}{(1+\chi_{0,\text{pp}}F_\text{s})^2}\right).
\end{align}
\end{subequations}
The iterative map for the triplet channel reads
\begin{align}
    f_\text{t}^F &= \frac{1}{2} (-\Lambda_\text{t} + \Gamma_\text{t}),
\end{align}
while the Jacobian coefficients for the triplet channel read
\begin{subequations}
\label{eq:J_F_t}
\begin{align}
    \fdv{f^F_\text{t}}{F_\text{d}} &= 0,\\
    \fdv{f^F_\text{t}}{F_\text{m}} &=0,\\
    \fdv{f^F_\text{t}}{F_\text{s}} &= 0,\\
    \fdv{f^F_\text{t}}{F_\text{t}} &= \frac{1}{2}\frac{1}{(1-\chi_{0,\text{pp}}F_\text{t})^2},\\
    \fdv{f^F_\text{t}}{G} &= -\frac{1}{2}\frac{F_\text{t}^2G}{(1-\chi_{0,\text{pp}}F_\text{t})^2}.
\end{align}
\end{subequations}
The iterative map for the Green's function reads
\begin{align}
    f_\text{G}^F = 1/(G_0^{-1}-\Sigma)\quad\text{with}\quad \Sigma=UG-\frac{U}{2}G^3(F_\text{d}-F_\text{m}),
\end{align}
while the Jacobian coefficients for the Green's function read
\begin{subequations}
\label{eq:J_F_G}
\begin{align}
    \fdv{f^F_\text{G}}{F_\text{d}} &= -UG^5/2,\\
    \fdv{f^F_\text{G}}{F_\text{m}} &= UG^5/2,\\
    \fdv{f^F_\text{G}}{F_\text{s}} &= 0,\\
    \fdv{f^F_\text{G}}{F_\text{t}} &= 0,\\
    \fdv{f^F_\text{G}}{G} &= G^2(U-3UG^2(F_\text{d}-F_\text{m})/2).
\end{align}
\end{subequations}
With \cref{eq:J_F_d,eq:J_F_m,eq:J_F_s,eq:J_F_t,eq:J_F_G}, we can build $J^F$ and $\Pi^F$ analogous to \cref{eq:J_general}. 
Note that the Jacobian $J^F$ does only depend on $F$, $G$, $U$ and $G_0$ but not on $\Lambda$. 
Further, since $F=0$ at $U=0$, we can calculate 
\begin{align}
\label{eq:Pi_F_U0}
    \Pi^F=\begin{pmatrix}
        \phantom{-}3/4 & 3/4 & -1/4 & -3/4 & 0\\
        \phantom{-}1/4 & 1/4 & \phantom{-}1/4 & -1/4 & 0\\
        -1/2 & 3/2 & \phantom{-}1/2 & \phantom{-}0 & 0 \\
        \phantom{-}0 & 0 & 0 & \phantom{-}1/2 & 0\\
        \phantom{-}0 & 0 & 0 & \phantom{-}0 & 1
    \end{pmatrix},
\end{align}
with the eigenvalues $\lambda^F=(-1/2,1/2,1,1,1)$ and (not normalized) eigenvectors 
\begin{align}
    \mathcal{S}^F=\begin{pmatrix}
    \phantom{-}1 &3 &3 &-1 &0\\
    -1 &1 &1 &\phantom{-}0 &0\\
    \phantom{-}2 &0 &0 &\phantom{-}1 &0\\
    \phantom{-}0 &2 &0 &\phantom{-}0 &0\\
    \phantom{-}0 &0 &0 &\phantom{-}0 &1\\
\end{pmatrix}.
\end{align}
Unlike the iteration in $\Phi$, the channels remain coupled even at $U=0$ and the physical fixed point is unstable, where the eigenvector that corresponds to the unstable eigenvalue couples density, magnetic and singlet channel. Another contrast to the $\Phi$ iteration is that the eigenvalues of $\Pi^F$ for $U\rightarrow\infty$ are $\lambda^F=(4,1/2,1,1,1)$ and therefore the physical fixed point is (for $p<0.5$) stable if $U\rightarrow \infty$.
\begin{figure}[!t]
    \centering
    \includegraphics[width=\textwidth]{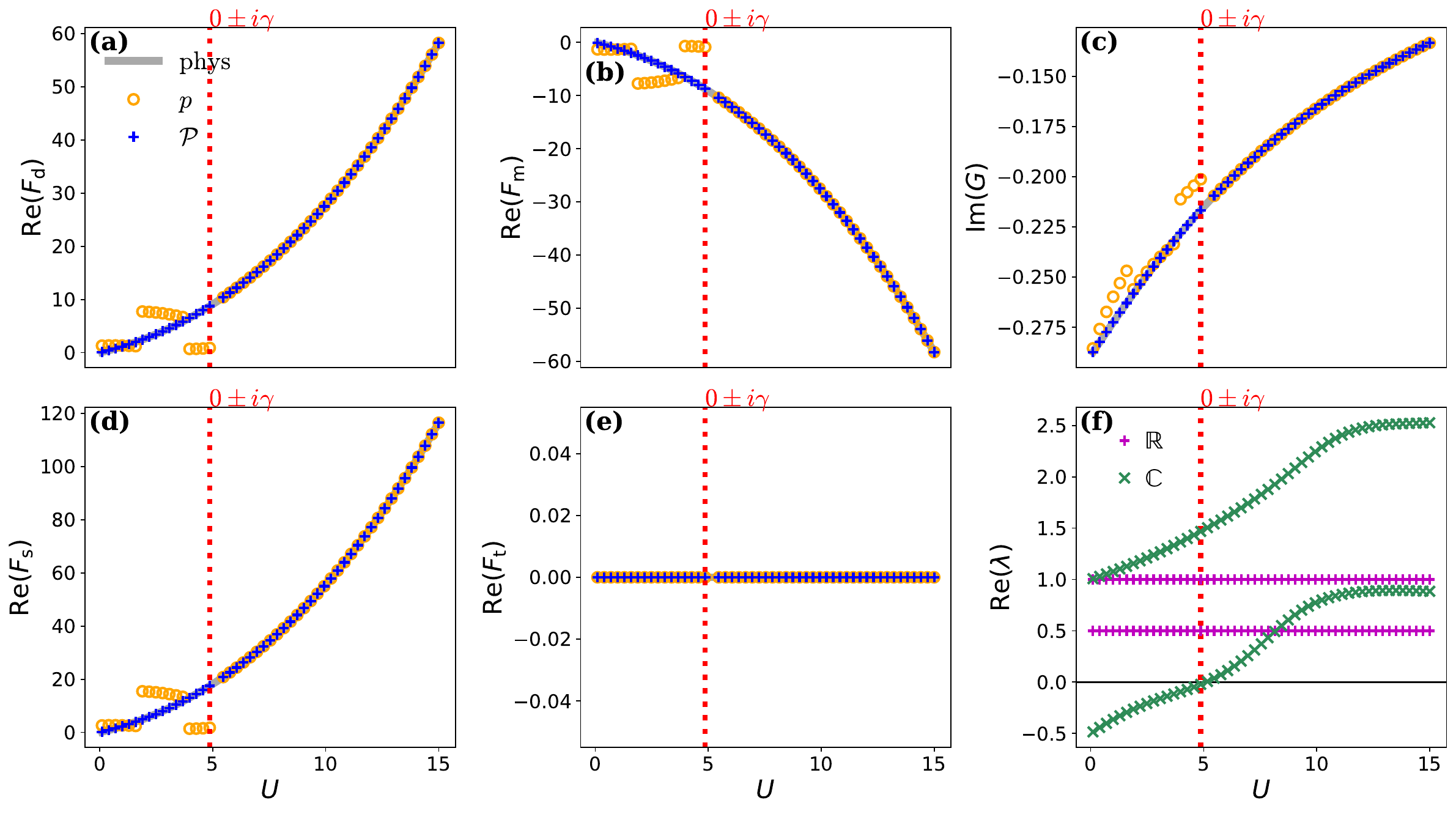}
    \caption{Parquet results for the ZP model analogous to \cref{fig:ZP_F_dmu0} out of ph-symmetry ($\mathrm{Re}(\delta\mu)=1$).}
    \label{fig:ZP_F_dmu1}
\end{figure}

Performing a calculation with the strong-coupling iteration for the ZP model at $\mathrm{Re}(\delta\mu)=1$ and increasing $U$ [\cref{fig:ZP_F_dmu1}], we find, similar to $\delta\mu=0$, that the physical fixed point is unstable for weak coupling and becomes stable for strong coupling when the damped iteration is used, while one always converges to the physical solution when using stabilization method. 
Inspecting the eigenvalues of $\Pi^F$, we find that there is only one instability, which is a complex eigenvalue whose real part crosses zero. 

For the HA, the strong-coupling iteration can be adapted analogously by inverting the BSEs~\eqref{eq:bse} to obtain $\Gamma_r$ as a functional of $F_r$:
\begin{subequations}
\label{eq:Gamma_F_HA}
\begin{align}
    \Gamma^{\nu\nu'\omega}_\text{d}[F,G]&=\left[F_\text{d}\cdot (\mathbb{1}-1/\beta^2\chi_{0,\text{ph}}\cdot F_\text{d})^{-1}\right]^{\nu\nu^\prime\omega},\\
    \Gamma^{\nu\nu'\omega}_\text{m}[F,G]&=\left[F_\text{m}\cdot (\mathbb{1}-1/\beta^2\chi_{0,\text{ph}}\cdot F_\text{m})^{-1}\right]^{\nu\nu^\prime\omega},\\
    \Gamma^{\nu\nu'\omega}_\text{s}[F,G]&=\left[(\mathbb{1}+1/\beta^2\chi_{0,\text{pp}}\cdot F_\text{s})^{-1} \cdot F_\text{s} \right]^{\nu\nu^\prime\omega},\\
    \Gamma^{\nu\nu'\omega}_\text{t}[F,G]&=\left[(\mathbb{1}-1/\beta^2\chi_{0,\text{pp}}\cdot F_\text{t})^{-1} \cdot F_\text{t} \right]^{\nu\nu^\prime\omega},
\end{align}
\end{subequations}
where the inversion and $\cdot$ multiplication are matrix operations defined in the fermionic Matsubara frequency space.
The functional $\Gamma_r$ is then used together with the parquet equations \cite{Thunstroem2018} to obtain the strong-coupling iteration:
\begin{subequations}
\label{eq:strong_coupling_iteration_HA}
\begin{align}
    f^{F,\nu\nu'\omega}_\text{d}&=\frac{1}{2}\left(\Gamma_\mathrm{d}^{\nu\nu'\omega}-\Lambda_\text{d}^{\nu\nu'\omega} \!-\!\tfrac{1}{2} \Gamma_\mathrm{d}^{\nu (\nu+\omega) (\nu'-\nu)}\!-\! \tfrac{3}{2} \Gamma_\mathrm{m}^{\nu (\nu+\omega) (\nu'-\nu)} \!+\! \tfrac{1}{2} \Gamma_\mathrm{s}^{\nu \nu' (-\omega -\nu -\nu')} \!+\! \tfrac{3}{2} \Gamma_\mathrm{t}^{\nu \nu' (-\omega -\nu -\nu')}\right),\\
    f^{F,\nu\nu'\omega}_\text{m}&=\frac{1}{2}\left(\Gamma_\mathrm{m}^{\nu\nu'\omega}-\Lambda_\text{m}^{\nu\nu'\omega} \!-\!\tfrac{1}{2} \Gamma_\mathrm{d}^{\nu (\nu+\omega) (\nu'-\nu)}\!+\! \tfrac{1}{2} \Gamma_\mathrm{m}^{\nu (\nu+\omega) (\nu'-\nu)} \!-\! \tfrac{1}{2} \Gamma_\mathrm{s}^{\nu \nu' (-\omega -\nu -\nu')} \!+\! \tfrac{1}{2} \Gamma_\mathrm{t}^{\nu \nu' (-\omega -\nu -\nu')}\right),\\
    f^{F,\nu\nu'\omega}_\text{s}&=\frac{1}{2}\left(\Gamma_\mathrm{s}^{\nu\nu'\omega}-\Lambda_\text{s}^{\nu\nu'\omega}\!+\!\tfrac{1}{2} \Gamma_\mathrm{d}^{\nu \nu' (-\omega-\nu-\nu')} \!-\! \tfrac{3}{2} \Gamma_\mathrm{m}^{\nu \nu' (-\omega-\nu-\nu')}
    \!+\! \tfrac{1}{2} \Gamma_\mathrm{d}^{\nu (-\nu'-\omega) (\nu' -\nu)} \!-\! \tfrac{3}{2} \Gamma_\mathrm{m}^{\nu (-\nu'-\omega) (\nu' -\nu)}\right),\\
    f^{F,\nu\nu'\omega}_\text{t}&=\frac{1}{2}\left(\Gamma_\mathrm{t}^{\nu\nu'\omega}-\Lambda_\text{t}^{\nu\nu'\omega}\!+\!\tfrac{1}{2} \Gamma_\mathrm{d}^{\nu \nu' (-\omega-\nu-\nu')} \!+\! \tfrac{1}{2} \Gamma_\mathrm{m}^{\nu \nu' (-\omega-\nu-\nu')}
    \!-\! \tfrac{1}{2} \Gamma_\mathrm{d}^{\nu (-\nu'-\omega) (\nu' -\nu)} \!-\! \tfrac{1}{2} \Gamma_\mathrm{m}^{\nu (-\nu'-\omega) (\nu' -\nu)}\right),
\end{align}
\end{subequations}
where the Jacobian is calculated analogously to the previous sections.

\section{Channel-resolved instability analysis}\label{app:decoupled_channels}
\begin{figure}[!t]
    \centering
    \includegraphics[scale=0.5]{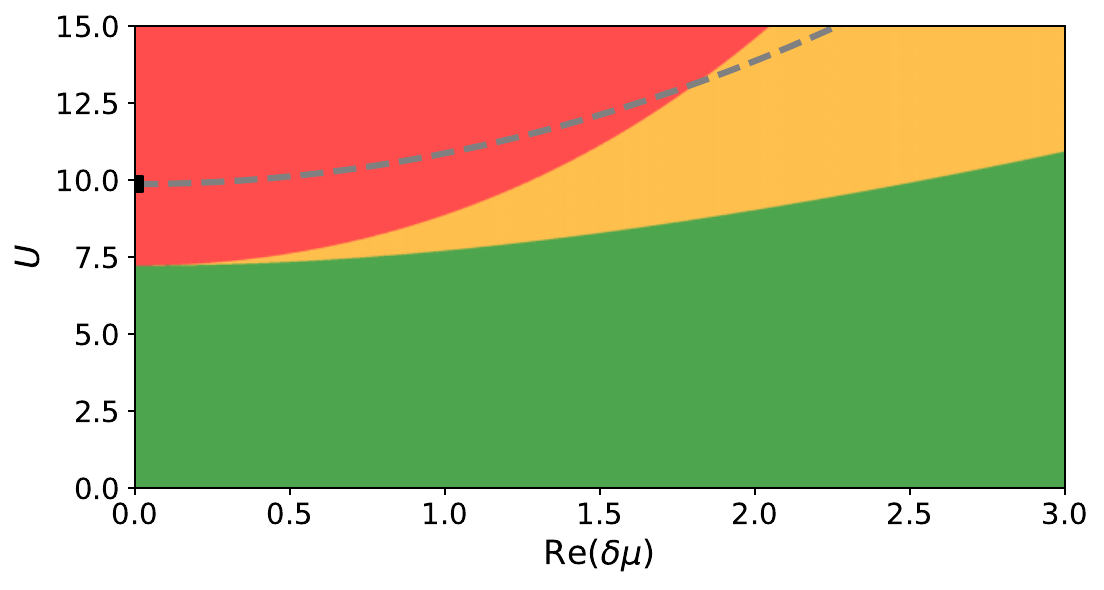}
    \caption{Stability plot for the ZP model if only the diagonal terms of $\Pi$ are considered. The presentation is analogous to \cref{fig:ZP_stability}.}
    \label{fig:ZP_stability_decoupled}
\end{figure}

\begin{figure}[!t]
    \centering
    \includegraphics[scale=0.6]{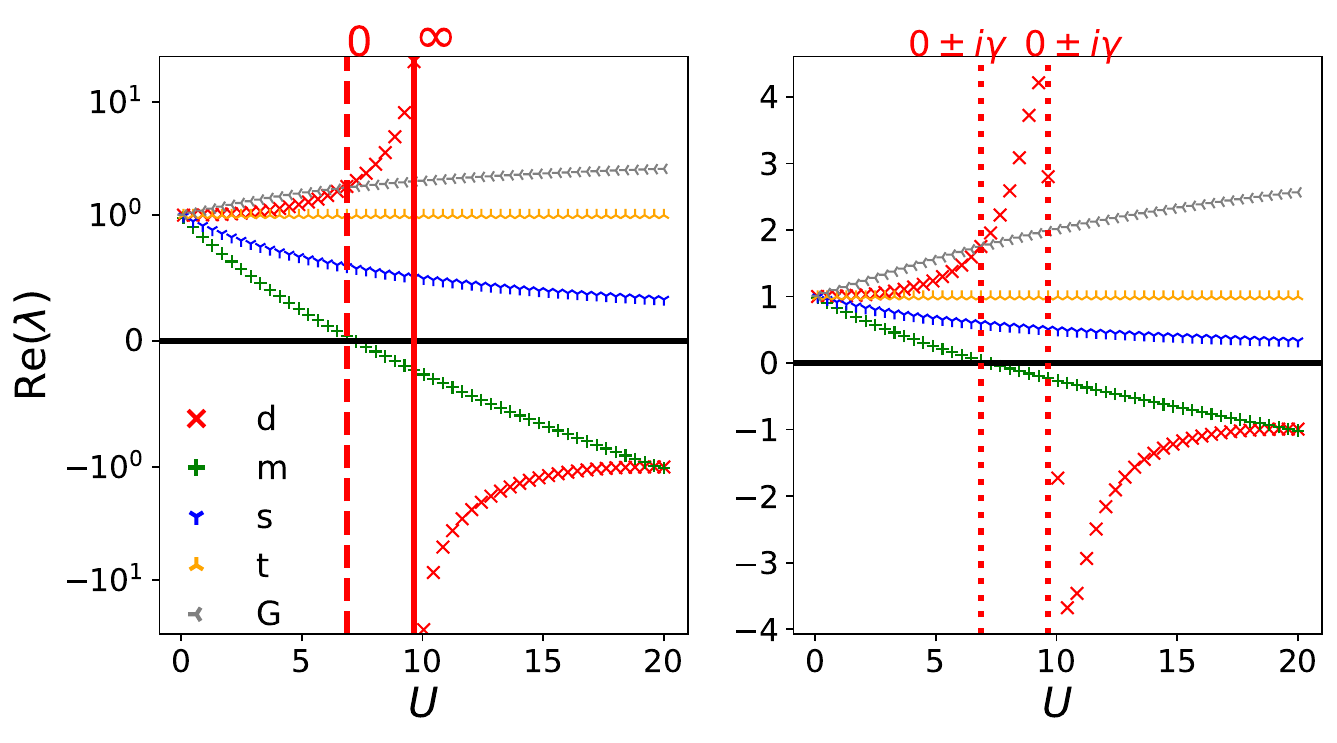}
    \caption{Diagonal terms of $\Pi$ for the ZP model over increasing $U$, once for ph-symmetry [$\mathrm{Re}(\delta\mu)=0$] (left) and once out of ph-symmetry [$\mathrm{Re}(\delta\mu)=1$] (right). The different colors and markers indicate to which channel the eigenvalue belongs.}
    \label{fig:ZP_Pi_decoupled}
\end{figure}

To investigate if an individual channel is responsible for the instability, we now analyze the eigenvalues of $\Pi$, where all channel-off-diagonal terms, i.e., $\Pi_{rr^\prime}$ with $r\neq r^\prime$, are set to zero. In \cref{fig:ZP_stability_decoupled}, we do the same investigation as in \cref{fig:ZP_stability} for the ZP model, but we consider only channel-diagonal terms in $\Pi$. Comparing these two situations, we find that, at ph-symmetry, the instability occurs at weaker coupling for the decoupled channels while it occurs at higher $U$ values when going out of ph-symmetry.
In \cref{fig:ZP_Pi_decoupled}, we show the individual eigenvalues of the decoupled $\Pi$ for the two considered paths $\text{Re}(\delta\mu)=0$ and $\text{Re}(\delta\mu)=1$.
We find that for both paths the number of instabilities is the same as in the coupled case while the $U$ values where they appear are shifted.
For $\text{Re}(\delta\mu)=0$, we find that the instability that stems from the vertex divergence in the density channel is unchanged while the zero crossing stems from the magnetic channel and occurs at lower $U$ values than for the coupled channels. 
At $\text{Re}(\delta\mu)=1$, we find two instabilities, one in the density and one in the magnetic channel, with a complex eigenvalue whose real part goes through zero. Both instabilities occur at different $U$ values as in the coupled case.

\begin{figure}[!t]
    \centering
    \includegraphics[scale=0.5]{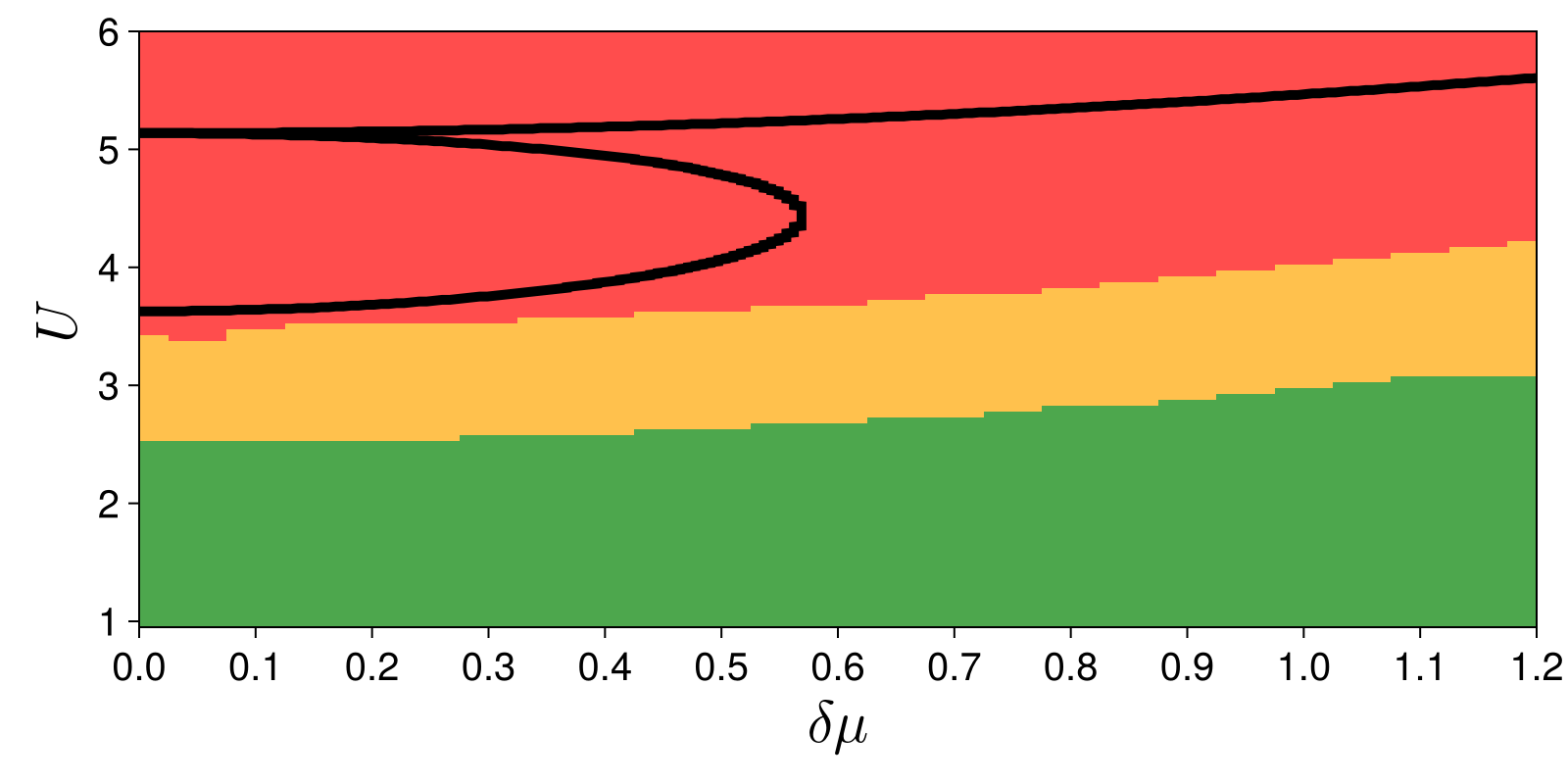}
    \caption{Stability plot for the HA if only the channel-diagonal terms of $\Pi$ are considered. The presentation is analogous to \cref{fig:HA_stability}.}
    \label{fig:HA_stability_decoupled}
\end{figure}

Performing the same investigation for the HA, i.e., taking into account only the channel-diagonal terms in $\Pi$, we find a very similar change as in the ZP model. As it can be seen in \cref{fig:HA_stability_decoupled}, the first instability at half-filling is before the first vertex divergence.
Specifically, by investigating the number of eigenvalues with negative real part in $\Pi$ for each channel for the two paths $\delta\mu=0,1$ we find, similar to the ZP model, that the first instability that occurs before the vertex divergence appears in the magnetic channel. Further, all channels that exhibit a divergence, i.e., density and singlet at $\delta\mu=0$ and only singlet at $\delta\mu=1$, have unstable eigenvalues after their divergence. Moreover, we also find for $\delta\mu=1$ unstable eigenvalues in the density channel that appear in the proximity of the pseudo-divergence of $\chi_{0,\text{ph}}^{-1}\chi_\text{d}$.
\begin{figure}[!t]
    \centering
    \includegraphics[scale=0.5]{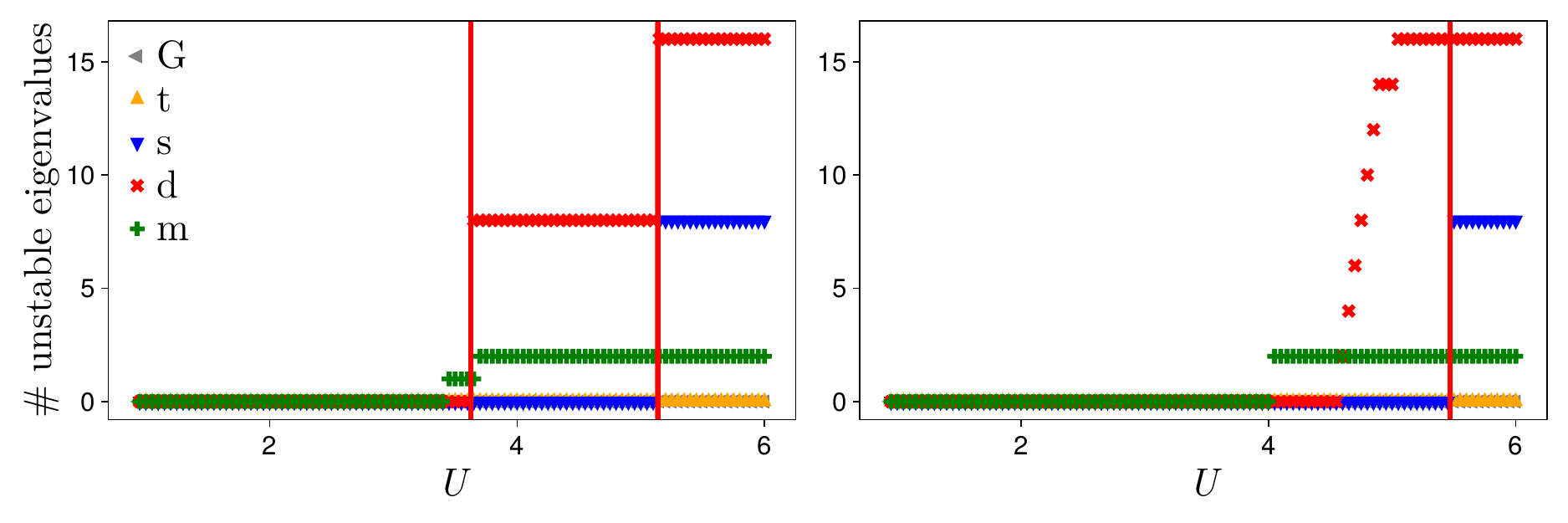}
    \caption{Number of eigenvalues with negative real part in $\Pi$ for the HA ($N_f=12, N_b=13$) if only the channel diagonal terms are considered over increasing $U$ for $\delta\mu=0$ (left) and $\delta\mu=1$ (right). The different colors and markers indicate to which channel the eigenvalue belongs, and the vertex divergences are marked by red solid lines.}
    \label{fig:HA_neg_Pi_decoupled}
\end{figure}

\begin{figure}[!t]
    \centering
    \includegraphics[scale=0.5]{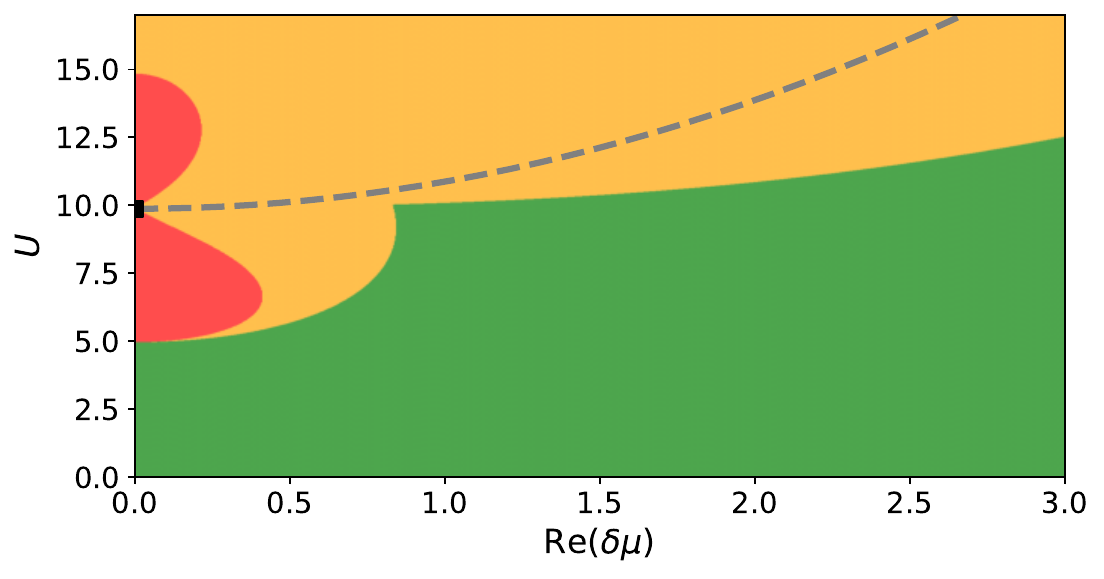}
    \caption{Stability plot for the ZP model with the strong-coupling parquet iteration if only the diagonal terms of $\Pi^F$ are considered. The presentation is analogous to \cref{fig:ZP_stability}.}
    \label{fig:ZP_F_stability_decoupled}
\end{figure}

\begin{figure}[!t]
    \centering
    \includegraphics[scale=0.6]{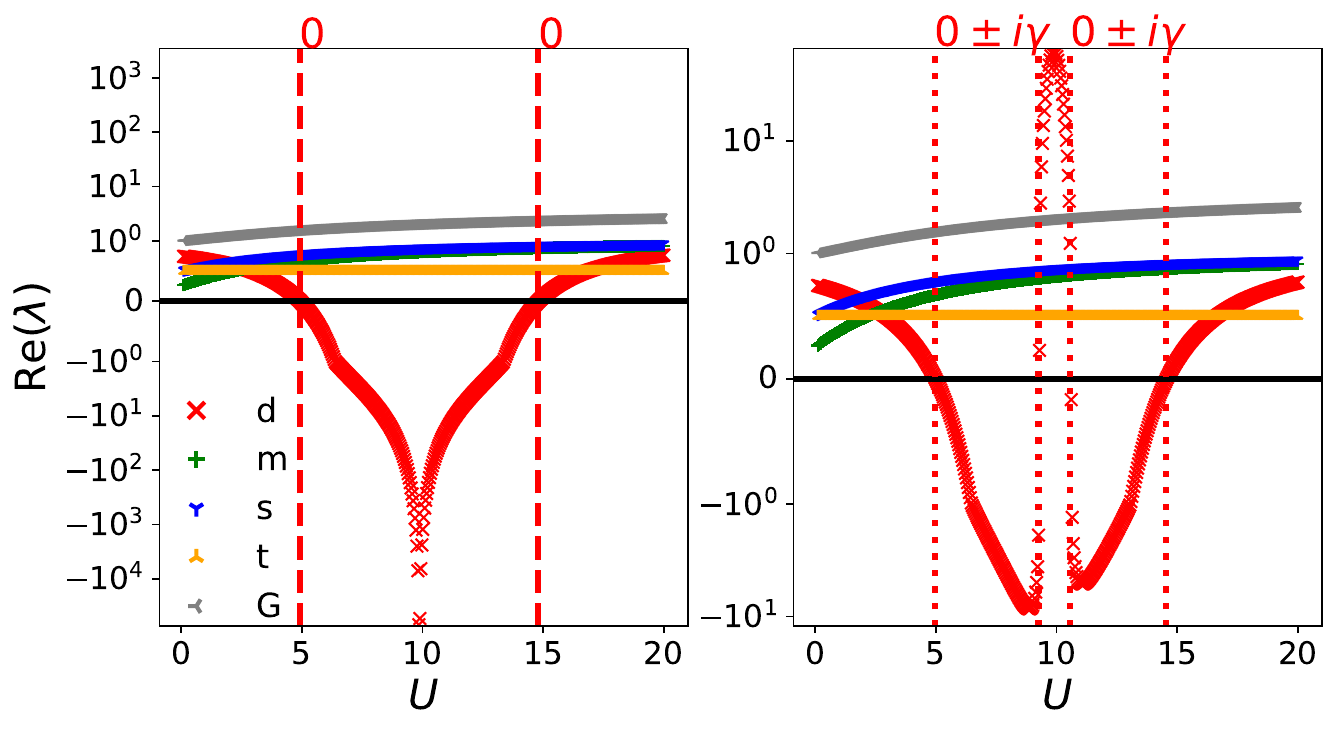}
    \caption{Diagonal terms of $\Pi^F$ for the ZP model over increasing $U$, once for $\mathrm{Re}(\delta\mu)=0$ (left) and once for $\mathrm{Re}(\delta\mu)=0.1$ (right). The different colors and markers indicate to which channel the eigenvalue belongs.}
    \label{fig:ZP_Pi_F_decoupled}
\end{figure}

Finally, we investigate the stability of the strong-coupling iteration for the decoupled channels. As it can be seen from \cref{fig:ZP_F_stability_decoupled}, the instability at weak interaction vanishes when only decoupled channels are considered. Only a small unstable region around the divergence in the density channel remains. In \cref{fig:ZP_Pi_F_decoupled}, it can be seen that all instabilities that do not originate from channel coupling originate from the density channel.

\section{Extended results for the Hubbard atom}\label{app:extended_results_HA}
\subsection{Negative eigenvalues} \label{app:n-eigenvalues}
\begin{figure}
    \centering
    \includegraphics[width=0.95\linewidth]{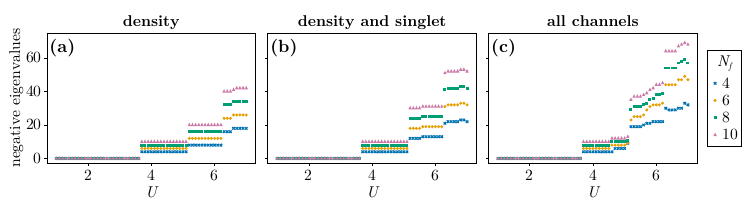}
    \caption{Number of negative eigenvalues of $\Pi=\mathbb{1}-J$ for different sub-Jacobians and different fermionic frequency boxes $N_f$ shown as a function of $U$ with $N_b=7$ for the HA at half-filling. (a) Only the density-density component of the full Jacobian is diagonalized. At the first vertex divergence ($U=3.628$), the number of negative eigenvalues jumps to $N_f$ and, after the second divergence ($U=5.137$), to $2N_f$. (b) The negative eigenvalues of the density and singlet channel sub-Jacobian with jumps to $N_f$ (first divergence) and to $3N_f$ (second divergence) and of the full Jacobian (c) are shown.}
    \label{fig:eigenvalues-Nf}
\end{figure}

\begin{figure}
    \centering
    \includegraphics[width=0.95\linewidth]{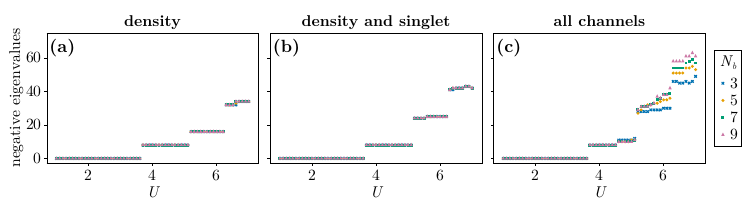}
    \caption{Number of negative eigenvalues of $\Pi=\mathbb{1}-J$ for different sub-Jacobians and different bosonic frequency boxes $N_b$ shown as a function of $U$ with $N_f=8$ for the HA at half-filling. (a) Only the density-density component of the full Jacobian is diagonalized. (b) The negative eigenvalues of the density and singlet channel sub-Jacobian and of the full Jacobian (c) are shown.}
    \label{fig:eigenvalues-Nb}
\end{figure}

In this section, we investigate the dependence of the number of negative eigenvalues in the $\Pi$ matrix (number of eigenvalues of the Jacobian with real part larger than 1) on the size of the fermionic and bosonic frequency boxes. The number of negative eigenvalues strongly depends on the number of fermionic frequencies included in the calculation [see Fig.~\ref{fig:eigenvalues-Nf}(c)], while it only weakly depends on the size of the bosonic frequency box [see Fig.~\ref{fig:eigenvalues-Nb}(c)] starting at larger values of $U$. This poses a problem since this means that, even though the unstable eigendirections originate from the Matsubara frequencies with small absolute value, we cannot treat these instabilities via a Jacobian on a smaller Matsubara grid than the actual parquet iteration is performed on. This is in contrast to the findings for the self-consistent perturbation theory in Ref.~\cite{Essl2026}. In the following, we will investigate the origin of this difference more closely.
Starting with the dependence on the fermionic frequency, we find that the number of negative eigenvalues matches precisely the size of the fermionic frequency box $N_f$ after the first divergence line ($U=3.628$). Moreover, after the second divergence line around $U=5.137$, there seems to be a similar pattern. 
To further analyze this pattern, we limit ourselves to the investigation of the density-density component of the Jacobian ($\frac{\delta f_\mathrm{d}^{\nu \nu' \omega}}{\delta \Phi_\mathrm{d}^{\tilde \nu \tilde \nu' \tilde \omega}}$) shown in panel~(a). Diagonalizing the corresponding sub-matrix of $\Pi$ leads to $N_f$ negative eigenvalues after the first, $2N_f$ negative eigenvalues after the second, and $4 N_f$ negative eigenvalues after the third divergence line (at $U=6.283$, for $\omega=\frac{2\pi}{\beta}$). Including the singlet channel in our investigation, we diagonalize the $\Pi$ matrix computed via the sub-Jacobian 
\begin{align}
    J = \begin{pmatrix}
        \frac{\delta f_\mathrm{d}^{\nu \nu' \omega}}{\delta \Phi_\mathrm{d}^{\tilde \nu \tilde \nu' \tilde \omega}} &\ \frac{\delta f_\mathrm{d}^{\nu \nu' \omega}}{\delta \Phi_\mathrm{s}^{\tilde \nu \tilde \nu' \tilde \omega}} \\
        \frac{\delta f_\mathrm{s}^{\nu \nu' \omega}}{\delta \Phi_\mathrm{d}^{\tilde \nu \tilde \nu' \tilde \omega}} &\ \frac{\delta f_\mathrm{s}^{\nu \nu' \omega}}{\delta \Phi_\mathrm{s}^{\tilde \nu \tilde \nu' \tilde \omega}},
    \end{pmatrix}
\end{align}
only including the d and s channels, and analyze the respective eigenvalues in panel~(b). Here, the number of negative eigenvalues after the first divergence line coincides with the previous case ($N_f$), but after the second divergence line the number of negative eigenvalues initially jumps to $3N_f$. This is clearly no coincidence and can be understood in the following way. In the density-density component of the Jacobian [\cref{eq:J_dd}], the following term can be found
\begin{align}
    -\Gamma_\mathrm{d}^{\nu \nu_1 \omega} \chi_{0,\text{ph}}^{\nu_1 \tilde \nu \omega}  \delta_{\nu', \tilde \nu'} \delta_{\omega, \tilde \omega}.
\end{align}
At a vertex divergence in $\Gamma_\mathrm{d}$, in $(\Gamma_\mathrm{d} \chi_{0,\text{ph}})^{\nu \tilde \nu \omega}$ a single eigenvalue becomes larger than one leading to an additional negative eigenvalue in $\mathbb{1}-J$. Now, the factor $\delta_{\nu', \tilde \nu'}$ basically produces a convolution of this negative eigenvalue leading to precisely $N_f$ additional eigenvalues. A similar analysis can be conducted for the singlet channel. Therefore, at the first vertex divergence (only in the density channel), $N_f$ negative eigenvalues are obtained, while at the second divergence (in $\Phi_\mathrm{d}$ and $\Phi_\mathrm{s}$) $2N_f$ additional negative eigenvalues are obtained ($N_f$ for the density and singlet components each). Since the high-dimensional Jacobian has many other components, these are only the dominant directions that mix with other components leading to deviations from this strict behavior past the second vertex divergence or even past the first divergence in panel~(c), where the Jacobian of the full parquet iteration is considered. Still, this gives an intuitive picture, why the number of negative eigenvalues is strongly dependent on the size of the fermionic frequency box.

In contrast to this behavior, the number of negative eigenvalues does not depend on $N_b$ for the sub-Jacobians with only the density [see panel~(a)] and density and singlet components [see panel~(b)] as can be seen in Fig.~\ref{fig:eigenvalues-Nb}. For the full parquet iteration, some dependence on $N_b$ can be observed in panel~(c). 

\subsection{Condition number} \label{app:condition-number}
\begin{figure}
    \centering
    \includegraphics[width=0.95\linewidth]{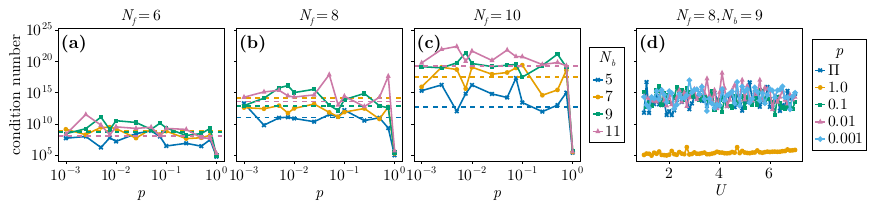}
    \caption{(a)--(c) Condition number of the eigenbasis $\mathcal{S}$ of $\Pi$ and $\mathbb{1}-p \Pi$ for various $N_f$ and $N_b$ as a function of $p$ for the HA at half-filling. The dashed lines indicate the condition number of $\mathcal{S}$ of $\Pi$. (d) Dependence of the condition number for various $p$ on $U$ for $N_f=8,N_b=9$.}
    \label{fig:condition-number}
\end{figure}

Here, we investigate the condition number of the eigenbasis $\mathcal{S}$ of the $\Pi$ matrix and $\mathbb{1}-p \Pi$. In panels~(a)--(c) in Fig.~\ref{fig:condition-number}, the condition number is shown for $U=4.0$, $\beta=1.0$, and half-filling for various sizes of the fermionic (and bosonic) frequency box $N_f$ ($N_b$) as a function of the damping $p$. The dashed lines indicate the condition number of the eigenbasis $\mathcal{S}$ of $\Pi$. It can be seen that the condition number strongly increases with the size of the fermionic frequency box, while for $N_b>N_f$ the size of the bosonic box does not seem to have a strong impact. Interestingly, the condition numbers for the eigenbasis of $\Pi$ and $\mathbb{1}-p \Pi$ seem to be very similarly independent of $p$, except where $p=1.0$. Then the condition number becomes much smaller in comparison. This becomes more obvious in panel~(d), where the condition number is shown for various values of $p$ as a function of $U$ for a fixed frequency box ($N_f=8,N_b=9$). Besides the much smaller condition number at $p=1.0$, no clear dependence on $U$ can be observed.
Why is this analysis important? Following Sec.~\ref{sec:sot}, the stabilization matrix is computed via $\mathcal{P}={\cal S}\mathcal{D} {\cal S}^{-1}$. Therefore, the eigenbasis $\mathcal{S}$ needs to be inverted for the computation of $\mathcal{P}$. If the condition number of $\mathcal{S}$ is very large, then the inversion becomes unstable leading to numerical results with $\mathcal{S} \mathcal{S}^{-1} \neq \mathbb{1}$. However, since $\mathcal{S}$ should be identical for $\Pi$ and $\mathbb{1}-p \Pi$ independently of $p$, we can choose the one with the smallest condition number to increase stability. Let us note that at half-filling the eigenbasis of $\mathbb{1}- \Pi$ has the lowest condition number, which can change, for example in the case out of half-filling.

A way to mitigate the problem of a large condition number using the Schur decomposition is presented in Ref.~\cite{Gievers2026Instabilities}.

\subsection{Additional results} \label{app:HA-additional-results}
Here, we extend Fig.~\ref{fig:HA-Nf-12-Nb-13-U-7} up to $U=11$. It can be seen that, after the third vertex divergence (red solid lines), the scheme does not converge as well as before, which is probably caused by the rapid continuous increase in the number of negative eigenvalues also between the divergences leading to a decrease of the convergence radius. Still, for some points, convergence up to $U=11$ can be reached. Therefore, further increasing the damping $p$ and/or $\Delta U_{\mathrm{init}}$ is expected to lead to better convergence in this challenging regime. We will further elaborate on this in App.~\ref{app:alpha-init}. 
\begin{figure}
    \centering
    \includegraphics[width=\linewidth]{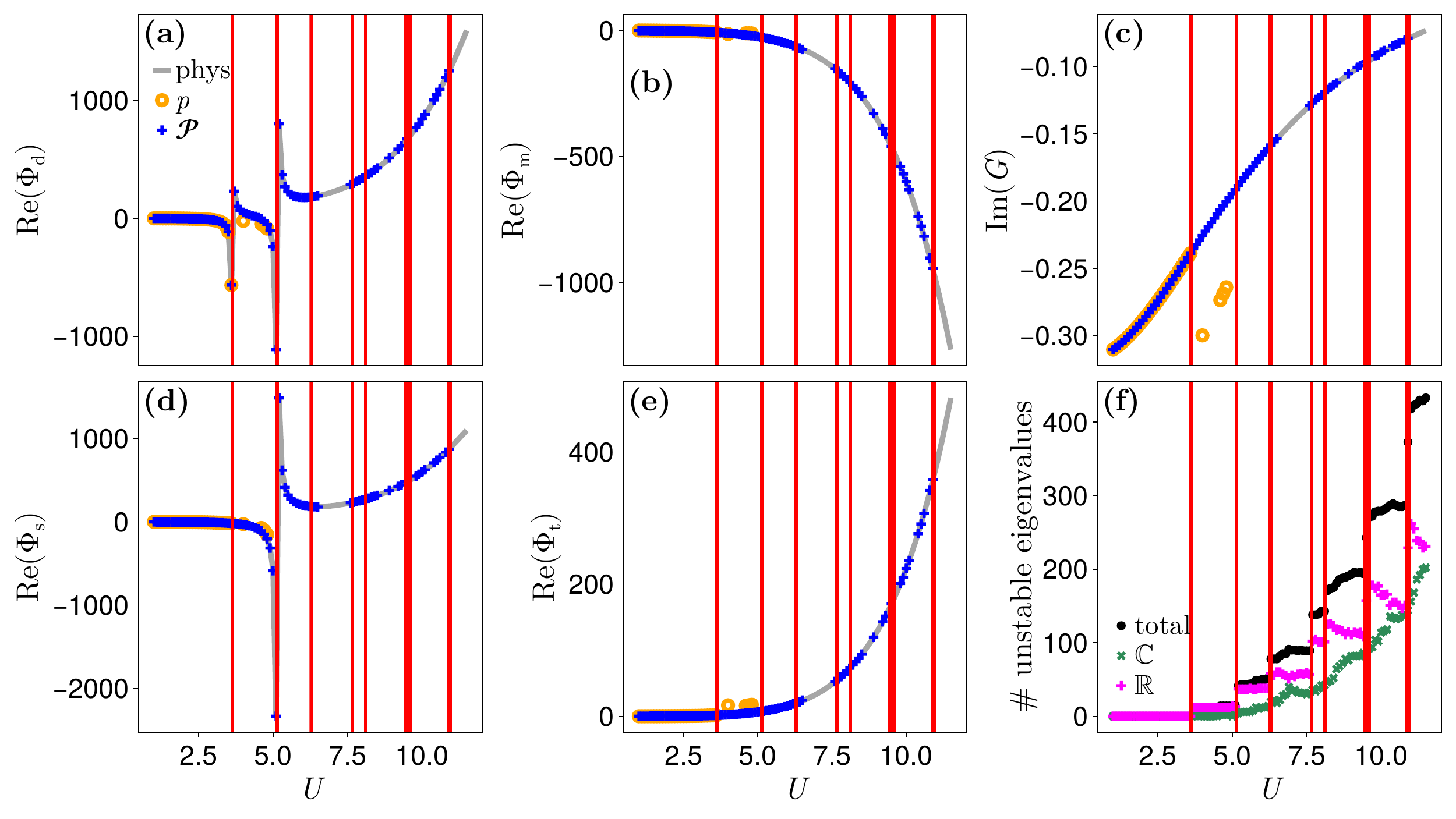}
    \caption{Parquet results for the HA as in \cref{fig:HA-Nf-12-Nb-13-U-7} but for a higher $U$ range to show how far we can converge with the stabilized iteration.}
    \label{fig:HA-Nf-12-Nb-13}
\end{figure}

\subsection{Influence of damping and input} \label{app:alpha-init}
\begin{figure}
    \centering
    \includegraphics[width=0.95\linewidth]{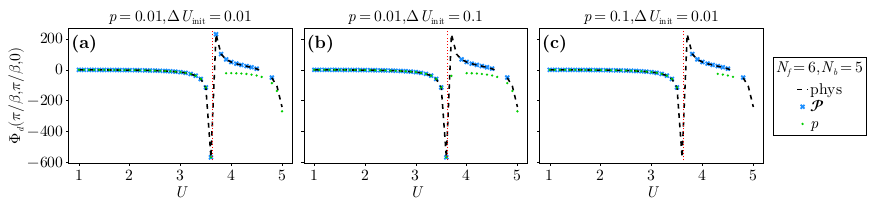}
    \caption{Comparison  for different damping values and initial inputs of full iterative parquet with (blue crosses) and without the stabilization method (green dots) for the HA at half-filling with $\beta=1$ and $N_f=6,N_b=5$. The component $\Phi_\mathrm{d}(\pi/\beta,\pi/\beta,0)$ is shown with respect to $U$. (a) $p=0.01, \Delta U_{\mathrm{init}}=0.01$ converges also very close to the vertex divergence. The cases (b) $p=0.01, \Delta U_{\mathrm{init}}=0.1$ and (c) $p=0.1, \Delta U_{\mathrm{init}}=0.01$ show less converged points in the vicinity of the vertex divergence.}
    \label{fig:sot-alpha-init-comparison}
\end{figure}

In the main text and the previous sections, it was mentioned that the stabilization method ensures local stability of the physical fixed point. However, since the parquet equations are high-dimensional coupled nonlinear equations, local stability does not imply global stability of the fixed point, meaning that convergence can also be dependent on the distance of the initial input from the fixed point. Moreover, the choice of damping $p$ is not only relevant for local stability, but can also change the global convergence radius. In Fig.~\ref{fig:sot-alpha-init-comparison}, we compare uniform damping values of $p=0.01,0.1$ and different initial inputs $\Delta U_{\mathrm{init}} = 0.01, 0.1$ with $U_{\mathrm{init}} = U - \Delta U_{\mathrm{init}}$ and $N_f=6$, $N_b=5$. In panel~(a), we show the case of $p=\Delta U_{\mathrm{init}}=0.01$, where the stabilization method converges very well, even in close vicinity of the vertex divergences. In panels~(b) and (c), the cases $p=0.01$, $\Delta U_{\mathrm{init}}=0.1$ and $p=0.1$, $\Delta U_{\mathrm{init}}=0.01$ are shown. In both of these cases, fewer points around the vertex divergence converge than in panel~(a). In panel~(b), convergence very close up to the divergence is possible, but after the divergence convergence fails for multiple points. In contrast, convergence only fails for a single point before and after the vertex divergence in panel~(c). 

\subsection{Frequency boxes} \label{app:box-size}
\begin{figure}
    \centering
    \includegraphics[width=0.95\linewidth]{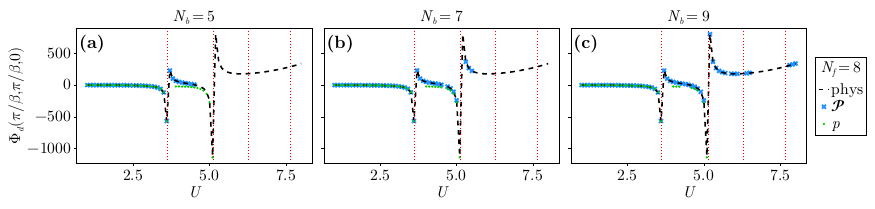}
    \caption{Comparison of full iterative parquet with (blue crosses) and without the stabilization method (green dots) for the HA at half-filling with $\beta=1$ and $N_f=8$ for different values of $N_b$. The component $\Phi_\mathrm{d}(\pi/\beta,\pi/\beta,0)$ is shown with respect to $U$. The stabilization method shows convergence past the first [$N_b=5$ in (a)], the second [$N_b=7$ in (b)] and fourth vertex divergence [$N_b=9$ in (c)].}
    \label{fig:sot-frequency-box-comparison}
\end{figure}

Due to the frequency shifts and the dependence of the vertex divergences on specific fermionic or bosonic frequencies, a minimum size of the regarding frequency boxes is required to converge past the vertex divergences. Past the first local divergence ($\omega=0,\nu=\pm \frac{\pi}{\beta}$) \cite{Thunstroem2018}, a frequency box of size $N_f=6, N_b=5$ is sufficient for convergence as can be seen in Fig.~\ref{fig:sot-alpha-init-comparison}. Increasing the fermionic box to $N_f=8$ in Fig.~\ref{fig:sot-frequency-box-comparison}(a) does not change the observed convergence behavior. Past the first divergence line, the stabilization method converges to the physical solution, but before the second divergence (first global for $\omega=0$ in singlet and density channels) is encountered, the method fails to converge. However, when increasing the bosonic frequency box to $N_b=7$ in panel~(b), the stabilization method also converges past this divergence, but fails to overcome the third vertex divergence (local, $\omega=\pm \frac{2\pi}{\beta}, \nu = \pm \frac{\pi}{\beta}$). By further increasing the bosonic box to $N_b=9$, we can also find converging points past the third and fourth (global, $\omega=\pm \frac{2\pi}{\beta}$) vertex divergence. In the case of the HA at half-filling, we find that the size of the bosonic box plays the determining role in whether the stabilization method can be converged past the first couple of vertex divergences, while increasing the size of the fermionic frequency box does not seem to have an influence. We suggest that this is primarily due to the fact that the first eight (four local and global each) divergences occur for the smallest fermionic frequencies $\nu=\pm \frac{\pi}{\beta}$ at increasing values of $\omega$ ($=0,\pm \frac{2\pi}{\beta},\pm \frac{4\pi}{\beta}, \pm \frac{6\pi}{\beta}$) \cite{Thunstroem2018} meaning that due to the frequency shifts in the parquet equations, information of vertex divergences occurring at for example $\omega \neq 0$ can be altered or lost. Therefore, larger bosonic frequency boxes are required to keep all the relevant information about the divergences during the channel or frequency transformations.

\bibliography{Bibliography.bib}

\end{document}